\documentclass[14pt]{article}
\pdfoutput=1
\usepackage{amsmath,amssymb,amsfonts,amsthm,bm,bbm,cancel,wasysym}
\usepackage{epsfig,graphics,graphicx,epstopdf,caption,subcaption}
\graphicspath{{Charts/}}
\usepackage{array,booktabs,colortbl,colordvi,multirow}
\usepackage{colordvi,color,xcolor}
\usepackage{hyperref}
\usepackage{rotating}
\usepackage{comment}
\usepackage{feynmf}

\usepackage{verbatim}
\usepackage{cite}
\usepackage{amsmath}
\usepackage{physics}
\usepackage{mathtools}
\usepackage{setspace}
\usepackage{url}
\usepackage[percent]{overpic}
\usepackage{slashed}
\usepackage{authblk}
\usepackage{xspace}
\usepackage{fullpage}
\usepackage[numbers,sort&compress]{natbib}
\usepackage{multirow}
\usepackage{array}
\newcolumntype{P}[1]{>{\centering\arraybackslash}p{#1}}

\def\ie{{\it i.e.}}
\def\eg{{\it e.g.}}

\newskip\zatskip \zatskip=0pt plus0pt minus0pt
\def\matth{\mathsurround=0pt}

\def\gsim{\mathrel{\mathpalette\atversim>}}
\def\atversim#1#2{\lower0.7ex\vbox{\baselineskip\zatskip\lineskip\zatskip
  \lineskiplimit 0pt\ialign{$\matth#1\hfil##\hfil$\crcr#2\crcr\sim\crcr}}}

\newif\ifdiagrams
\diagramstrue
\ifdiagrams

\else
  \excludecomment{fmffile}
\fi

\begin{document}


\begin{flushright}
\today
\end{flushright}
\vspace*{5mm}

\renewcommand{\thefootnote}{\fnsymbol{footnote}}
\setcounter{footnote}{1}
\begin{center}

{\Large {\bf Lepton Flavor Portal Matter}}\\

\vspace*{0.75cm}

{\bf George N. Wojcik~\footnote{gwojcik@wisc.edu}, Lisa L.~Everett~\footnote{leverett@wisc.edu}, Shu Tian Eu~\footnote{eu@wisc.edu}, 
 Ricardo Ximenes~\footnote{dossantosxim@wisc.edu}}

\vspace{0.5cm}

{Department of Physics, University of Wisconsin-Madison, Madison, WI 53706, USA}

\end{center}
\vspace{.5cm}

\begin{abstract}
 
\noindent  
The paradigm of portal matter represents a well-motivated extension to models with kinetic mixing/vector portal dark matter. In previous work, we constructed a simple leptonic portal matter model in which the portal matter fields could mediate a new physics correction to the anomalous magnetic moment of the muon consistent with the observed discrepancy between the measured value for this quantity and the SM prediction. Here, we present a version of this mechanism by constructing a model with an extended dark gauge sector in which SM and portal matter fields exist as members of the same dark gauge multiplets, which provides a natural extension of simple portal matter models. We find a rich phenomenology in this extended model, including nontrivial novel characteristics that do not appear in our earlier minimal construction, and discuss current experimental constraints and future prospects for this model. We find that a multi-TeV muon collider has excellent prospects for constraining or measuring the crucial parameters of this model.
\end{abstract}

\renewcommand{\thefootnote}{\arabic{footnote}}
\setcounter{footnote}{0}
\thispagestyle{empty}
\vfill
\newpage
\setcounter{page}{1}

\section{Introduction}
Estimated to compose roughly 80\% of the matter content of the universe, the presence of dark matter (DM) is clearly indicated through a plethora of cosmological and astrophysical data. However, DM's characteristics remain a mystery apart from its gravitational interactions, leading to a variety of extensions of the Standard Model (SM) featuring DM candidates \cite{Arcadi:2017kky, Kawasaki:2013ae,Graham:2015ouw,Battaglieri:2017aum,Alexander:2016aln, Bernal:2017kxu, Ferreira:2020fam}. As the parameter space for WIMP DM \cite{Arcadi:2017kky} continues to be constrained by experimental searches\footnote{See, however \cite{Leane:2018kjk}.}, other thermal DM paradigms have proliferated that can readily avoid these constraints and easily open up new viable parameter spaces for dark matter. For dark matter in the sub-GeV mass range (which might avoid stringent direct detection constraints from nuclear recoils \cite{PandaX-II:2020oim,XENON:2018voc,LUX:2016ggv,SuperCDMS:2017mbc,DarkSide:2018kuk}), a popular framework is that of vector portal/kinetic mixing \cite{Holdom:1985ag,Holdom:1986eq,Pospelov:2007mp,Izaguirre:2015yja,Essig:2013lka,Curtin:2014cca}. In the simplest realization of this setup, the SM gauge group is augmented by a local Abelian $U(1)_D$ symmetry, under which an SM singlet DM candidate is charged, while the SM particle content is entirely neutral under the new group. Contact between the dark sector and the SM occurs via small kinetic mixing between $U(1)_D$ and $U(1)_Y$, the SM hypercharge, via a term in the action
\begin{align}\label{eq:EpsilonDef}
    \frac{\epsilon}{2 c_w} B^{\mu \nu} X_{\mu \nu},
\end{align}
where $B^{\mu \nu}$ is the usual SM hypercharge field strength tensor, $X_{\mu \nu}$ is the field strength tensor for the dark photon field, and $c_w$ is the cosine of the Weinberg angle. For a massive dark photon (where this mass might be achieved by a Higgs or Stuckelberg mechanism), this mixing then leads to a coupling between the dark photon and SM fields with non-zero hypercharge proportional to $\epsilon$. For a sub-GeV DM candidate and a dark photon of comparable mass, this model will reproduce the \emph{Planck} \cite{Planck:2018vyg} observation of the DM relic abundance for $\epsilon \sim 10^{-(3-5)}$, depending on model specifics.

As noted by Holdom \cite{Holdom:1985ag,Holdom:1986eq}, finite and calculable kinetic mixing of this magnitude can be generated at one loop if there exist additional fields in the theory which are charged under both $U(1)_D$ and $U(1)_Y$, and the $U(1)_D$ and $U(1)_Y$ charges of the additional fields are such that the ultraviolet divergence from each field's loop contribution is cancelled by the others. Recently there has been interest in the theory and phenomenology of these ``portal matter'' fields \cite{Rizzo:2018vlb,Rueter:2019wdf,Rueter:2020qhf,
Wojcik:2020wgm,Rizzo:2022qan,Wojcik:2022rtk,Rizzo:2022jti,Rizzo:2022lpm,Carvunis:2022yur,Wojcik:2022woa,Rizzo:2023qbj}, in scenarios in which they are light enough to be kinematically accessible at present or near-future experiments. It was noted in \cite{Rizzo:2018vlb} that if the portal matter is fermionic, the most phenomenologically viable scenario would be for the portal matter to be charged under $U(1)_D$, have identical SM quantum numbers to some SM field, and be vector-like under the SM gauge group and $U(1)_D$. Such fields have distinctive experimental signatures in collider experiments \cite{Verma:2022nyd,Guedes:2021oqx}, dominantly decaying into highly boosted jets (for QCD-charged portal matter) or leptons (for leptonic portal matter), plus a dark photon or dark Higgs, in contrast to channels featuring the emission of an electroweak boson that dominate vector-like fermion decay widths in scenarios without the $U(1)_D$ symmetry. Furthermore, these new portal matter fields may play a role in a variety of other unresolved questions in physics, such as the recently-observed $W$ mass anomaly \cite{Rizzo:2022jti,CDF:2022hxs}. 

One intriguing possibility is that the portal matter is purely leptonic in nature. As a concrete example, in \cite{Wojcik:2022woa} we suggested a minimal portal matter construction in which loops of portal matter fields with dark photons and dark Higgs bosons can account for the discrepancy between the experimental measurement of the anomalous magnetic moment of the muon \cite{Muong-2:2021ojo,Muong-2:2006rrc} and the theoretical expectation \cite{Aoyama:2020ynm,Aoyama:2012wk,Aoyama:2019ryr,Czarnecki:2002nt,Gnendiger:2013pva,Davier:2017zfy,Keshavarzi:2018mgv,Colangelo:2018mtw,Hoferichter:2019mqg,Davier:2019can,Keshavarzi:2019abf,Kurz:2014wya,Melnikov:2003xd,Masjuan:2017tvw,Colangelo:2017fiz,Hoferichter:2018kwz,Gerardin:2019vio,Bijnens:2019ghy,Colangelo:2019uex,Blum:2019ugy,Colangelo:2014qya},
\begin{align}\label{eq:delta-a}
    \Delta a_\mu = a_\mu^{\textrm{exp}}-a_\mu^{\textrm{SM}} =  (251 \pm 59) \times 10^{-11}.
\end{align}
In \cite{Wojcik:2022woa} two pairs of leptonic portal matter fields (weak isospin singlets and doublets) were introduced. The weak isospin doublet and singlet fields are mixed by a Yukawa coupling to the SM Higgs field, and the dominant contribution to $\Delta a_\mu$ is directly proportional to this coupling, $y_{LE}$. Somewhat remarkably, the beyond Standard Model (BSM) contribution to $\Delta a_\mu$ is to an excellent approximation \emph{independent} of the details of the sub-GeV DM sector, and \emph{only} dependent on the portal matter masses and their couplings to the SM and dark Higgs fields. For portal matter masses of $\sim 1 \; \textrm{TeV}$, the observed $\Delta a_\mu$ can be accommodated for intriguingly small values of $y_{LE}$, roughly on par with the $b$ quark or $\tau$ lepton Yukawa couplings, and for larger $y_{LE} \sim O(1)$, portal matter masses of several TeV 
are potentially permitted.

While the minimality of the leptonic portal matter construction of \cite{Wojcik:2022woa} is in some respects attractive, like many minimal portal matter models, it suggests the possibility of a UV extension. As noted in \cite{Rizzo:2018vlb}, the $U(1)_D$ charge assignments in a minimal realization of the portal matter paradigm are essentially arbitrarily chosen to satisfy the need for finite and calculable kinetic mixing, and the fact that portal matter fields share quantum numbers with SM fields immediately suggests that both portal and SM matter might be embedded in multiplets of a larger dark sector gauge group, which, if it is semisimple, will automatically yield finite and calculable kinetic mixing. Therefore a number of more ``UV-complete'' models of portal matter have recently been developed to explore the model building possibilities of these types of  constructions, such as \cite{Rueter:2019wdf,Rizzo:2022lpm,Wojcik:2020wgm,Wojcik:2022rtk}. 

This paper expands the work of  \cite{Wojcik:2022woa} in this direction by proposing an extended dark sector gauge group. Noting that the necessary $y_{LE}$ values to account for the observed $\Delta a_\mu$ in our minimal construction can be of the same magnitude as the $\tau$ Yukawa couplings, and that by restricting our model to the second and third generations we can avoid stringent precision constraints on new physics couplings to electrons, we choose as our semisimple extended dark sector the group $SU(2)_A \times SU(2)_B$ and associate it with a flavor symmetry in the second and third lepton generations. This extended construction possesses several appealing aspects: it is anomaly-free, and, because the dark gauge group is semi-simple, kinetic mixing between any $U(1)$ subgroups of it (including the group $U(1)_D$ which we identify with the group of the same name in minimal portal matter models) and the SM are finite and calculable. Furthermore, in contrast to the case of \cite{Wojcik:2022woa}, the multiplet structure we choose relates some of the Higgs Yukawa couplings between isospin singlet and doublet portal matter directly with observed SM lepton masses. Constraints on flavor-changing currents are significantly mitigated with the addition of a $Z_2$ global discrete symmetry to the theory.

Upon analyzing the viable parameter space of this model, we find a rich phenomenology, including two qualitatively different scenarios which can both address the anomalous magnetic moment of the muon, based on the $Z_2$ parity of the particle that is identified as the muon. The first, which we call Scenario A, closely resembles the minimal model of \cite{Wojcik:2022woa}, with $U(1)_D$-charged portal matter that dominantly decays into muons and a contribution to the muon magnetic moment which is independent of any gauge couplings or other parameters of the underlying simplified dark matter model. In the other case, which we denote as Scenario B, the portal matter does \emph{not} directly mix with the muons, and instead decays dominantly to $\tau$ leptons-- there are meanwhile $U(1)_D$-\emph{neutral} vector-like leptons that dominantly decay into the muons. Because loops involving heavier $\textrm{TeV}$-scale dark sector gauge fields now dominate the calculation, the $g-2$ correction in this scenario bears a greater similarity to the model of \cite{CarcamoHernandez:2019ydc}, in which a heavy $Z'$ acts to generate the muon magnetic moment anomaly rather than a sub-GeV dark photon.

The remainder of the paper is laid out as follows. In Section \ref{sec:model}, we outline the construction of our model and identify various important features, as well as presenting notation that we shall use for the remainder of the work. In Section \ref{sec:g-2}, we explore the parameter space for which the constructions of Scenario A and Scenario B can recreate the observed $g-2$ anomaly. In Section \ref{sec:pheno}, we outline existing constraints and future prospects on both Scenario A and Scenario B, focusing on collider phenomenology and the usefulness of a multi-TeV muon collider in constraining or observing crucial model parameters. In Section \ref{sec:conclusions}, we summarize and discuss our findings and mention directions for future work.

\section{Model Description}\label{sec:model}
We will consider the augmentation of the SM gauge group $G_{SM}=SU(3)_{C}\otimes SU(2)_{L} \otimes U(1)_{Y}$ by a dark gauge group $G_D$, which we take to be of the form
\begin{equation}
     G_D=SU(2)_A \otimes SU(2)_B,
\end{equation}
and further augment the theory with a global $Z_2$ symmetry. Turning now to the fermionic matter content of the model, the SM quarks are taken to be singlets with respect to $G_D$ and $Z_2$, as is one generation of the SM leptons. However, the other two generations of SM leptons will arise as the light degrees of freedom from mixing with sets of leptonic portal matter states. These states have nontrivial quantum numbers with respect to $G_D$ and/or $Z_2$ as well as the electroweak gauge group, and are chosen to satisfy gauge anomaly constraints.   Because $G_D$ is semisimple, any kinetic mixing featuring the gauge fields associated with $G_D$ will be finite and calculable. The list of these fields is given in Table [\ref{Fcontent}]. 

\begin{table}[h!]
\centering
\begin{tabular}{ |P{1.6cm}||P{0.8cm}|P{1.1cm}|P{1.1cm}|P{0.9cm}|||P{1.6cm}||P{0.8cm}|P{1.1cm}|P{1.1cm}|P{0.9cm}| }
 \hline
 \multicolumn{10}{|c|}{Lepton Matter Content} \\
 \hline
 LH fields& $G_{SM}$ & $SU(2)_A$ & $SU(2)_B$ & $Z_2$ &RH Fields& $G_{SM}$ & $SU(2)_A$ & $SU(2)_B$ & $Z_2$
 \\
 \hline
 $~~L_L$ & $\vb{L}$ & $\vb{1}$ & $\vb{1}$ & $+1$
 &
 $~~e_R$ & $\vb{e}$ & $\vb{1}$ & $\vb{1}$ & $+1$
 \\
 $~~\Psi_L$   & $\vb{L}$    & $\vb{2}$ &   $\vb{2}$ & $+1$ 
 &
 $~~\Psi_R$ &   $\vb{e}$  & $\vb{2}$  & $\vb{2}$ & $+1$ 
 \\
 $~~V_L$ & $\vb{e}$ & $\vb{1}$ &  $\vb{3}$ & $+1$
 &
 $~~V_R$    & $\vb{L}$ & $\vb{1}$ &  $\vb{3}$ & $+1$ 
 \\
 $~~S_L$ &   $\vb{L}$ & $\vb{1}$ & $\vb{1}$ & $-1$ 
&
 $~~S_R$ & $\vb{e}$  & $\vb{1} $ & $\vb{1}$ & $-1$
 \\
 \hline
\end{tabular}
\caption{The lepton field content and corresponding representations with respect to $G_D$ and $G_{SM}$. Here, $\mathbf{L}$ and $\mathbf{e}$ denote the $G_{SM}$ charge assignments of the usual SM electroweak doublet and singlet leptons. }
\label{Fcontent}
\end{table}

The breaking of $G_D$ occurs in two stages: the first to the usual $U(1)_D$ of minimal portal matter constructions, which occurs at $O({\rm TeV})$ scales, and the second the breaking of $U(1)_D$, which occurs at sub-GeV scales. The SM Higgs boson $H$ is taken to be a singlet with respect to $G_D$, and thus new scalar fields must be introduced to achieve the required symmetry breaking pattern. More precisely, we introduce the dark scalar $\Phi$, which is a bidoublet with respect to $SU(2)_A\otimes SU(2)_B$,  and two real scalar triplets $\Delta_{A,B}$, which are triplets with respect to $SU(2)_{A,B}$, respectively. In addition, $\Phi$ is even and $\Delta_{A,B}$ are odd with respect to $Z_2$. The dark scalar field content is summarized in Table \ref{Scontent}.
\begin{table}[h!]
\centering
\begin{tabular}{ |P{1.6cm}||P{0.8cm}|P{1.1cm}|P{1.1cm}|P{0.9cm}|}
 \hline
 \multicolumn{5}{|c|}{Dark Scalar Content} \\
 \hline
 Fields& $G_{SM}$ & $SU(2)_A$ & $SU(2)_B$ & $Z_2$ 
 \\
 \hline
 $\Phi$ & $\vb{1}$ & $\vb{2}$ & $\vb{2}$ & $+1$
 \\
 $\Delta_A$   & $\vb{1}$    & $\vb{3}$ &   $\vb{1}$ & $-1$ 
 \\
 $\Delta_B$ & $\vb{1}$ & $\vb{1}$ & $\vb{3}$ & $-1$
 \\
 \hline
\end{tabular}
    \caption{The dark scalar field content of this model. All the dark scalars are singlets under $G_{SM}$.}
    \label{Scontent}
\end{table}
Writing the overall scales of the vacuum expectation value (vevs) of $\Phi$ and $\Delta_{A,B}$ as $v_\Phi$ and $v_{\Delta_{A,B}}$, schematically, we have
\begin{equation}
        SU(2)_A \otimes SU(2)_B \otimes Z_2 \xRightarrow[\Phi]{v_\Phi\sim 1~\text{TeV}} U(1)_D \otimes Z_2  \xRightarrow[\Delta_A,\Delta_B]{v_{\Delta_{A,B}}
        \sim 1~\text{GeV}} Z'_2 \nonumber.
\end{equation}
We note the appearance of a residual unbroken $Z_2'$, which arises from a combination of the unbroken part of $U(1)_D$ and the global $Z_2$ symmetry, as discussed in greater detail in Appendix \ref{Sappendix}. We will see that while the conserved $Z_2^\prime$ leads to a number of intriguing features, it will ultimately need to be broken to generate viable lepton mixing. We will comment later in this paper on possible extensions of this scenario that allow for $Z_2'$ breaking.

As discussed in \cite{Wojcik:2022woa}, we  will not include terms that couple the SM Higgs boson to the dark sector at the renormalizable level, as such interactions are strongly constrained by \eg~the invisible Higgs width (this issue is generic to portal matter constructions). In this case, the electroweak symmetry breaking and the dark symmetry breaking are decoupled. Focusing on the dark gauge group breaking, the most general renormalizable scalar potential that respects $SU(2)_A\otimes SU(2)_B\otimes Z_2$ is given by
\begin{align}
    V(\Phi,\Delta_{A,B})&=-\mu^2_{1}\Tr(\Phi^\dagger \Phi)-\frac{\mu_{2}^2}{2}\left[\Tr(\widetilde {\Phi}^\dagger \Phi)+\text{h.c.}\right]-\mu_{3}^2 \Tr(\Delta_A^2) -\mu_4^2 \Tr(\Delta_B^2) \nonumber \\
    +&\lambda_1\Tr(\Phi^\dagger \Phi)^2+\frac{\lambda_2}{2}\left[\Tr(\widetilde {\Phi}^\dagger \Phi)^2+\text{h.c.}\right]+\frac{\lambda_3}{2} \left[\Tr(\Phi^\dagger \Phi)\Tr(\widetilde {\Phi}^\dagger \Phi) + \text{h.c.} \right]+\lambda_4 \Tr (\Delta^4_B) \label{eq:scalarpot} \\
    +&\lambda_5 \Tr(\Phi^\dagger \Phi) \Tr(\Delta^2_B)+\frac{\lambda_6}{2}\left[\Tr(\widetilde {\Phi}^\dagger \Phi)\Tr(\Delta^2_B) + \text{h.c.} \right]+\lambda_7 \abs{\Tr(\widetilde {\Phi}^\dagger \Phi)}^2+\lambda_8 \Tr(\Phi^\dagger \Delta_{A} \Phi \Delta_B) \nonumber \\
    +& \frac{\lambda_9}{2}\left[\Tr(\widetilde{\Phi}^\dagger \Delta_A \Phi \Delta_B)+\text{h.c.} \right]+\lambda_{10}\Tr(\Delta_A^4)+\lambda_{11} \Tr(\Phi^\dagger \Phi) \Tr(\Delta_A^2) \nonumber \\ +& \frac{\lambda_{12}}{2}\left[\Tr(\widetilde {\Phi}^\dagger \Phi)\Tr(\Delta^2_A)+\text{h.c.} \right]+\lambda_{13}\Tr(\Delta^2_A)\Tr(\Delta^2_B), 
    \nonumber
\end{align}
in which $\widetilde{\Phi}_{i\alpha}= -\epsilon_{ij}\Phi^{*j\beta}\epsilon_{\beta \alpha}$.

As shown in detail in Appendix \ref{Sappendix}, to achieve the two-stage breaking of $G_D$ through the minimization of Eq.~(\ref{eq:scalarpot}), the vevs of the dark scalar fields $\Phi$ and $\Delta_{A,B}$ take the form:
\begin{equation}\label{eq:vevs}
\langle \Phi\rangle = v_\Phi \left (\begin{array}{cc} \cos\theta_\Phi & 0\\ 0 & \sin\theta_\Phi \end{array} \right ),\qquad \langle \Delta_{A,B} \rangle =  \left (\begin{array}{cc} 0 & v_{\Delta_{A,B}}\\ v_{\Delta_{A,B}} & 0 \end{array} \right ).
\end{equation}
We define $v_\Delta = \sqrt{v_{\Delta_A}^2+v_{\Delta_B}^2}$ and $\tan\theta_\Delta = v_{\Delta_A}/v_{\Delta_B}$, where $0\leq \theta_\Delta\leq \pi/2$, and take $r_\Delta=v_\Delta/v_\Phi$. Note that $r_\Delta \sim 10^{-3} \ll 1$, and hence the symmetry breaking analysis can be performed perturbatively in $r_\Delta$.

We begin by describing the gauge boson masses and eigenstates. Here we label the gauge fields of $SU(2)_A\otimes SU(2)_B$ as $W_{A,B}$, with coupling constants $g_{A,B}$. These gauge couplings are related to the elementary dark charge $e_D$ of the $U(1)_D$ gauge group by $e_D=g_A \cos \theta_D = g_B \sin \theta_D$, which allows us to exchange the inputs $(g_A,g_B)$ for $(e_D,\theta_D)$. The mass eigenstates are given by 
\begin{align}\label{eq:gauge-sector-defs}
    Z_D&=\cos \theta_D W^z_{A}+\sin \theta_D W^z_{B}, \qquad \,
     M_{Z_D}= \sqrt{2} v_\Phi e_D \csc(2 \theta_D), \nonumber
    \\
    A_D&=-\sin \theta_D W^z_A + \cos \theta_D W^z_B, \qquad \, 
    m_{A_D}=\frac{1}{\sqrt{2}}\,r_\Delta \sin 2\theta_D M_{Z_D}, 
    \\
    W_{l}^{\pm}&=\cos \theta_W W_{A}^\pm + \sin \theta_W W^\pm_B, \qquad 
    ~M_{W_l}=\sin \theta_{lh} M_{Z_D}, \nonumber
    \\
    W^{\pm}_h&=- \sin \theta_W W^\pm_A+\cos \theta_W W^\pm_B, \quad~ 
    M_{W_h}=\cos \theta_{lh} M_{Z_D}, \nonumber
\end{align}
in which $\tan 2 \theta_W = \sin 2 \theta_\Phi \tan 2 \theta_D$ and $\cos 2\theta_{lh}=\cos 2\theta_D \sqrt{1+\sin^2 \theta_\Phi \tan^2 \theta_D}$. We note that we can exchange the inputs $(e_D,\theta_D)$ for $(M_{Z_D},\theta_{lh})$. The field $A_D$ is the dark photon, which has its mass controlled by $r_\Delta$, while the remaining gauge boson masses scale with $v_\Phi$. We note that $Z_D$ and $A_D$ are even with respect to $Z_2^\prime$, while $W_{l,h}$ are odd.

The details of the scalar mass spectrum are provided in Appendix \ref{Sappendix}; here we briefly summarize the relevant results. There are 14 real degrees of freedom in the scalar sector (8 from $\Phi$ and 6 from $\Delta_{A,B}$). 6 of these degrees of freedom are eaten via the Higgs mechanism in the breaking of $SU(2)_A\otimes SU(2)_B$, leaving 8 physical scalar fields. We denote the 8 physical scalars as 
\begin{equation}\label{eq:h-list}
\{h_1,h_2,h_3,h_{deg}^1,h_{deg}^2,h_4,h_5, h_6
\},
\end{equation} 
in which $(h_{deg}^1,h_{deg}^2)$ form a degenerate subspace. Of these scalars, all but $h_3$ have their masses controlled by $v_\Phi$, while the mass of $h_3$ is governed by $r_\Delta v_\Phi$.
We will denote the mass of $h^{1,2}_{deg}$ as $M_{h^\pm}$; this can be understood by considering the quantities $h^\pm =(h^1_{deg}\pm i h^2_{deg})/\sqrt{2}$, in which the ``$\pm$" superscript denotes the charges with respect to the unbroken $U(1)_D$ that results after the first stage of symmetry breaking. In the forthcoming phenomenological analysis, we will use $M_{h^\pm}$ as an input parameter in the scalar sector  and parameterize the remaining scalar masses as
\begin{equation}\label{eq:boson-masses}
M_{h_{1,2,4}}=r_{1,2,4}M_{h^\pm}, \;\; 
m_{h_3}=r_\Delta r_3M_{h^\pm}, \;\;
M_{h_5}=\cos\theta_M M_{h^\pm}, \;\; M_{h_6}=\sin\theta_M M_{h^\pm}, 
\end{equation}
in which the $r_i$ and $\theta_M$ are $O(1)$ parameters; their detailed definitions are provided in Appendix \ref{Sappendix}. As the minimum is CP-conserving and $Z_2^\prime$-conserving, the scalars have well-defined CP and $Z_2^\prime$ quantum numbers $({\rm CP},Z_2^\prime)$, as summarized in Table \ref{SMcontent}.

\begin{table}[h!]
\centering
\begin{tabular}{ |P{1.6cm}||P{0.8cm}|P{1.1cm}|P{3.6cm}|}
 \hline
 \multicolumn{4}{|c|}{Scalars Mass Eigenstates} \\
 \hline
 Eigenstate& CP & $Z'_2$ & Mass (In units of $M_{h^\pm}$)
 \\
 \hline
 $h_1$ & $+1$ & $+1$ & $M_{h_1}=r_1 $ 
 \\
 $h_2$   & $+1$    & $+1$ &   $M_{h_2}=r_2$ 
 \\
 $h_3$ & $+1$ & $+1$ & $~~m_{h_3}=r_\Delta r_3$ 
 \\
 $~~h_{deg}^{1}$ & $+1$ & $+1$ & $1$
 \\
 \hline
 $~~h_{deg}^{2}$ & $-1$ & $+1$ & $1$
 \\
 $h_4$ & $-1$ & $+1$ & $M_{h_4}=r_4$
 \\
 \hline
 $h_5$ & $+1$ & $-1$ & $M_{h_5}=\cos \theta_M $
 \\
 $h_6$ & $+1$ & $-1$ & $M_{h_6}=\sin \theta_M $\\
 \hline
\end{tabular}
\caption{Scalar mass eigenstates listed with their CP and $Z_2^\prime$ quantum numbers, and their masses.
}
\label{SMcontent}
\end{table}

Turning now to the fermion sector, the Yukawa interactions among the $SU(2)_A\otimes SU(2)_B\otimes Z_2$-charged fermions as listed in Table \ref{Fcontent} to the dark scalars $\Phi$ and $\Delta_{B}$ as well as to the SM Higgs $H$ take the form
\begin{align} \mathcal{L}_{Y}=&y_{H}\left[\Tr(\overline{\Psi_L}H\Psi_{R})+\text{h.c.} \right]+y_{HV}\left[\Tr(\overline{V_L}HV_{R})+\text{h.c.} \right]+y_{HS}\left[\Tr(\overline{S_L}HS_{R})+\text{h.c.} \right]\nonumber \\ +&y_P\left[\Tr(\overline{\Psi_L}\Phi V_{R})+\text{h.c.} \right]+\tilde{y}_P\left[\Tr(\overline{\Psi_L}\widetilde{\Phi} V_{R})+\text{h.c.} \right]+y_{P'}\left[\Tr(\overline{V_L}\Phi^{\dagger}\Psi_{R})+\text{h.c.} \right]  \\
    +&\tilde{y}_P'\left[\Tr(\overline{V_L}\widetilde{\Phi}^{\dagger}\Psi_{R})+\text{h.c.} \right]+y_{SE}\left[\Tr(\overline{V_L}S_{R}\Delta_{B})+\text{h.c.} \right]+y_{SL}\left[\Tr(\overline{S_L}V_{R}\Delta_{B})+\text{h.c.} \right]. \nonumber
\end{align}
For simplicity, we will take all Yukawa couplings to be real. We will further find it convenient to write the Yukawa couplings as 
\begin{equation}
    y_P=y_L \cos \theta_L, \quad \tilde{y}_P=y_L \sin \theta_L, \quad y'_P=y_E \cos \theta_E, \quad \text{and} \quad \tilde{y}'_P=y_E \sin \theta_E,
\end{equation}
and define
\begin{align}\label{eq:m-plus-minus}
    M^+_{L}&=v_\Phi y_L \cos(\theta_L-\theta_\Phi), \qquad M^-_{L}=v_\Phi y_L \sin (\theta_L+\theta_\Phi), \\ M^+_{E}&=v_\Phi y_E \cos(\theta_E-\theta_\Phi),\qquad M^-_{E}=v_\Phi y_E |\sin(\theta_E+\theta_\Phi)|. \nonumber
\end{align}
Reflected above is the fact that by applying a series of chiral rotations to the fermions, we are capable of ensuring that $y_L$, $y_E$, $y_{SL}$, $y_{SE}$, $y_{HS}$, $\cos(\theta_L-\theta_\Phi)$, $\cos(\theta_E-\theta_\Phi)$, and $\sin(\theta_L + \theta_\Phi)$ are all positive without loss of generality, while $y_{H}$, $y_{HV}$, and $\sin(\theta_E + \theta_\Phi)$ can be either positive or negative-- so we must take the absolute value of  $\sin(\theta_E + \theta_\Phi)$ to get the physical mass $M_E^-$. The signs of $y_{H}$, $y_{HV}$, and $\sin(\theta_E + \theta_\Phi)$ will have significant phenomenological implications later, so it is important to be aware that they remain undetermined under our conventions.
For notational purposes, the electric-charged components of the matter fields will be written as
\begin{equation}
    \Psi_{L,R}=\mqty(\Psi_{L,R_{11}} & \Psi^+_{L,R} \\ \Psi^-_{L,R} & \Psi_{L,R_{22}} ) \quad \text{and} \quad  V_{L,R}=\mqty(V^0_{L,R}/\sqrt{2} & V^+_{L,R} \\ V^-_{L,R} & -V^0_{L,R}/\sqrt{2} ),
\end{equation}
and for the electric-neutral components we will use a similar definition, but with the inclusion of the superscript $N$. In the electric-charged fermion sector, we will organize the components of these fields in the following two sets: $X_{F,(L,R)}=(X^{(1)}_{F,(L,R)},X_{F,(L,R)}^{(2)})$, in which
\begin{equation}
     X^{(1)}_{F,(L,R)}=\left (\Psi_{{(L,R)}_{11}}, \Psi_{{(L,R)}_{22}}, V^0_{(L,R)} \right), \qquad X^{(2)}_{F,(L,R)}=\left(S_{(L,R)}, \Psi^{+}_{(L,R)}, V^{+}_{(L,R)}, \Psi^{-}_{(L,R)},V^{-}_{(L,R)} \right).
     \label{eq:xpmdefchg}
\end{equation}
The electric-charged fermion mass matrix $M_F$ is composed of two blocks, $M^{(1)}_F$ and $M^{(2)}_F$, that do not mix with each other. These two blocks take the form
\begin{align}\label{eq:mf-block-def}
    M^{(1)}_F&=
    \mqty(\frac{y_H v}{\sqrt{2}} & 0 & M^+_{L} \\ 0 & \frac{y_H v}{\sqrt{2}} & -M^-_{L} \\ M^+_{E} & \mp M^-_{E} & \frac{y_{HV} v}{\sqrt{2}} ),
    \\
    M^{(2)}_F&=\mqty(\frac{y_{HS} v}{\sqrt{2}} & 0 & y^{\Delta}_{SL} r_\Delta v_\Phi  & 0 & y^{\Delta}_{SL} r_\Delta v_\Phi  \\
    0 & \frac{y_H v}{\sqrt{2}} & M^+_{L} & 0 & 0 \\ y^{\Delta}_{SE}  r_\Delta v_\Phi  & M_{E}^+ & \frac{y_{HV}v}{\sqrt{2}} & 0 & 0 \\ 0 & 0 & 0 & \frac{y_H v}{\sqrt{2}} & M^-_{L} \\ y^{\Delta}_{SE} r_\Delta v_\Phi  & 0 & 0 & \pm M^-_{E} & \frac{y_{HV}v}{\sqrt{2}}),
\end{align}
in which $v$ is the usual SM electroweak Higgs vev, and we have introduced the 
shorthanded notation $y^\Delta_{SL,SE}=(y_{SL,SE}/2) \cos \theta_\Delta$. 
Upon diagonalizing these matrices, to leading order in $r_H=v/v_\Phi \sim  10^{-2} \ll 1$ and $r_\Delta$, the left and right-handed mass eigenstates for the $Z'_2$ positively charged fermions are
\begin{align}
   L^b_{L}&=\frac{1}{\sqrt{2}}\frac{M^-_L}{M^0_{L}}\Psi_{L_{11}}+\frac{1}{\sqrt{2}}\frac{M^+_L}{M^0_{L}}\Psi_{L_{22}}, \qquad e^b_{R}=\pm\frac{1}{\sqrt{2}}\frac{ M^-_E}{M^0_{E}}\Psi_{R_{11}}+\frac{1}{\sqrt{2}}\frac{M^+_E}{M^0_{E}}\Psi_{R_{22}},\nonumber
     \\
   L^0_{L}&=-\frac{1}{\sqrt{2}}\frac{M^+_L}{M^0_{L}}\Psi_{L_{11}}+\frac{1}{\sqrt{2}}\frac{M^-_L}{M^0_{L}}\Psi_{L_{22}}, \quad L^0_{R}=V^0_R, \\
   E^0_{L}&=V^0_L, \qquad E^0_{R}=-\frac{1}{\sqrt{2}}\frac{M^+_E}{M^0_{E}}\Psi_{R_{11}}\pm \frac{1}{\sqrt{2}}\frac{ M^-_E}{M^0_{E}}\Psi_{R_{22}}, \nonumber
\end{align}
with masses
\begin{equation}\label{eq:neutral-block-fermion-masses}
    m_b =\frac{M^+_E M^+_L\pm M^-_E M^-_L}{\sqrt{\left({M^-_E}^2+{M^+_E}^2 \right) \left({M^-_L}^2+{M^+_L}^2 \right)}} \frac{y_H v}{\sqrt{2}}, \quad  M^0_{L}=\sqrt{\frac{(M^-_L)^2+(M^+_L)^2}{2}}, \quad M^0_{E}=\sqrt{\frac{(M^-_E)^2+(M^+_E)^2}{2}},
\end{equation}
respectively, while for the $Z'_2$ negatively charged fermions, the left and right-handed mass eigenstates to leading order are
\begin{align}
    L^a_{L}&=S_{L}, \qquad e^a_{R}=S_{R}, \qquad
    L^\pm_L=\Psi^\pm_L, \qquad L^\pm_R=V^\pm_R, \qquad
    E^\pm_L=V^\pm_L, \qquad 
    E^\pm_R=\Psi^\pm_R,
\end{align}
with masses 
\begin{equation}\label{eq:pm-block-fermion-masses}
m_a=\frac{y_{HS}v}{\sqrt{2}},\qquad 
M^\pm_{L},\qquad \quad M^\pm_{E},
\end{equation}
respectively. The charged lepton mass eigenstates and their $Z_2^\prime$ charges are summarized in Table \ref{LME}.

\begin{table}[h!]
\centering
\begin{tabular}{ |P{1.6cm}||P{0.8cm}|P{1.1cm}|||P{1.6cm}||P{0.8cm}|P{1.6cm}| }
 \hline
 \multicolumn{6}{|c|}{Lepton Mass Eigenstates} \\
 \hline
 LH fields& $G_{SM}$ & $Z'_2$ &RH Fields& $G_{SM}$ & $Z'_2$
 \\
 \hline
 $L^{b}_{L}$ & $\vb{L}$ & $+1$
 &
 $e^b_{R}$ & $\vb{e}$  & $+1$
 \\
 $L_{L}^0$ & $\vb{L}$  & $+1$
 &
 $L_{R}^0$ &   $\vb{L}$ & $+1$ 
 \\
 $E_{L}^0$   & $\vb{e}$   & $+1$
 &
 $E_{R}^0$    & $\vb{e}$ & $+1$ 
 \\
 \hline
 $L^a_{L}$ &   $\vb{L}$ & $-1$ 
&
 $e^a_{R}$ & $\vb{e}$  & $-1$
 \\
 $L^\pm_L$ & $\vb{L}$ & $-1$
 &
 $L^\pm_R$ & $\vb{L}$ & $-1$
 \\
 $E^\pm_L$ & $\vb{e}$ & $-1$
 &
 $E^\pm_R$ & $\vb{e}$ & $-1$
 \\
 \hline
\end{tabular}
\caption{List of charged fermion mass eigenstates and their corresponding quantum numbers with respect to $G_{SM}$ and $Z_2^\prime$.}
\label{LME}
\end{table}

For the electrically charge-neutral fermions, in analogy with Eq.~(\ref{eq:xpmdefchg}), we can define 
\begin{equation}
X^N_{F,L}=(\Psi^N_{L_{11}}, \Psi^N_{L_{22}}, S^{N}_L,\Psi^{N+}_L,\Psi^{N-}_L), \qquad X^{N}_{F,R}=(V^{N0}_R, V^{N+}_R, V^{N-}_R). 
\end{equation}
In this basis, the mass matrix for the electric charge-neutral fermions $M_F^N$ thus takes the form
\begin{equation}
M^N_F=\mqty(M^+_{L}/\sqrt{2} & 0 & 0 \\ -M^-_{L}/2 & 0 & 0 \\ 0 &  y^\Delta_{SL} r_\Delta v_\Phi  &  y^\Delta_{SL} r_\Delta v_\Phi  \\ 0 & M^+_{L} & 0 \\ 0 & 0 & M^-_{L}), 
\end{equation}
which again shows the separation into blocks according to the $Z_2^\prime$ charges. The electrically-neutral mass eigenstates are given by
\begin{equation}
    \nu_{bL}=\frac{1}{\sqrt{2}}\frac{M^-_L}{M^0_{L}}\Psi^N_{L_{11}}+\frac{1}{\sqrt{2}}\frac{M^+_L}{M^0_{L}}\Psi^N_{L_{22}}, \qquad
     N^0_{L}=-\frac{1}{\sqrt{2}}\frac{M^+_L}{M^0_{L}}\Psi^N_{L_{11}}+\frac{1}{\sqrt{2}}\frac{M^-_L}{M^0_{L}}\Psi^N_{L_{22}}, \qquad 
     N^0_{R}=V^{N0}_{R},
\end{equation}
and
\begin{equation}
    \nu_{aL}=S^N_L, \qquad 
    N^\pm_{L} =\Psi^{N\pm}_L, \qquad 
    N^\pm_{ R}=V^{N\pm}_R,
\end{equation}
with masses
\begin{equation}
    m_{\nu_a}=m_{\nu_b}=0, \quad M_{N^0}=M^0_{L}, \quad M_{N^\pm}=M^0_L.
\end{equation}
As we have not introduced right-handed neutrino singlet counterparts to balance the degrees of freedom among the left-handed and right-handed fermions, the active neutrinos $\nu_{aL}$ and $\nu_{bL}$ remain massless. For the purposes of this study, the issue of generating nonzero neutrino masses and observable mixing angles does not have a strong impact for the implications for the muon $g-2$ or the portal matter collider phenomenology, and thus we will neglect it in this work. We will comment later about the need for $Z_2^\prime$ breaking and its implications for the generation of neutrino masses within this framework.

A straightforward one-loop computation quickly allows us to explicitly verify that in this model, the kinetic mixing coefficient between $A_D$ and the $U(1)_Y$ gauge field $B$ is finite, calculable, and consistent with the $O(10^{-(3-5)})$ range we would require for a sub-GeV dark matter model. We arrive at
\begin{align}
    \epsilon = \frac{e_D e}{6 \pi^2} \log \bigg(\frac{M_L^+ M_E^+}{M_L^- M_E^-}\bigg) \approx (5.3 \times 10^{-3}) e_D \log \bigg(\frac{M_L^+ M_E^+}{M_L^- M_E^-}\bigg).
\end{align}
Returning to the electrically charged leptons, we see that there are two light states, which are massless in the limit that the electroweak vev $v\rightarrow 0$. These light states are $e_a$ and $e_b$, with masses $m_a$  and $m_b$, respectively. $e_a$ is negatively charged with respect to $Z_2^\prime$, while $e_b$ is positively charged with respect to $Z_2^\prime$. The third light SM lepton is a gauge singlet with respect to $G_D$, and thus does not mix with the additional portal matter states; we will assume throughout this work that this state is the electron for the sake of concreteness and simplicity. We can therefore envision two general cases, depending on whether we identify $e_a$ or $e_b$ with the muon. To be more precise, we have the following two scenarios:
\begin{itemize}
    \item {\bf Scenario A}: the $Z_2^\prime$-negative $e_a$ is the muon, while $e_b$ is the $\tau$ lepton.
    \item {\bf Scenario B}: the $Z_2^\prime$-positive $e_b$ is the muon, while $e_a$ is the $\tau$.
\end{itemize}
We note that in Scenario B, the muon mass is controlled by the Yukawa coupling $y_H$ and the mass eigenstate is dominated by the $G_D$-bidoublet fermions $\Psi_{L,R}$. In contrast, in  Scenario A, the muon mass is governed by $y_{HS}$ and the muon eigenstate is dominantly given by the $SU(2)_A\otimes SU(2)_B$ singlet $S_{L,R}$.

For our phenomenological studies of this model, it will be useful to have a compact list of parameters that we can use to uniquely specify a point in parameter space. For clarity, we shall present that here-- we have selected a parameterization that maximizes the number of parameters with immediate physical interpretation (\ie masses, relative coupling strengths) and uses entirely quantities which are explicitly defined in this Section. In the purely gauge and scalar sector, we have
\begin{align}\label{eq:param-scals}
    (e_D, \lambda_1, M_{Z_D}, M_{h_{1,2,4}}, M_{h^\pm}, \theta_M, \theta_D, \theta_\Delta, \theta_{lh}),
\end{align}
Since we are principally concerned with the phenomenology of the theory at $O(\textrm{TeV})$ energy scales, there will be no need for us to uniquely specify the masses of the dark photon $A_D$ or the dark Higgs $h_3$, which are both of sub-GeV scales. Furthermore, of the couplings in the potential of Eq.~(\ref{eq:scalarpot}), we need only retain the quartic coupling $\lambda_1$, since all other terms are either expressible in terms of masses or the angles given above, or only enter into scalar self-coupling interactions that we shall not explore in detail in this work.
To specify the fermion sector we shall have the masses and Yukawa couplings
\begin{align}\label{eq:param-ferms}
    (M_L^\pm, M_E^\pm, |y_{HV}|, |y_{H}|, |y_{SL}|, |y_{SE}|).
\end{align}
Finally, as mentioned earlier in this Section, there are a handful of nontrivial signs for parameters that we must specify. For a complete specification, we need only include the signs
\begin{align}\label{eq:param-signs}
    \textrm{sign}(y_{HV}, y_{H}, \sin(\theta_E + \theta_\Phi), \sin(2 \theta_\Phi)).
\end{align}
We note that although the sign of $\sin(2 \theta_\Phi)$ must be independently specified, its magnitude is always determined in terms of other parameters. In all of our subsequent computations in this work, we shall employ the parameterization of Eqs.(\ref{eq:param-scals}-\ref{eq:param-signs}) to specify points in parameter space of this model.

\section{Muon $g-2$}\label{sec:g-2}

In this section, we will present the results of the computation of the new physics contribution to the anomalous magnetic moment of the muon in  Scenario A and Scenario B. In both scenarios, the dominant contributions arise at the one-loop level from loops of the new scalar and gauge bosons with heavy intermediate  fermions, as in \cite{CarcamoHernandez:2019ydc,Wojcik:2022woa}. If the SM Higgs mediates strong enough couplings between the isosinglet and isodoublet vector-like fermions, then the resultant chirality flip in the internal heavy fermion line can give rise to a chirally-enhanced contribution to $\Delta a_\mu$ which can fully account for the observed discrepancy.
Other contributions, such as that stemming from the coupling of the dark photon to the muon through kinetic mixing (as discussed in, \eg, \cite{Pospelov:2008zw}), are heavily suppressed, and can generally contribute to the final result at no more than the few percent level while remaining consistent with experimental constraints.\footnote{We find that it is in principle possible for the dark photon loop as in \cite{Pospelov:2008zw} to account for as much as $\lesssim 10\%$ while still remaining barely consistent with current limits on dark photon mass and kinetic mixing (\eg, \cite{NA64:2021xzo}), but this contribution in most regions of parameter space is dramatically more subdued.} In this analysis, we shall therefore focus exclusively on the dominant contributions in our analysis.
While the two scenarios both exhibit a similar mechanism by which the dominant contribution to $\Delta a_\mu$ is realized, the specifics of the scenarios are markedly different. As the scenarios differ in which light fermion is identified with the muon, the specific diagrams that contribute to $\Delta a_\mu$ are in fact entirely different in the two scenarios, and they must be addressed separately.

\subsection{$g-2$ in Scenario A}\label{sec:g-2-A}

In Scenario A, we identify the muon with the light fermion appearing in the block $M_F^{(2)}$ of the fermion mass matrix in Eq.~(\ref{eq:mf-block-def}), while the $\tau$ is identified with the light mass eigenstate appearing in block $M_F^{(1)}$. In turn, this allows us to use Eqs.(\ref{eq:neutral-block-fermion-masses}) and (\ref{eq:pm-block-fermion-masses}) fix the Yukawa couplings $y_H$ and $y_{HS}$ to
\begin{align}
    |y_H| = \frac{\sqrt{2} m_\tau}{v} \sqrt{\frac{({M_L^+}^2+{M_L^-}^2)({M_E^+}^2+{M_E^-}^2)}{(M_L^+ M_E^+ \pm M_L^- M_E^-)^2}}, \;\;\; y_{HS} = \frac{\sqrt{2} m_\mu}{v},
\end{align}
where we note that in our sign conventions, $y_{HS}$ is always positive, but $y_{H}$ may be either positive or negative. Recalling Eq.~(\ref{eq:m-plus-minus}) and the discussion surrounding it in Section \ref{sec:model}, we recall that the $+(-)$ sign in our expression is applied when $\sin(\theta_E + \theta_\Phi)$ is positive (negative). As discussed in Section \ref{sec:model}, this relative sign cannot be eliminated by chiral phase rotations, and hence the selection of a positive or negative sign here represents a selection between two distinct points in the model parameter space.

\begin{figure*}[h!]
\centering
\subfloat{
\begin{fmffile}{feyngraph3}
  \begin{fmfgraph*}(80,100)
    \fmfbottom{f2,f1}
    \fmftop{g}
    \fmf{boson,t=1.5}{g,v}
    \fmf{fermion}{f2,v1}
    \fmf{fermion}{v1,v}
    \fmf{fermion} {v,v2}
    \fmf{fermion}{v2,f1}
    \fmfv{l.a=-20,l.d=8,l=$\gamma$}{g}
    \fmfv{l.a=-10,l=$\mu^-$}{f1}
    \fmfv{l.a=-170,l=$\mu^-$}{f2}
    \fmfv{l.a=120, d=10,l=$F^\pm$}{v1}
    \fmfv{l.a=60, d=10,l=$F^\pm$}{v2}
    \fmffreeze
    \fmf{boson,label=$A_D$}{v1,v2}
  \end{fmfgraph*}
  \end{fmffile}
}\hspace{15mm}
\subfloat{
\begin{fmffile}{feyngraph4}
  \begin{fmfgraph*}(80,100)
    \fmfbottom{f2,f1}
    \fmftop{g}
    \fmf{boson,t=1.5}{g,v}
    \fmf{fermion}{f2,v1}
    \fmf{fermion}{v1,v}
    \fmf{fermion} {v,v2}
    \fmf{fermion}{v2,f1}
    \fmfv{l.a=-20,l.d=8,l=$\gamma$}{g}
    \fmfv{l.a=-10,l=$\mu^-$}{f1}
    \fmfv{l.a=-170,l=$\mu^-$}{f2}
    \fmfv{l.a=120, d=10,l=$F^\pm$}{v1}
    \fmfv{l.a=60, d=10,l=$F^\pm$}{v2}
    \fmffreeze
    \fmf{dashes,label=$h_3/h^\mp$}{v1,v2}
  \end{fmfgraph*}
  \end{fmffile}
}\hspace{15mm}
\subfloat{
\begin{fmffile}{feyngraph5}
  \begin{fmfgraph*}(80,100)
    \fmfbottom{f2,f1}
    \fmftop{g}
    \fmf{boson,t=1.5}{g,v}
    \fmf{fermion}{f2,v1}
    \fmf{fermion}{v1,v}
    \fmf{fermion} {v,v2}
    \fmf{fermion}{v2,f1}
    \fmfv{l.a=-20,l.d=8,l=$\gamma$}{g}
    \fmfv{l.a=-10,l=$\mu^-$}{f1}
    \fmfv{l.a=-170,l=$\mu^-$}{f2}
    \fmfv{l.a=120, d=10,l=$F^0$}{v1}
    \fmfv{l.a=60, d=10,l=$F^0$}{v2}
    \fmffreeze
    \fmf{dashes,label=$h_5/h_6$}{v1,v2}
  \end{fmfgraph*}
  \end{fmffile}
}
\caption{ Feynman diagrams contributing to $\Delta a_\mu$ in Scenario A, with $F^0=\{L^0,E^0 \}$, $F^\pm=\{L^\pm,E^\pm \}$.}\label{fig-Afeyn}
\end{figure*}
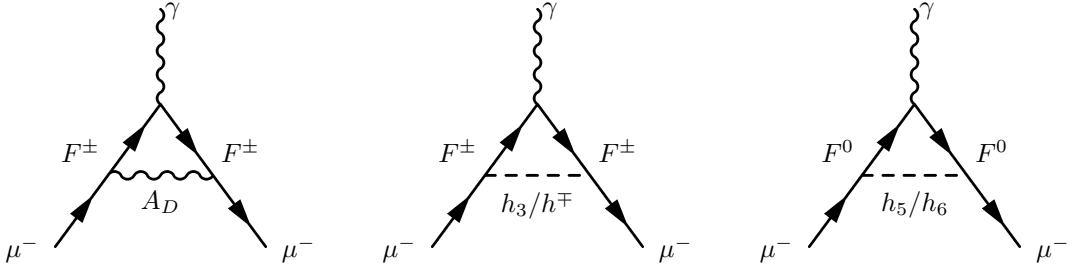

Scenario A features all of the same contributions to $\Delta a_\mu$ that our minimal construction in \cite{Wojcik:2022woa} does: Namely chirally-enhanced contributions from one-loop diagrams with dark photons and the dark Higgs, with intermediate portal matter fermions. In contrast to the minimal model presented in that work, however, the chirally enhanced contributions here feature the contributions from the exchange of heavy hidden gauge bosons and scalars, in addition to the contributions of the dark photon and dark Higgs that are associated with the $U(1)_D$ gauge group. Keeping only the chirally enhanced contribution
\begin{align}\label{eq:amu-1}
    a_\mu = a_\mu^{A_D h_D} + a_\mu^{h_5 h_6} + a_\mu^{h_+,h_-},
\end{align}
where $a_\mu^X$ represents the contribution to $a_\mu$ from one-loop diagrams featuring gauge or scalar bosons $X$. For convenience, we have also depicted the relevant diagrams in Figure \ref{fig-Afeyn}. Using the general expressions of \cite{Leveille:1977rc}, we find the contributions
\begin{align}\label{eq:amu-2}
    a_\mu^{A_D h_D} &\approx \mp \sigma_\tau y_{SL} y_{SE}\frac{m_\mu m_\tau}{16 \pi^2} \cos^2(\theta_\Delta) \frac{M_L^0 M_E^0}{M_L^+ M_L^- M_E^+ M_E^-},\nonumber\\
    a_\mu^{h_5 h_6} &\approx \frac{y_{SL} y_{SE}}{16 \pi^2} m_\mu (m_\tau \sigma_\tau a^{h5,h6}_{\tau} + m_{HV} a^{h5,h6}_{HV}),\\
    a_\mu^{h+,h-} &\approx \frac{y_{SL} y_{SE}}{16 \pi^2} m_\mu (m_\tau \sigma_\tau a^{h\pm}_{\tau} + m_{HV} a^{h+,h-}_{HV}),\nonumber\\
    \sigma_\tau &\equiv \textrm{sign}[y_H (M_L^+ m_E^+ \pm M_L^- M_E^-)],  \nonumber
\end{align}
where $ m_{HV} \equiv y_{HV} v/{\sqrt{2}}$, and
\begin{align}\label{eq:amu-3}
    a^{h5,h6}_{\tau} &\equiv M_L^0 M_E^0 \int_{0}^1 dx \, x^3\frac{- M_{h^\pm}^2 c_\Delta^2 (1-x) P_x(M_{h^\pm}^2,{M_L^0}^2+{M_E^0}^2) - Q_x}{P_x(M_{h_5}^2,{M_L^0}^2) P_x(M_{h_6}^2,{M_L^0}^2) P_x(M_{h_5}^2,{M_E^0}^2) P_x(M_{h_6}^2,{M_E^0}^2)},\nonumber\\
    a^{h5,h6}_{HV} &\equiv \int_{0}^1 dx \, (1-x)x^2\frac{M_{h_5}^2 M_{h_6}^2 (1-x) P_x(M_{h^+}^2,{M_L^0}^2+{M_E^0}^2) + M_{h^\pm}^2 s_\Delta^2 Q_x}{P_x(M_{h_5}^2,{M_L^0}^2) P_x(M_{h_6}^2,{M_L^0}^2) P_x(M_{h_5}^2,{M_E^0}^2) P_x(M_{h_6}^2,{M_E^0}^2)},\nonumber\\
    a^{h+ h-}_\tau &\equiv \frac{-2 s_\Delta^2 M_L^0 M_E^0}{M_L^+ M_E^+ \pm M_L^- M_E^-} \int_{0}^1 dx \, \bigg( \frac{ M_L^+ M_E^+}{P_x(M_{h^\pm}^2, {M_L^+}^2) P_x(M_{h^\pm}^2, {M_E^+}^2)} \pm (M_{L,E}^+ \rightarrow M_{L,E}^-)\bigg),\\
    a^{h+, h-}_{HV} &\equiv s_\Delta^2 \int_{0}^1 dx \, (1-x)x^2  \bigg( \frac{M_{h^+}^2}{P_x(M_{h^\pm}^2,{M_L^+}^2)P_x(M_{h^\pm}^2, {M_E^+}^2)}+(M_{L,E}^+ \rightarrow M_{L,E}^-)\bigg),\nonumber\\
    P_x(A,B) &\equiv (1-x)A + x B, \;\;\; Q_x \equiv {M_L^0}^2 {M_E^0}^2 x^2 - M_{h_5}^2 M_{h_6}^2 (1-x)^2, \;\;\; (s_\Delta, c_\Delta) \equiv (\sin \theta_\Delta, \cos \theta_\Delta). \nonumber
\end{align}
Notably, because these expressions are invariant under $M_{L,E}^+ \leftrightarrow M_{L,E}^-$ and $M_L^\pm \leftrightarrow M_E^\pm$, for the purposes of our numerical explorations here we can assume that the lightest portal matter field is the isodoublet with $+1$ charge under $U(1)_D$, with mass $M_L^+$. Numerically, we shall also find that although the various physical signs in our model, namely those of $y_{HV}$, $y_{H}$, and $\sin(\theta_E + \theta_\Phi)$ have physical effects, certain choices are overwhelmingly favored in order to reproduce the observed $g-2$ anomaly. First, because the signs of $y_{HV}$ and $y_{H}$ are arbitrary, we may always select them so that the $y_{HV}$ and $y_H$ contributions to $\Delta a_\mu$ interfere constructively and are of the appropriate sign to be consistent with experiment. This is, however, a physical assumption, and other possibilities exist-- in particular one of the two terms may actually contribute \emph{negatively} to $\Delta a_\mu$ and be compensated by a larger contribution of the other. In practice, this choice only has notable qualitative effects in the case that $y_{HV} > y_{H}$, where we shall explore it, and otherwise we shall assume that $y_{HV}$ and $y_{H}$'s signs are such that both terms contribute to $\Delta a_\mu$ in the direction consistent with the anomaly. The second physical sign selection we make, the sign of $\sin(\theta_E + \theta_\Phi)$ has a slightly more subtle effect on our results. The sign of $\sin(\theta_E + \theta_\Phi)$ leaves the $y_{HV}$ terms of $\Delta a_{\mu}$ invariant, but if $\sin(\theta_E + \theta_\Phi) < 0$ various contributions to the $y_H$ terms will interfere destructively, severely reducing the magnitude of this contribution relative to the $\sin(\theta_E + \theta_\Phi) > 0$ case. When applicable, we will discuss the significance of our $\sin(\theta_E + \theta_\Phi)$ sign choice.

In total, then, the chirally enhanced contributions to $\Delta a_\mu$ will be dependent on fixed SM parameters, plus a comparatively small subset of the model parameters that we have outlined in Eqs.(\ref{eq:param-scals}-\ref{eq:param-signs}): The masses of the portal matter fermions (where we always take $M_L^+$ to be the lightest) and the two heavy scalars $h_5$ and $h_6$ (or equivalently, the mass of the heavy scalar $h_\pm$ and the angle $\theta_M$, defined in Eq.~(\ref{eq:boson-masses})), the angle $\theta_\Delta$ (defined below Eq.~(\ref{eq:vevs})), the Yukawa coupling $y_{HV}$, and the product of Yukawa couplings $y_{SL} y_{SE}$, plus the sign choices of $\sin(\theta_E + \theta_\Phi)$, $y_{H}$, and $y_{HV}$. There are several points to be made about these parameters which might guide our numerical explorations of the model's parameter space. First, as is apparent from Eqs.(\ref{eq:amu-1}-\ref{eq:amu-3}), when the masses of the heavy scalars are non-negligible both the Yukawa couplings $y_H \sim m_\tau/v$ and $y_{HV}$ can play a role in the predicted value of $g-2$. This is particularly significant because while $y_H$ is determined entirely by the value of the $\tau$ mass and the relative masses of the various portal matter fermions, $y_{HV}$ can in principle be assigned any (perturbative) value we like-- apart from affecting some branching fractions of heavier vector-like fermions to lighter ones, its only observable phenomenological consequence lies in the $g-2$ anomaly. We shall therefore consider two scenarios in our numerical investigations here: First, that $y_{HV} \ll y_{H}$ and so does not significantly contribute to the $g-2$ anomaly (perhaps, for example, if its value were related to the electron mass in a larger model of lepton flavor), and second that $y_{HV}$ can take on up to $O(1)$ values.

Second, we can intuitively see that the new physics contribution to $\Delta a_\mu$ will be somewhat sensitive to the relative values of the masses of the portal matter fermions -- if we set the lightest portal matter mass at some value, larger mass splittings between the portal matter fermions will generally suppress $\Delta a_\mu$ as the higher masses of the heavier portal matter fields will suppress their contributions in loops. For our numerical analysis, we shall consider a ``typical'' selection of $O(1)$ proportional mass splittings between portal matter fermions in most cases-- $M_L^-/M_L^+ = 1.3$, $M_E^+/M_L^+ = 1.5$, and $M_E^-/M_L^+ = 1.8$ to get a sense of the numerics of a reasonable point in parameter space, as well as occasionally considering a ``maximal'' case in which all portal matter masses are degenerate and therefore the value of $\Delta a_\mu$ is maximized. Of course, the kinetic mixing between the dark and visible sectors will be weakly (logarithmically) sensitive to the mass splitting between the portal matter states, which suggests that requiring a kinetic mixing of $\gsim 10^{-(4-5)}$ likely merits at least an $O(1)$ mass splitting reminiscent of our typical case, and certainly renders the fully mass-degenerate scenario unfeasible for producing a viable dark matter model. The mass-degenerate numerics, where depicted, should not be considered a realistic scenario, therefore, but more of an absolute bound on the magnitude of $\Delta a_\mu$ that might be achieved in the model.

Finally, we note that while the new particle masses and the $\theta_\Delta$ angle are both essentially arbitrary in the absence of additional experimental evidence, we can estimate limits on the Yukawa couplings from partial wave unitarity. Following the analysis of \cite{Allwicher:2021rtd} (generalized to incorporate scalars in the adjoint representations of $SU(2)$ rather than simply in the fundamental representation), we find that partial wave unitarity requires that
\begin{align}\label{eq:yL-yE-unitarity}
    y_{SL}^2,y_{SE}^2 \leq 16 \pi, \;\;\; 3 y_{SE}^2 + 5 y_{SL}^2 + \sqrt{(3 y_{SE}^2 + 5 y_{SL}^2)^2 - 28 y_{SE}^2 y_{SL}^2} < 32 \pi,
\end{align}
such that
\begin{align}\label{eq:yL-yE-unitarity2}
    y_{SL} y_{SE} \leq \frac{16 \pi}{7} (\sqrt{15}-2 \sqrt{2}) \sim 7.5,
\end{align}
and
\begin{align}\label{eq:yHV-unitarity}
    y_{HV}^2 \leq \frac{8 \pi}{3}.
\end{align}
While these couplings might be further constrained in future experiments, in particular from muon collider monophoton searches in the case of $y_{SL,SE}$ (see Section \ref{sec:boson-pheno}) and  branching fractions of heavier vector-like fermions (in the event of their discovery) for $y_{HV}$, it is useful to keep in mind these bounds which maintain perturbative unitarity.

\begin{figure}
    \centerline{\includegraphics[width=3.5in]{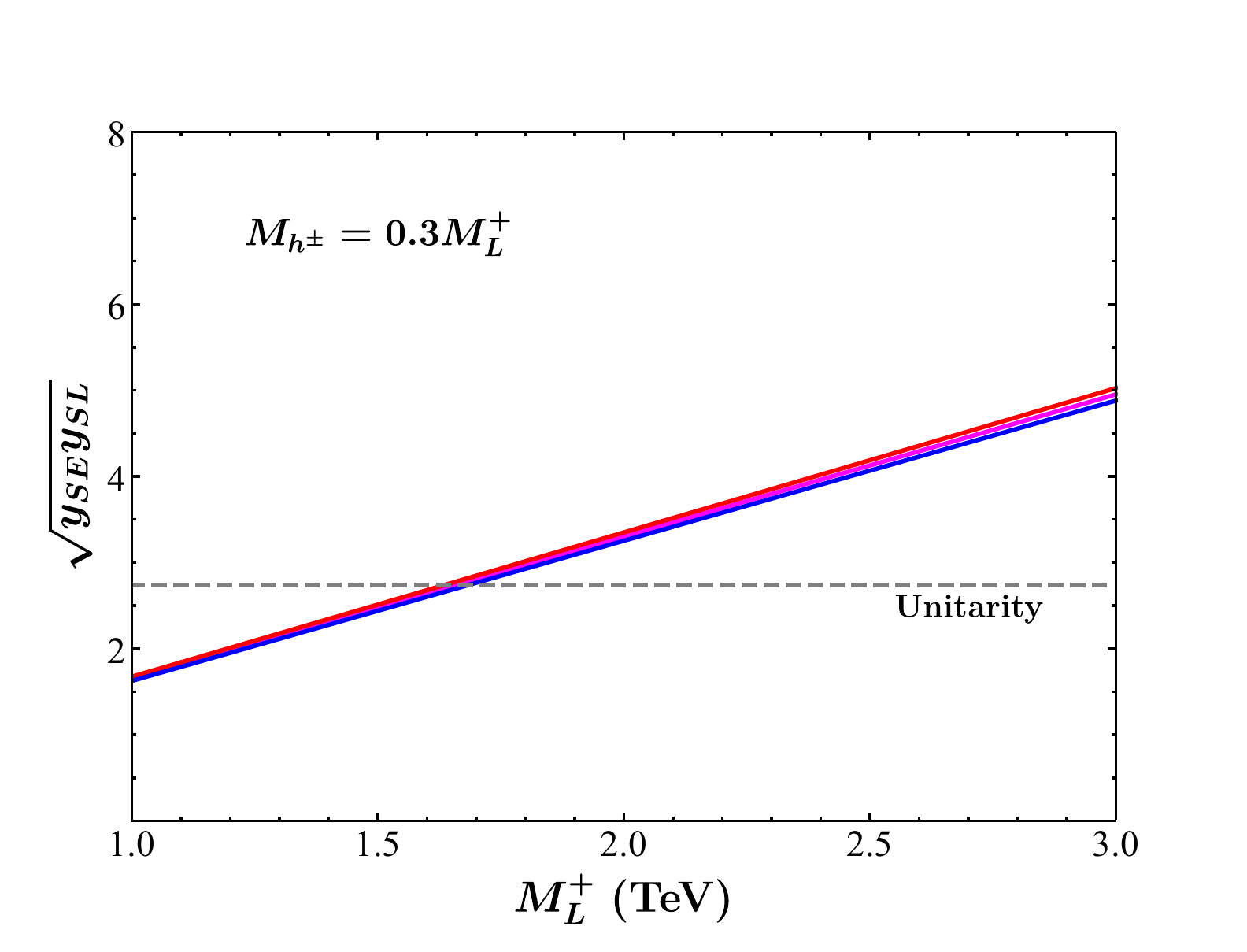}
    \hspace{-0.75cm}
    \includegraphics[width=3.5in]{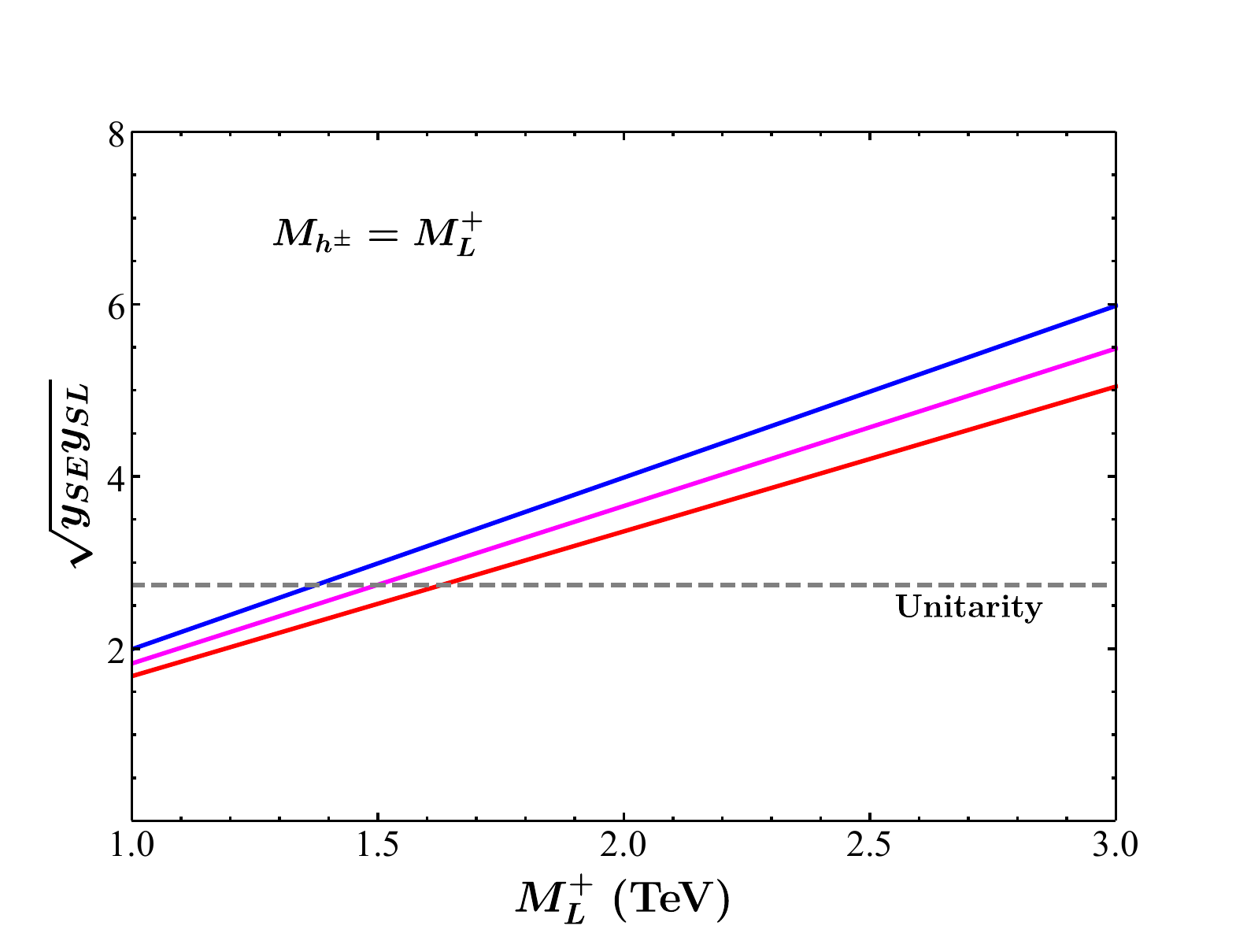}}
    \vspace*{-0.75cm}
    \centering
    \includegraphics[width=3.5in]{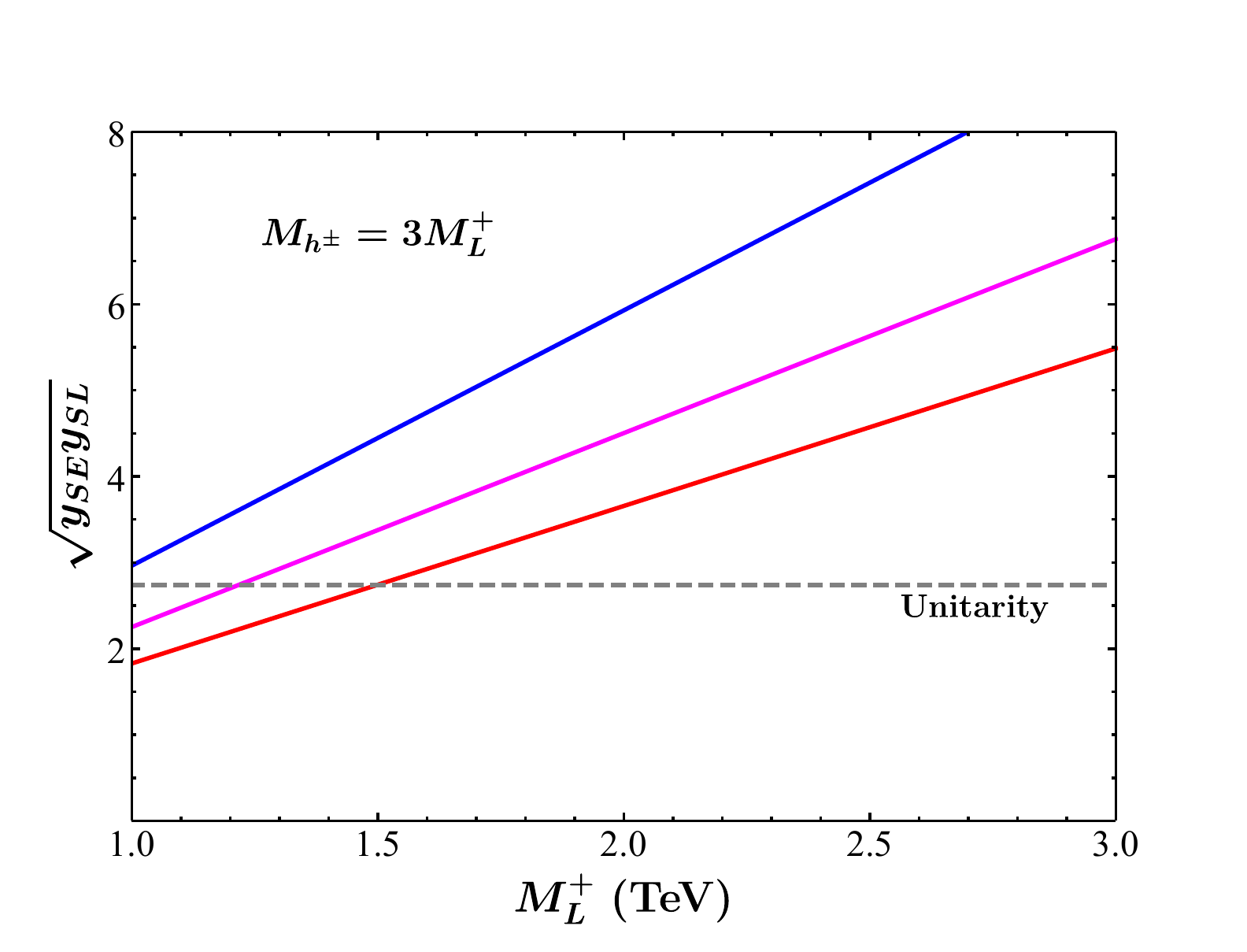}
    \caption{The necessary coupling $\sqrt{y_{SL} y_{SE}}$ in order to recreate the observed $g-2$ anomaly as a function of $M_L^+$ in Scenario A, assuming $M_L^- = 1.3 M_L^+$, $M_E^+ = 1.5 M_L^+$, and $M_E^- = 1.8 M_L^+$, where different line colors denote $\theta_\Delta = 3\pi/8$, $\theta_M = \pi/8$ (Blue), $\theta_\Delta = \theta_M = \pi/4$ (Magenta), $\theta_\Delta = \theta_M = \pi/8$ (Red). We have taken $M_{h^\pm}$ to have values relative to $M_L^+$ as shown. The unitarity bound shown is $y_{SL} y_{SE} \approx 7.5$.}
    \label{fig2}
\end{figure}

\begin{figure}
    \centerline{\includegraphics[width=3.5in]{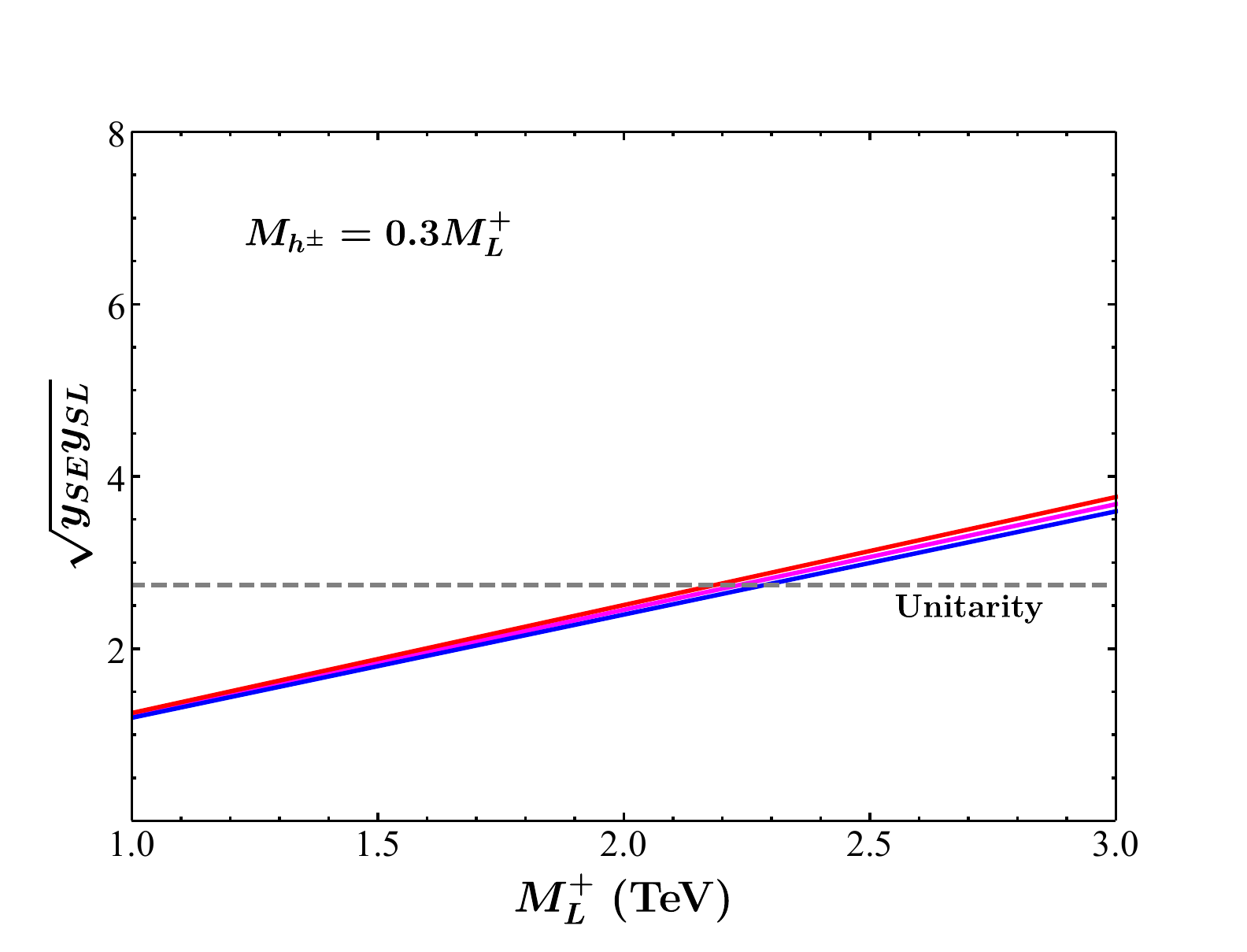}
    \hspace{-0.75cm}
    \includegraphics[width=3.5in]{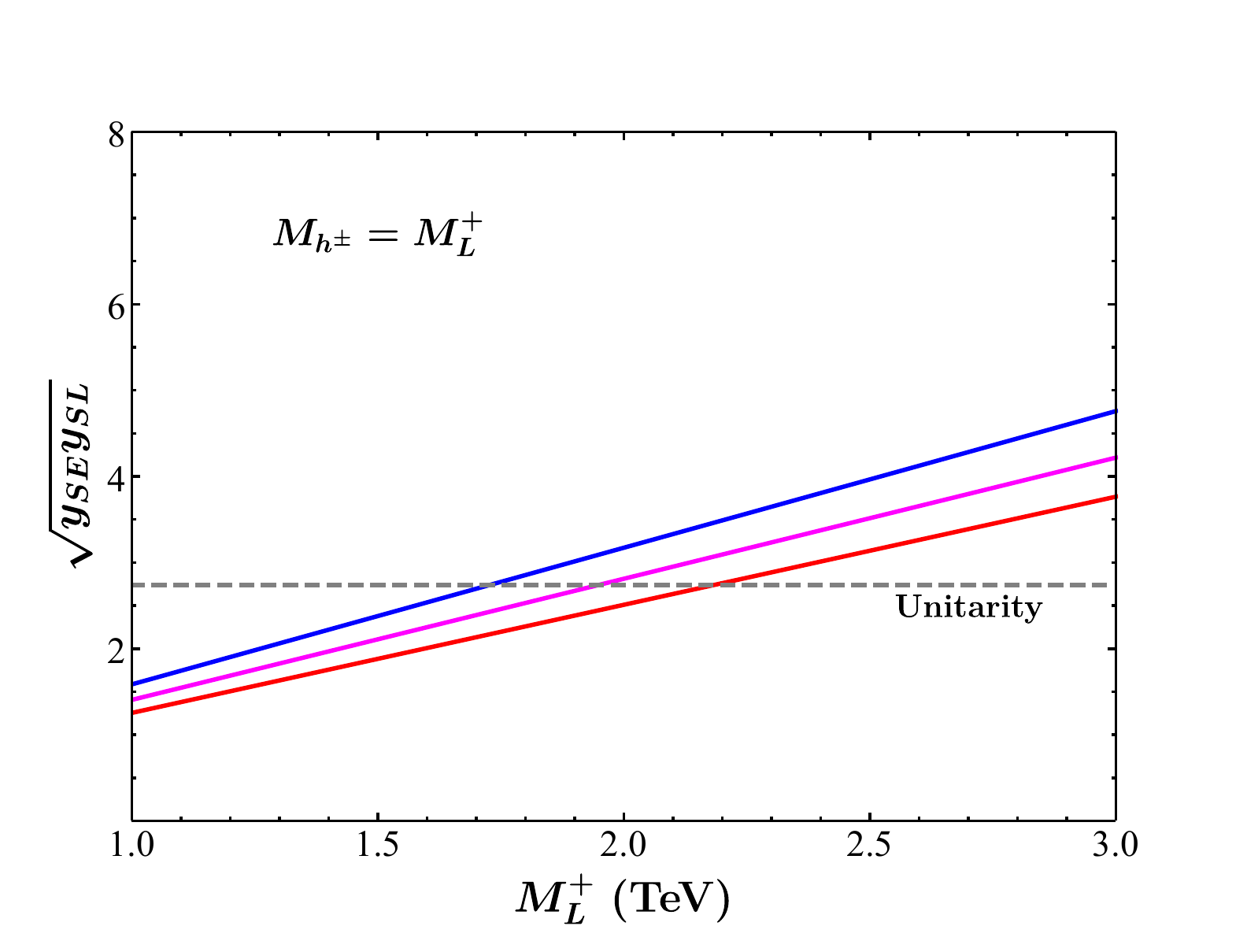}}
    \vspace*{-0.75cm}
    \centering
    \includegraphics[width=3.5in]{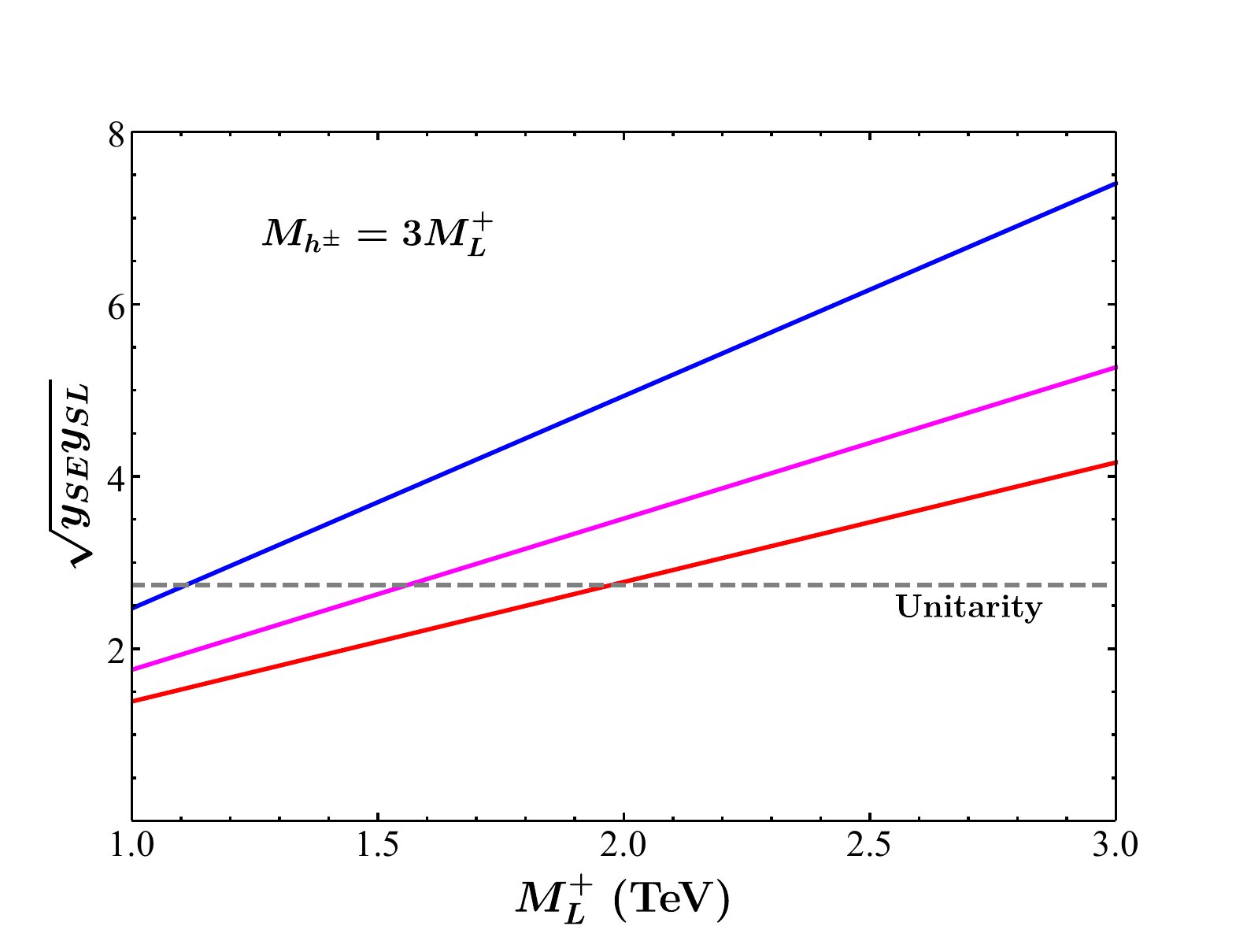}
    \caption{As in Figure \ref{fig2}, but with all portal masses assumed to be degenerate in order to near-maximize the new physics contribution to $g-2$ in Scenario A.}
    \label{fig2-degen}
\end{figure}

With the above observations in mind, we can now move on to considering a probe of some benchmark points in the model parameter space, beginning with the case in which $y_{HV} \ll y_{H}$, and hence the terms $a^{h_5,h_6}_{HV}$ and $a^{h+ h-}_{HV}$ do not contribute appreciably to $\Delta a_\mu$.
This scenario has the appealing consequence that the $g-2$ correction is specifically related to another observable quantity, namely the $\tau$ mass. In turn, this relationship significantly limits the portal matter mass range for which the model in this regime can reproduce the observed $g-2$ anomaly. In Figures \ref{fig2} and \ref{fig2-degen}, we have plotted the necessary value of the product $y_{SL} y_{SE}$ in order to fully account for the observed $g-2$ anomaly in the $y_{HV} \ll y_{H}$ scenario as a function of $M_L^+$, assuming that $y_H$ is of the appropriate sign to make the resultant $\Delta a_\mu$ consistent with experiment, and taking $\sin(\theta_E + \theta_\Phi)>0$ (when $\sin(\theta_E + \theta_\Phi)<0$, the destructive interference among various $\Delta a_\mu$ corrections renders it unrealistic that the observed anomaly can be reproduced, so we do not consider that case here). In Figure \ref{fig2}, we consider a typical $O(1)$ mass splitting between the various portal matter masses, while in Figure \ref{fig2-degen}, consider the case in which the portal matter masses are all degenerate (or at least have insignificant mass splitting), which will generally maximize the value of $\Delta a_\mu$. In both cases, we find that for $M_L^+ \gsim 1 \; \textrm{TeV}$, roughly on par with existing LHC constraints on leptonic portal matter (we shall discuss this and other collider constraints in Section \ref{sec:fermion-pheno}), a modest $O(1)$ enhancement of the new physics $g-2$ contribution from the product $y_{SL} y_{SE}$ is necessary. This product quickly approaches the perturbative unitarity bound on the product $y_{SL} y_{SE}$-- we see for the points in parameter space considered in Figure \ref{fig2} that $M_L^+$ must generally be lighter than $\sim 1.5 \textrm{TeV}$ in order to address the $g-2$ anomaly in this scenario, while in the more maximal case considered in Figure \ref{fig2-degen} we instead see that the lightest portal matter masses might be as large as $\sim 2.2 \; \textrm{TeV}$ before the unitarity bound is saturated. Consulting the study of \cite{Guedes:2021oqx}, we see that portal matter masses within this range are potentially within the reach of the $\sqrt{s} = 27 \; \textrm{TeV}$ HE-LHC, and likely also within the discovery reach of a muon collider with a maximum center-of-mass energy greater than $\sim 4 \; \textrm{TeV}$ \cite{OsmanAcar:2021plv}.

\begin{figure}
    \centerline{\includegraphics[width=3.5in]{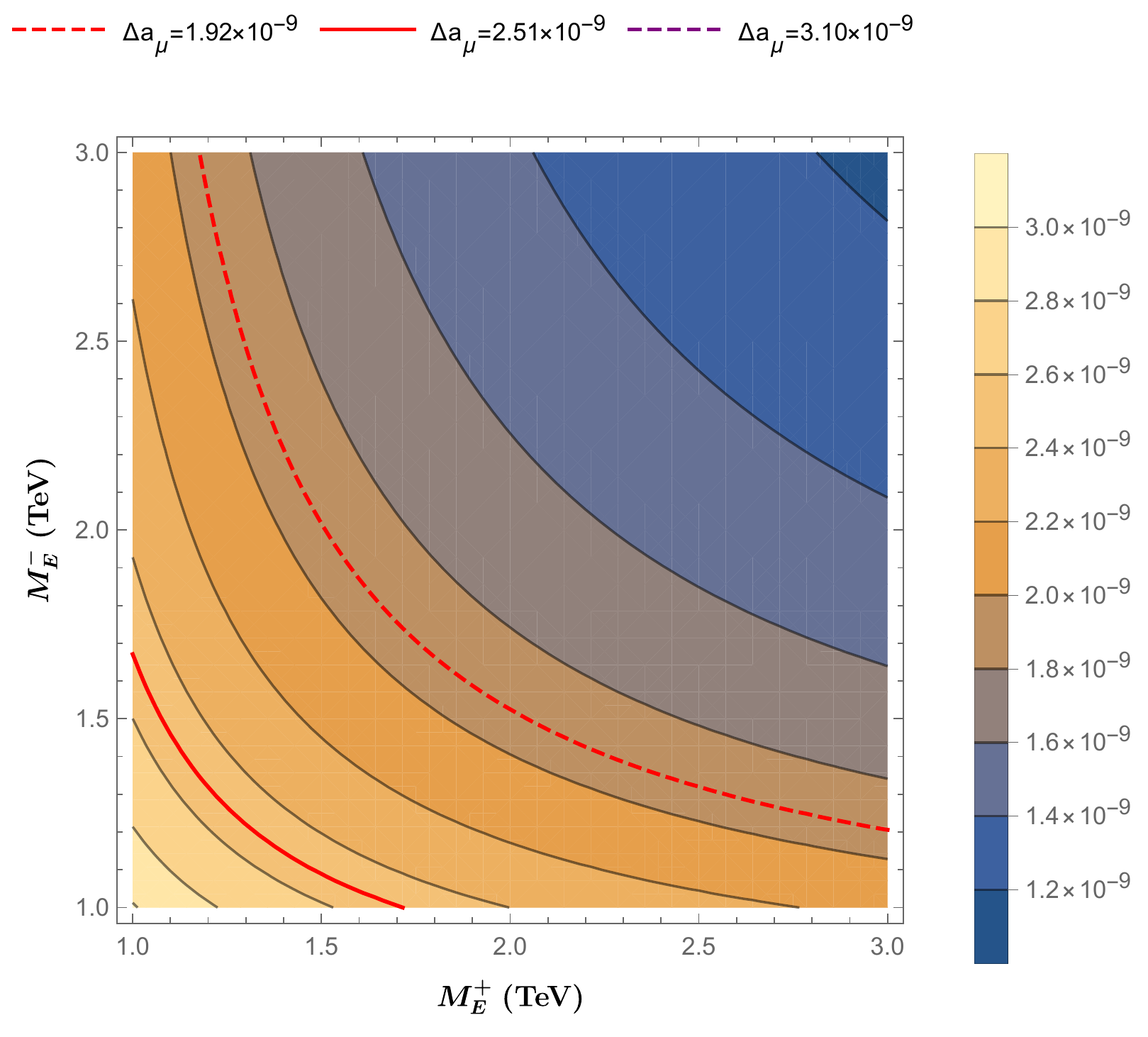}
    \hspace{-0.4cm}
    \includegraphics[width=3.5in]{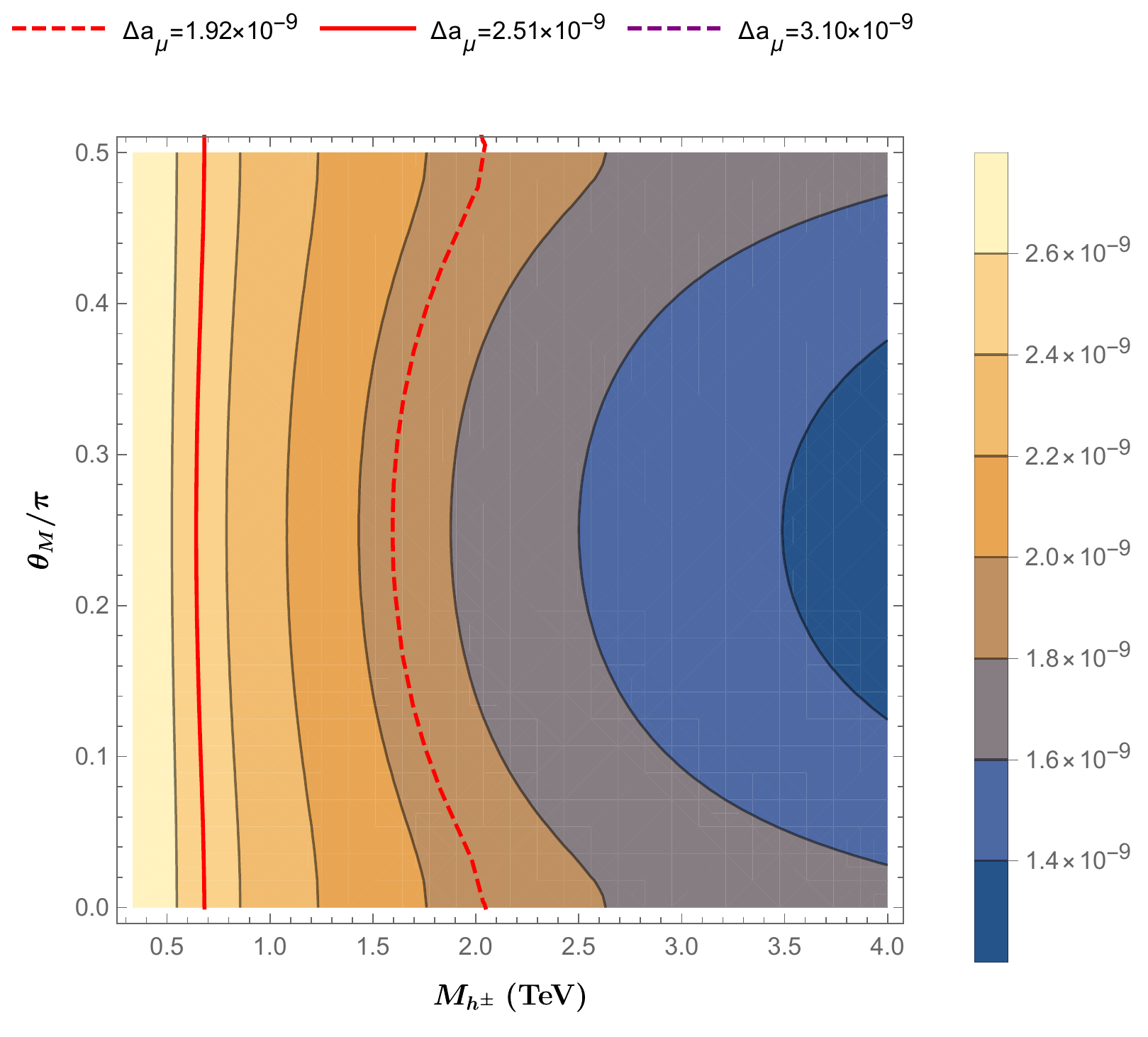}}
    \vspace*{-0.25cm}
    \centering
    \includegraphics[width=3.5in]{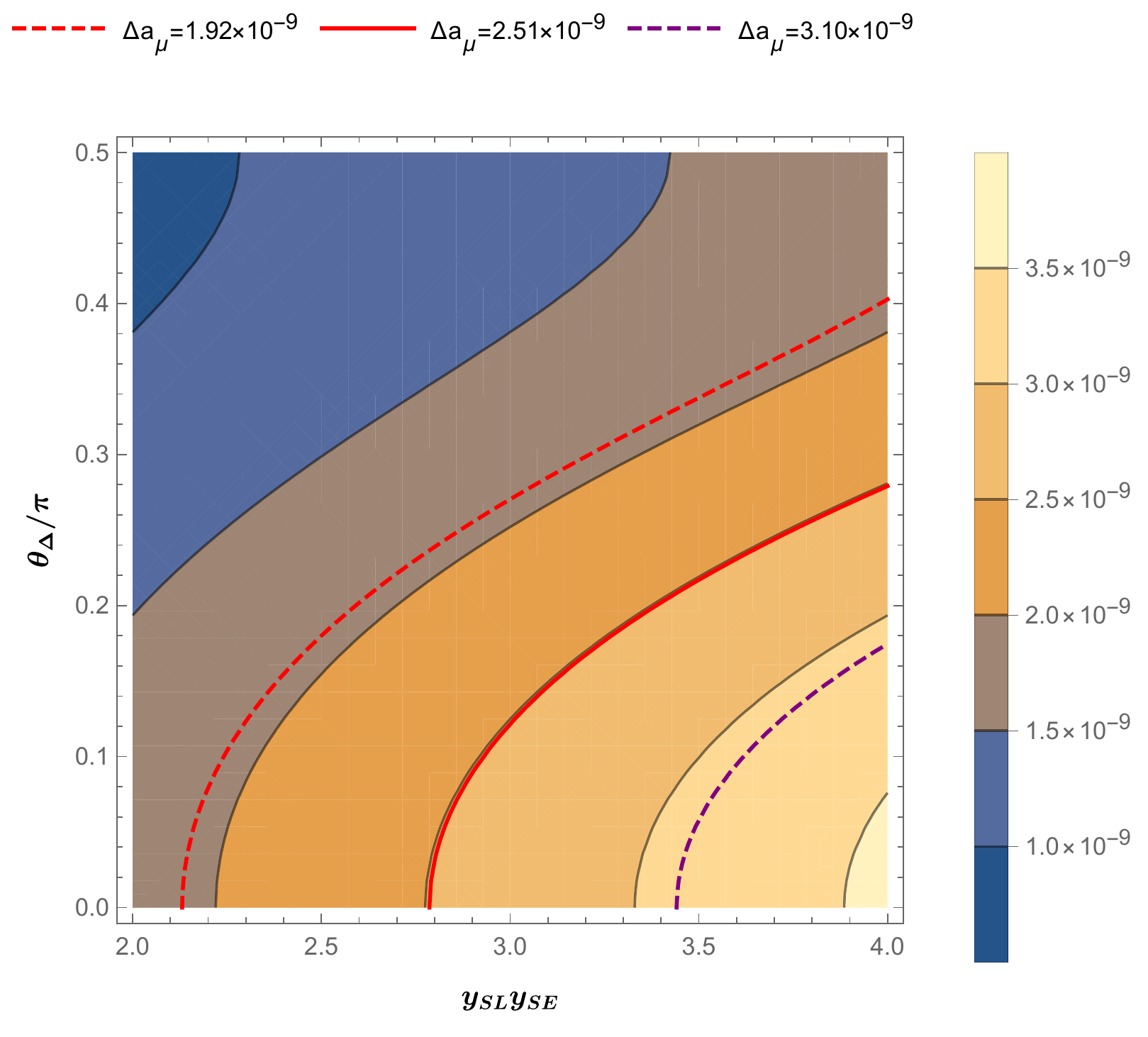}
    \caption{Contour plots of $\Delta a_\mu$ in Scenario A, with $y_{HV} \ll y_{H}$,$M_L^+ = 1 \; \textrm{TeV}$ and $y_{SL} y_{SE} = 3$.
    (Top Left): contour plot as a function of $M_E^+$ and $M_E^-$, assuming $M_{h^\pm} = 1 \; \textrm{TeV}$ and $\theta_M = \pi/4$, $M_L^- = 1.3 \; \textrm{TeV}$, and $\theta_\Delta = \pi/4$. (Top Right): Contour plot as a function of $M_{h^\pm}$ and $\theta_M$, for $M_L^- = 1.3 \; \textrm{TeV}$, $M_E^+ = 1.5 \; \textrm{TeV}$, $M_E^- = 1.8 \; \textrm{TeV}$, and $\theta_\Delta = \pi/4$. (Bottom): Contour plot as a function of $y_{SL} y_{SE}$ and $\theta_\Delta$, for $M_{h^\pm} = 1 \; \textrm{TeV}$, $\theta_M = \pi/4$, and $M_L^- = 1.3 \; \textrm{TeV}$, $M_E^+ = 1.5 \; \textrm{TeV}$, $M_E^- = 1.8 \; \textrm{TeV}$.  Note that points with $\theta_\Delta \neq \pi/4$ are unphysical 
    (see Appendix \ref{Sappendix}), but due to a lack of sensitivity of $\Delta a_\mu$ to $\theta_M$, the plot remains illustrative of the $\theta_\Delta$ behavior of $\Delta a_\mu$ at physical points in parameter space.}
    \label{fig3}
\end{figure}

We can get a better feel for the different parameters' effects on $\Delta a_\mu$ in the $y_{HV} \ll y_{H}$ scenario by plotting the results as functions of several different variables. In Figure \ref{fig3}, we depict contour plots of $\Delta a_\mu$ along several different variables. We can see several important behaviors from these plots. For example, we see that achieving the observed $\Delta a_\mu$ in this regime for $\gsim 1 \; \textrm{TeV}$ portal matter will require modest $y_{SL} y_{SE} \gsim O(\textrm{a few})$ enhancements of the $y_H$ contribution, and will generally be limited to scenarios in which the other portal matter fields and the scalars $h_5$ and $h_6$ are $\lesssim \textrm{a few TeV}$, certainly not much greater than $2-3 \;\textrm{TeV}$. Given that these masses are all dictated in the action by the same scale $v_\Phi$, this requirement of only modest $O(1)$ splittings between them is encouraging-- it suggests that the favored parameter space for this model requires no careful tuning of the individual heavy particle masses.

\begin{figure}
    \centerline{\includegraphics[width=3.5in]{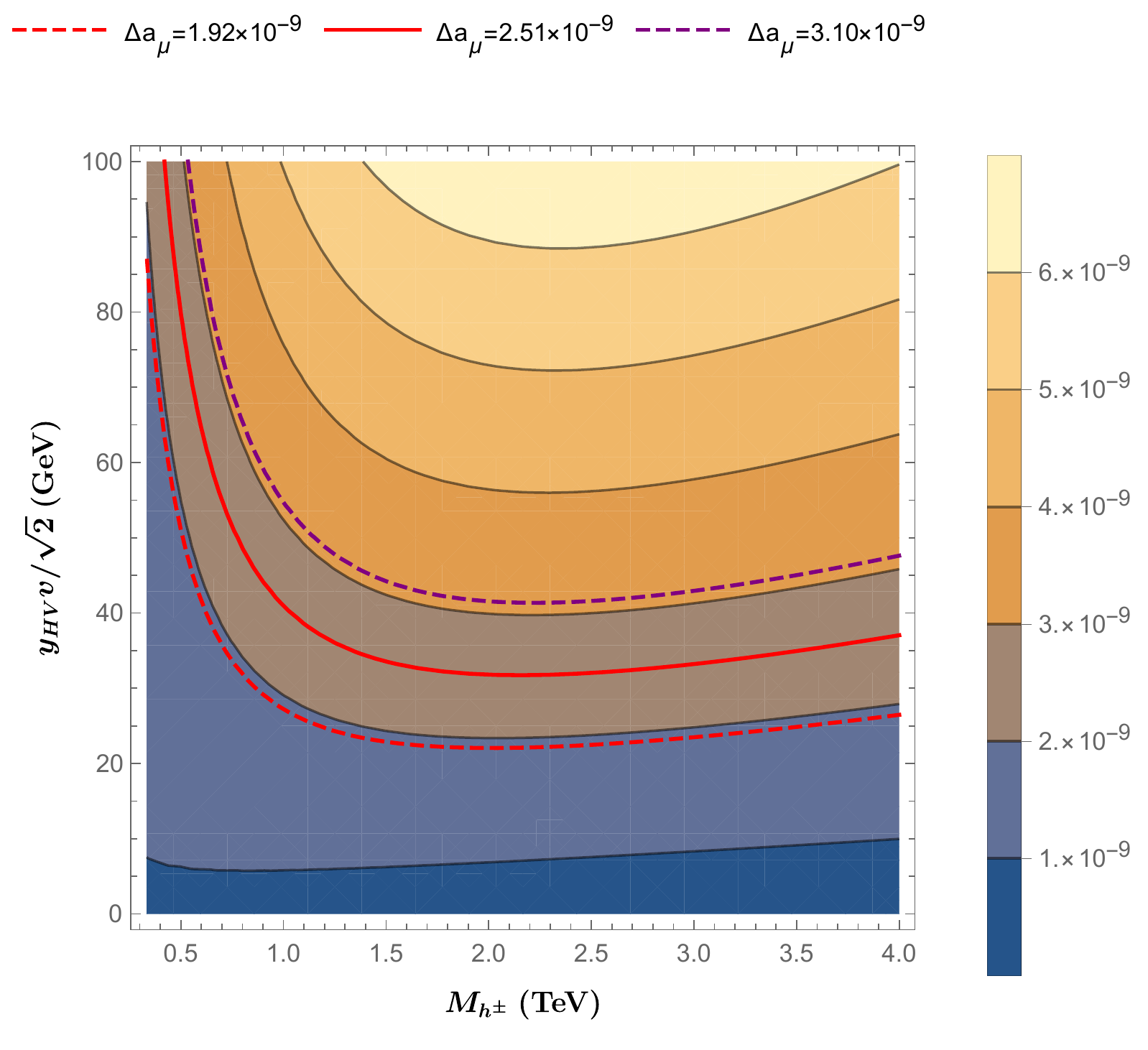}
    \hspace{-0.4cm}
    \includegraphics[width=3.5in]{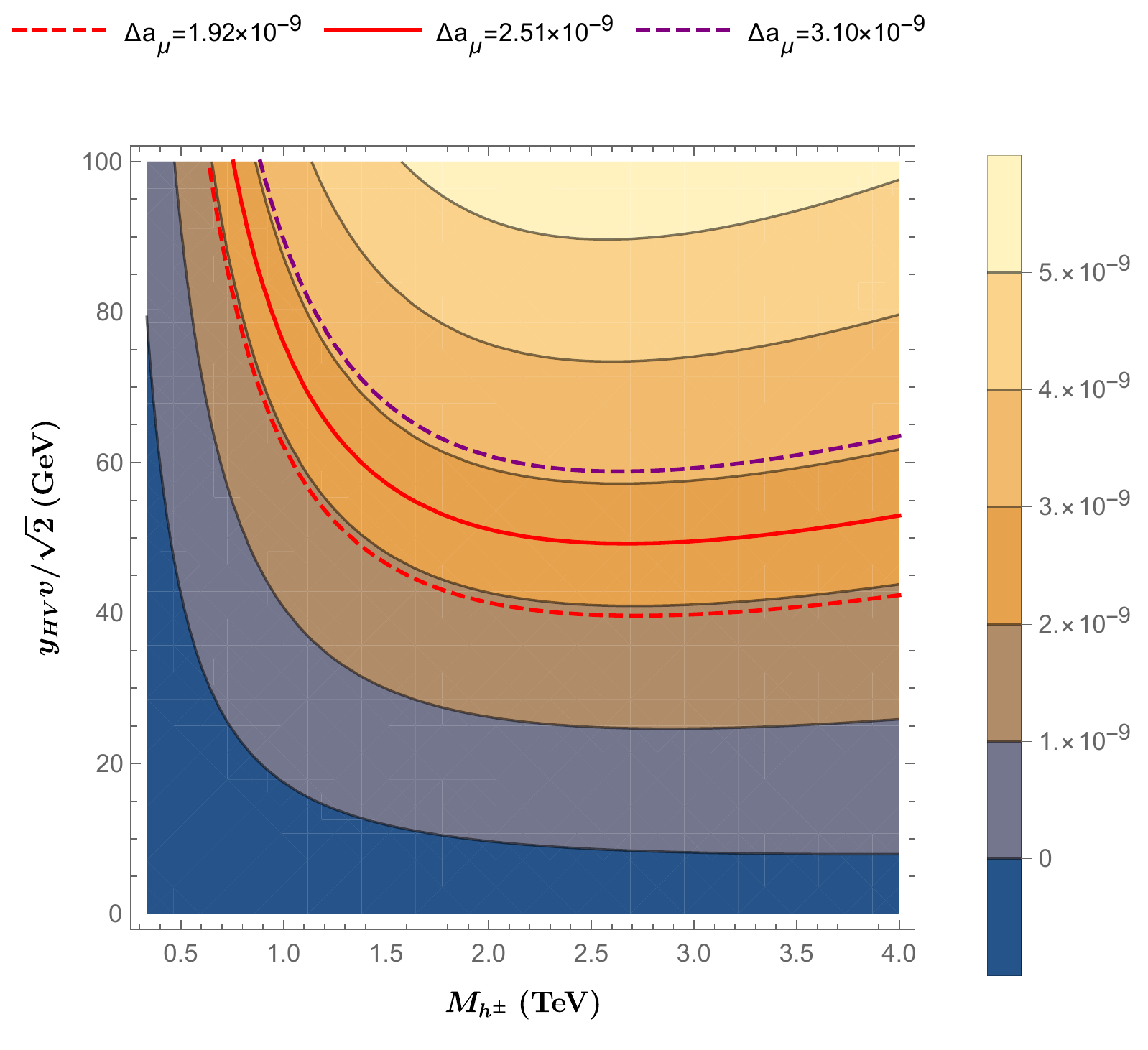}}
    \vspace*{-0.25cm}
    \centerline{\includegraphics[width=3.5in]{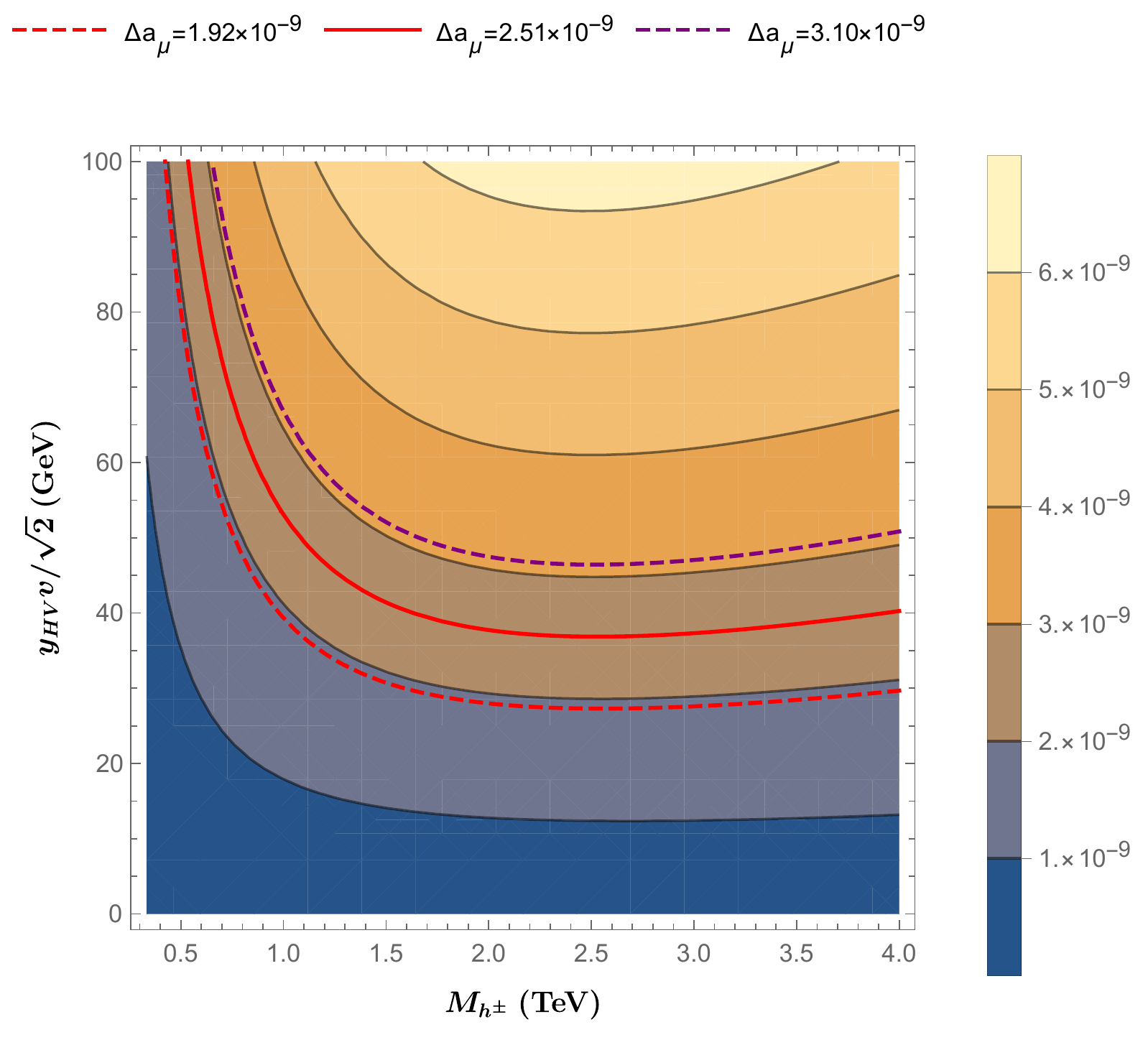}
    \hspace{-0.4cm}
    \includegraphics[width=3.5in]{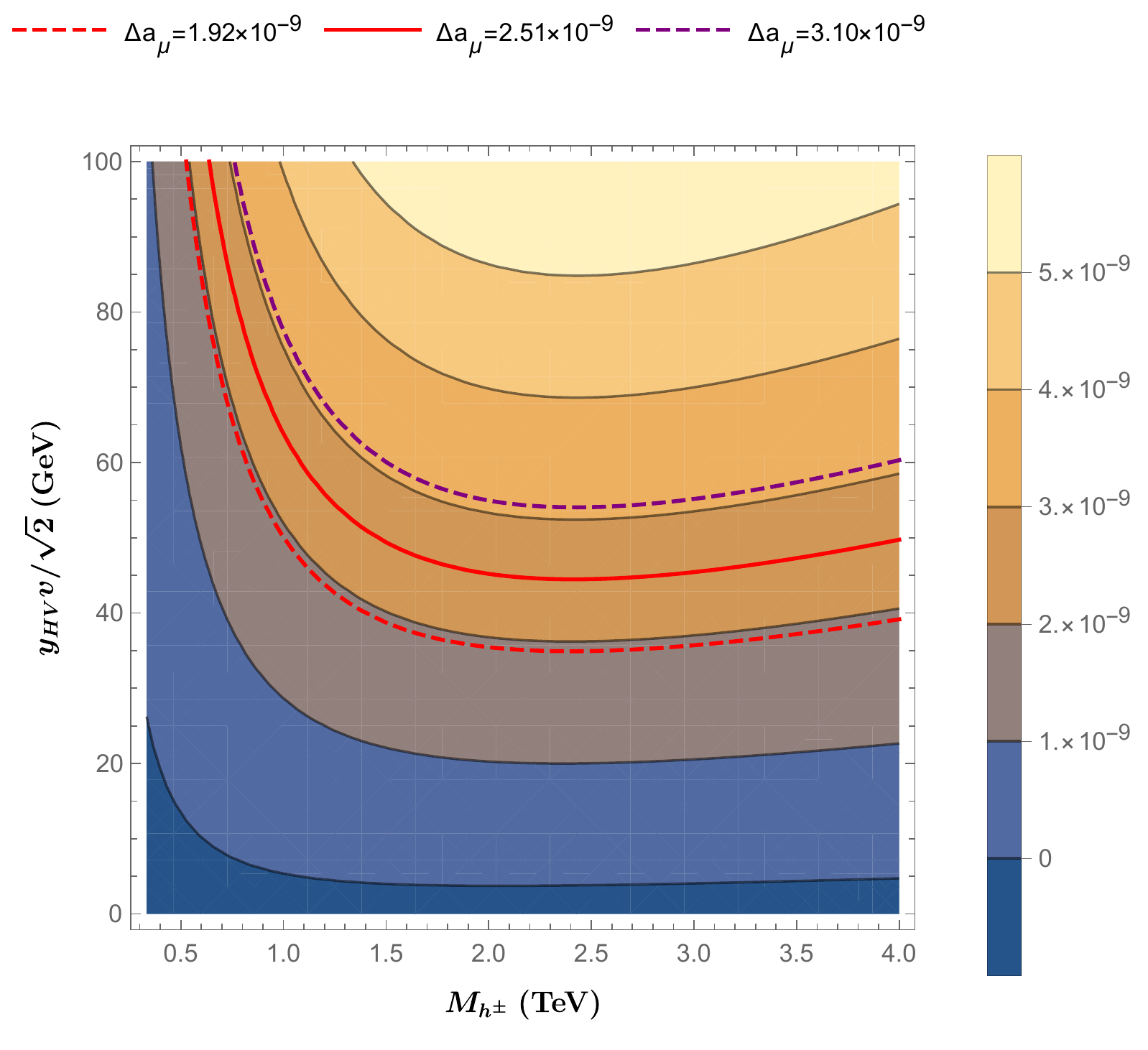}}
    \caption{Contour plots of $\Delta a_\mu$ in Scenario A, for $y_{SL} y_{SE} = 1$, $M_L^+ = 1 \; \textrm{TeV}$, $M_L^- = 1.3 \; \textrm{TeV}$, $M_E^+ = 1.5 \; \textrm{TeV}$, $M_E^- = 1.8 \; \textrm{TeV}$, $\theta_\Delta = \pi/4$, and $M_{h_5}=M_{h_6}=M_{h^\pm}/\sqrt{2}$. 
    (Top Left): Assuming the $y_{HV}$ and $y_{H}$ terms interfere constructively, and $\sin (\theta_E + \theta_\Phi) > 0$. (Top Right): 
    the $y_{HV}$ and $y_{H}$ terms interfere destructively, and $\sin (\theta_E + \theta_\Phi) > 0$. (Bottom Left):  
    the $y_{HV}$ and $y_{H}$ terms interfere constructively, and $\sin (\theta_E + \theta_\Phi) < 0$. (Bottom Right):  
    the $y_{HV}$ and $y_{H}$ terms interfere destructively, and $\sin (\theta_E + \theta_\Phi) < 0$.}
    \label{fig4}
\end{figure}

Next, we consider the scenario in which the coupling $y_{HV}$ is comparable to or greater than $y_H$. As noted before, because $y_{HV}$'s magnitude is not dictated by any other model parameter, it is feasible for it to be as large as $O(1)$, in which case $y_{HV} v/\sqrt{2} \sim 200 \; \textrm{GeV}$. In Figure \ref{fig4}, we depict contour plots of $\Delta a_\mu$ as a function of $m_{HV} = y_{HV} v/\sqrt{2}$ and $M_{h^+} = \sqrt{M_{h_5}^2 + M_{h_6}^2}$ for various choices of physical signs in the model, assuming that the lightest portal matter field has a mass of $1 \; \textrm{TeV}$.  We note that as $M_{h_5}$ and $M_{h_6}$ both approach 0, the required $y_{HV}$ value to reproduce the observed $\Delta a_\mu$ diverges, since in the limit that these masses both vanish the $y_{HV}$ term of the $\Delta a_\mu$ contribution vanishes. However, past a low-mass regime in which the observationally consistent $y_{HV}$ value depends strongly on the particular value of $M_{h^+}$, which generally ends by $M_{h^+} \gsim 2 \; \textrm{TeV}$ for all sign conventions we consider here, $\Delta a_\mu$ becomes only weakly dependent on the specific value of the scalar mass. Notably, we see that for $y_{SL} y_{SE} = 1$, the observed value of $\Delta a_\mu$ can be recreated for $m_{HV} \sim 20-50 \; \textrm{GeV}$ as long as the $h_5$ and $h_6$ scalar bosons have masses $\gsim 0.5-1 \; \textrm{TeV})$, depending on the signs of various parameters within the model. Since these values of mass terms are well below $1 \; \textrm{TeV}$, the scale of our portal matter mass, this suggests that our perturbative treatment $y_{HV} v \ll v_\Phi$ is, like our other perturbative expansions, numerically justified here.

\begin{figure}
    \centerline{\includegraphics[width=3.5in]{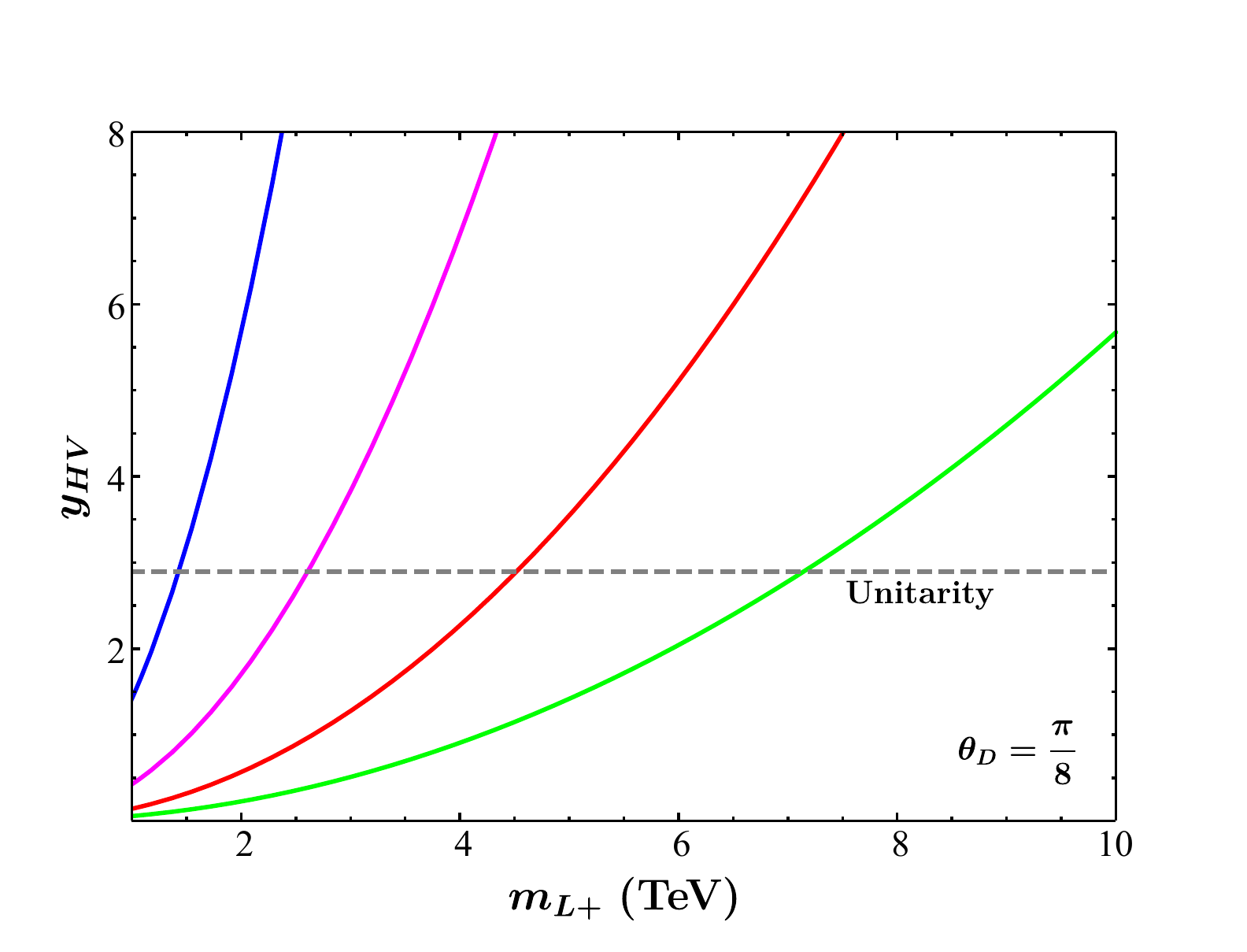}
    \hspace{-0.75cm}
    \includegraphics[width=3.5in]{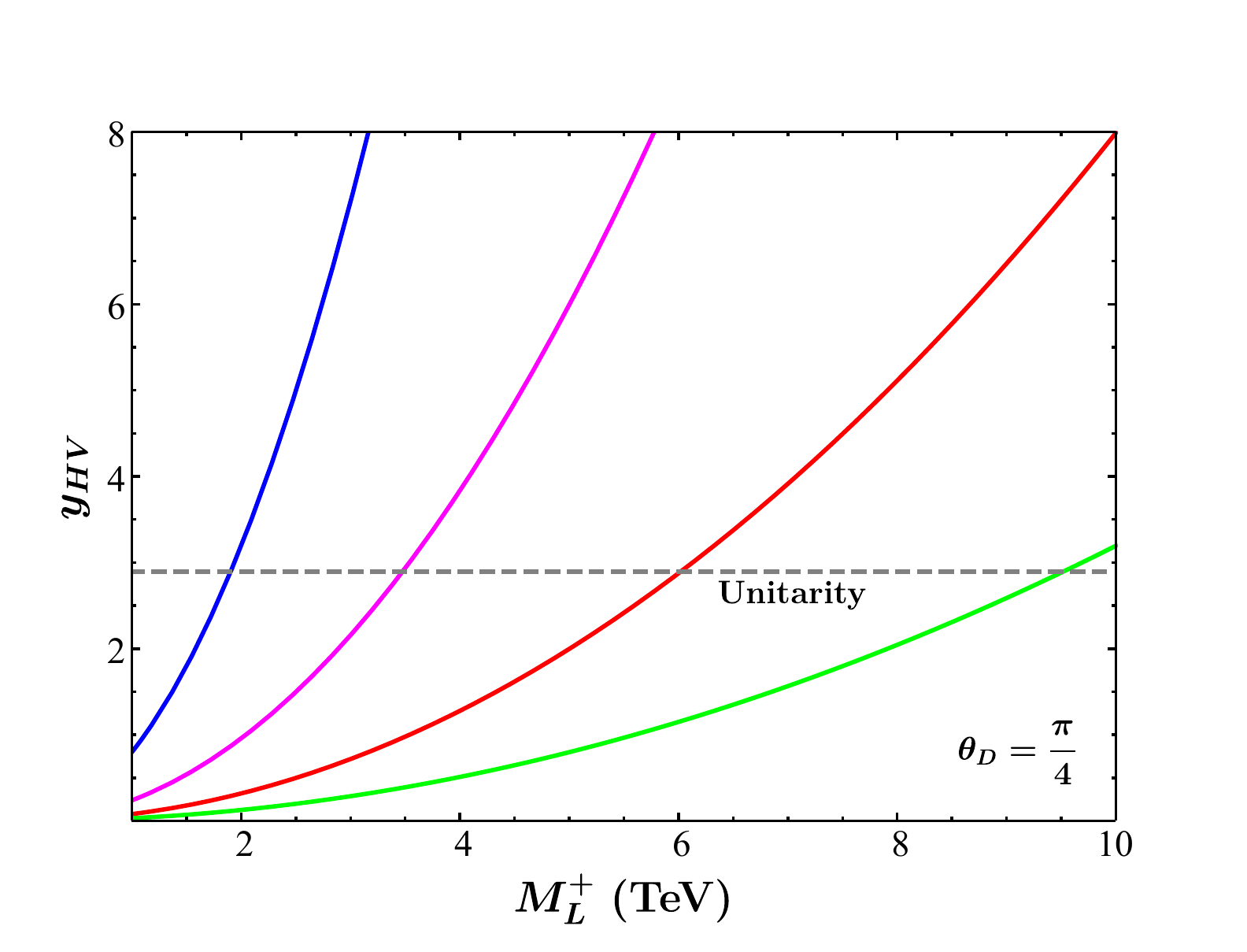}}
    \vspace*{-0.5cm}
    \centering
    \includegraphics[width=3.5in]{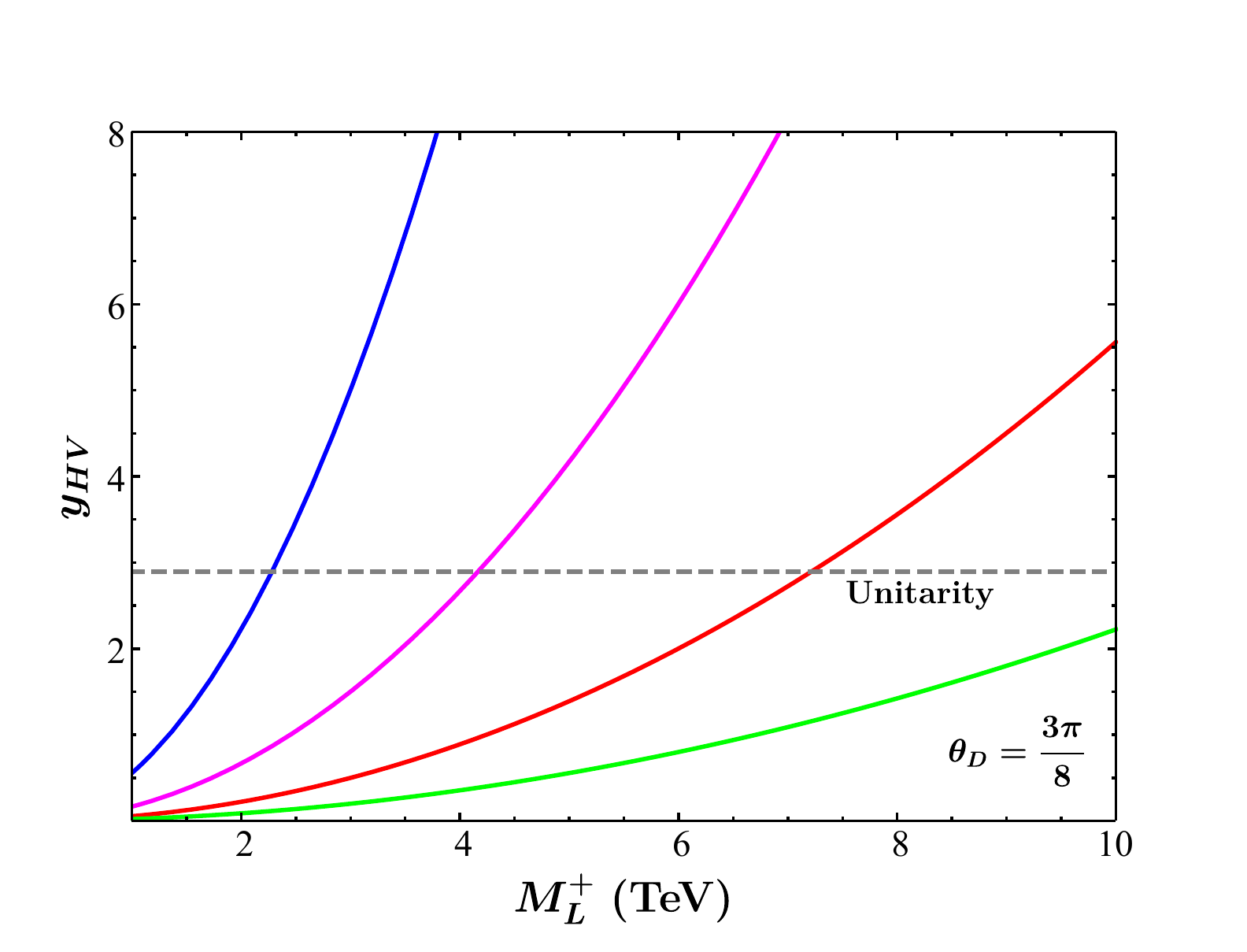}
    \caption{Plots of the $y_{HV}$ coupling necessary to reproduce $\Delta a_\mu$ as a function of $M_L^+$ in Scenario A, for  $M_L^-/M_L^+ = 1.3$, $M_E^+/M_L^+ = 1.5$, $M_E^-/M_L^+ = 1.3$. 
    The perturbative unitarity bound $y_{HV}^2 = 8 \pi/3$ is depicted as a gray dashed line, while we have taken 
    $y_{SL} y_{SE} = 0.3$ (Blue), $1$ (Magenta), $3$ (Red), and the partial wave unitarity bound $\approx 7.5$ (Green). We have also taken $\theta_\Delta = \pi/8$ (Top Left), $\pi/4$ (Top Right), and $3 \pi/8$ (Bottom). Because when $y_{HV} \sim O(1)$, the $y_{HV}$ term dominates the $y_{H}$ term, this result is insensitive to choices of the sign of $y_{H}$ and $\sin(\theta_E+\theta_\Phi)$.}
    \label{fig5}
\end{figure}

\begin{figure}
    \centerline{\includegraphics[width=3.5in]{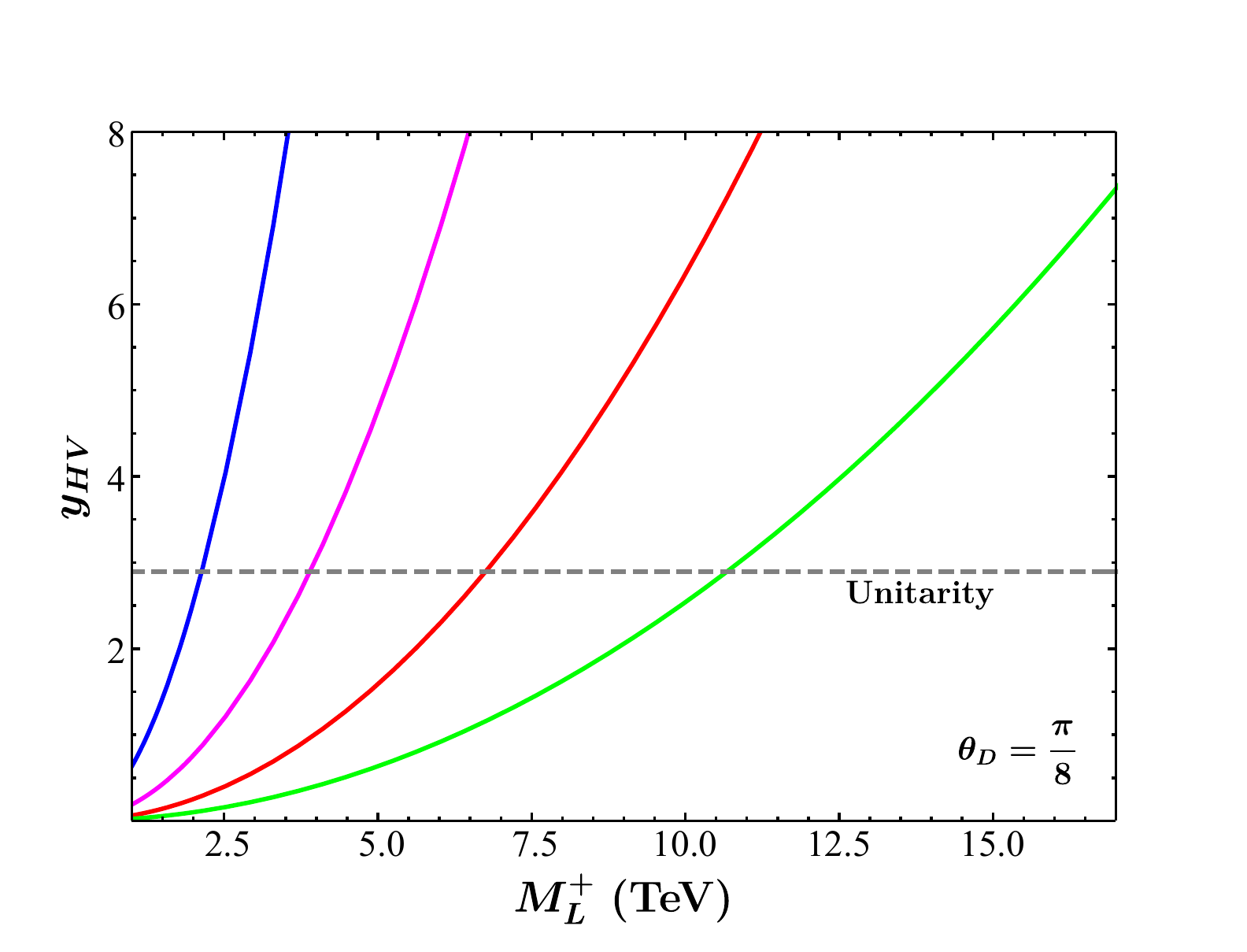}
    \hspace{-0.75cm}
    \includegraphics[width=3.5in]{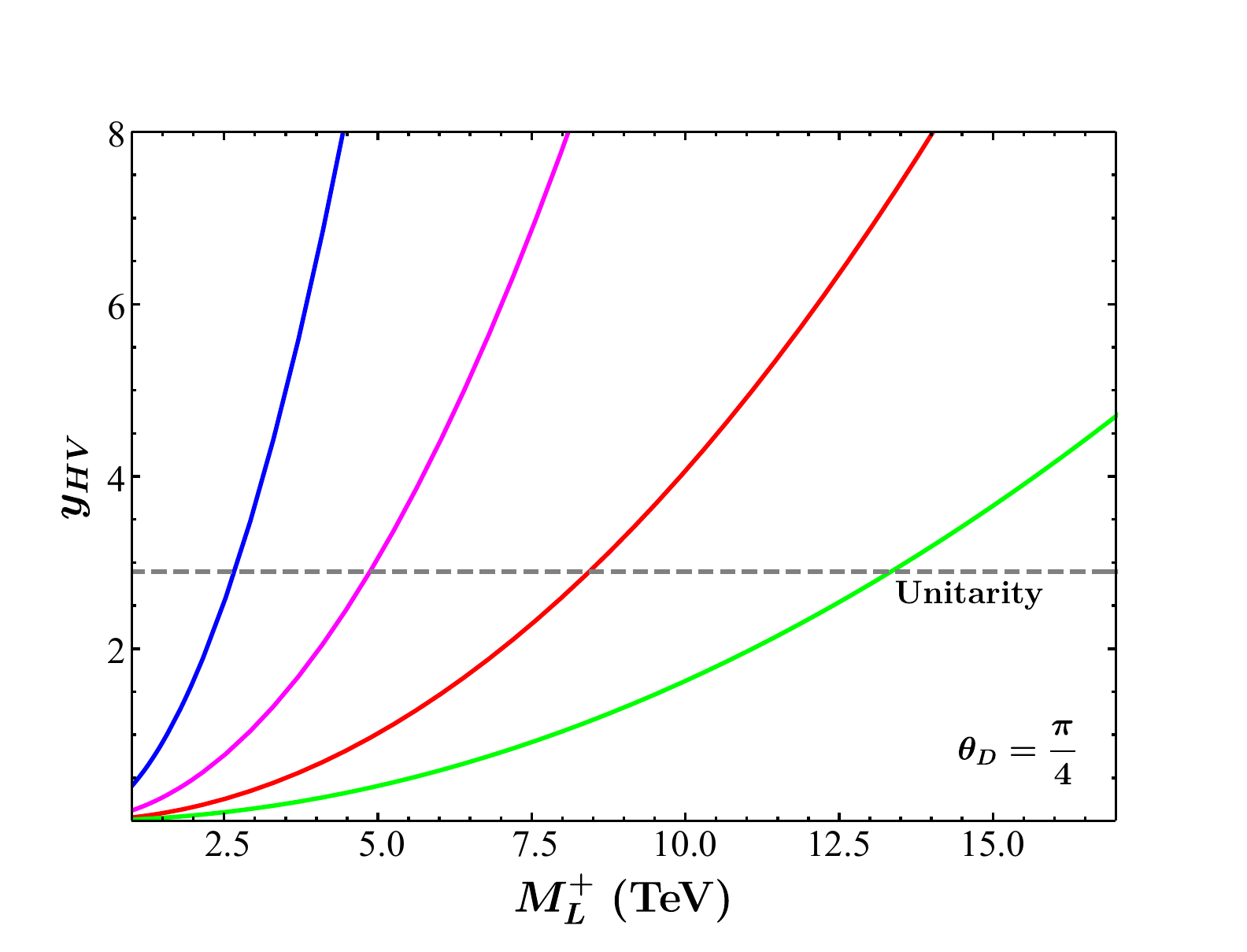}}
    \vspace*{-0.5cm}
    \centering
    \includegraphics[width=3.5in]{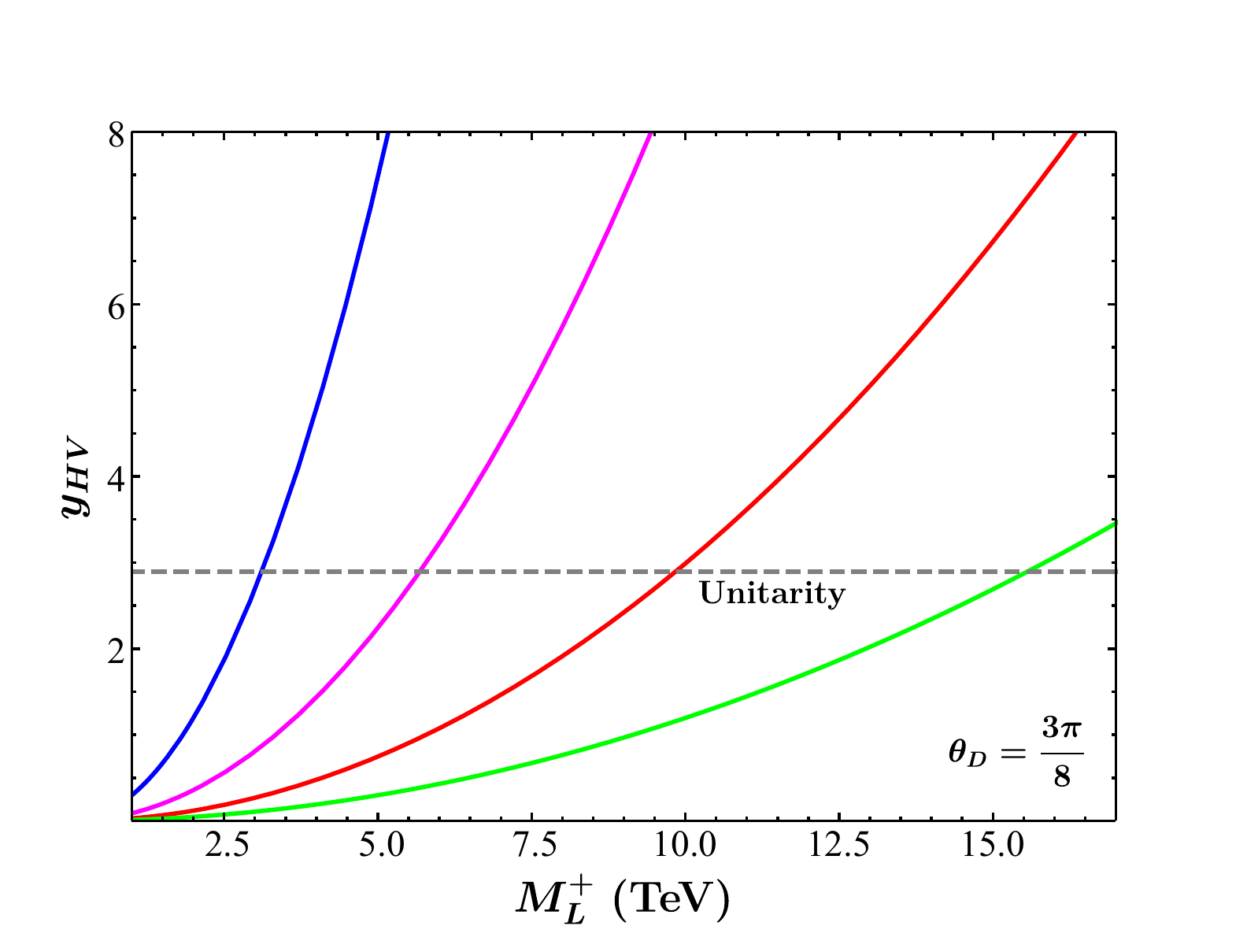}
    \caption{As in Figure \ref{fig5}, but assuming all portal matter masses are degenerate in order to estimate a near-maximal effect on $\Delta a_\mu$.}
    \label{fig5-degen}
\end{figure}

As will be discussed later in Section \ref{sec:fermion-pheno}, a constraint on $y_{HV}$ may be identified by measuring the branching fractions of certain vector-like lepton decays via the emission of electroweak bosons, since this parameter contributes to the partial widths of both $U(1)_D$-uncharged and $U(1)_D$-charged vector-like leptons decaying into states which involve other vector-like leptons. However, given that these measurements are predicated on the successful discovery of not only the lightest portal matter field, but also the heavier vector-like leptons that may possess the relevant decay channels, the principal constraint on $y_{HV}$ currently (and for the foreseeable future) can come only from perturbative unitarity, even as the Yukawa couplings $y_{SL}$ and $y_{SE}$ and the mass of the lightest portal matter field can be constrained by monophoton searches at a muon collider and vector-like lepton pair production, respectively. It is therefore of some interest to explore how large the scale of the portal matter masses can be before it becomes impossible to reproduce the observed $\Delta a_\mu$ with perturbative $y_{HV}$, namely before $y_{HV}$ saturates the partial wave unitarity bound of Eq.~(\ref{eq:yHV-unitarity}). In Figures \ref{fig5} and \ref{fig5-degen}, we take a sample spectrum of portal matter masses relative to one another ($O(1)$ separation in the case of Figure \ref{fig5} and complete mass degeneracy in the case of Figure \ref{fig5-degen}) and plot the necessary $y_{HV}$ in order to match observation as a function of the lightest portal matter mass, for various choices of the parameters $\theta_\Delta$ and $y_{SL} y_{SE}$, along with the perturbative unitarity bound on $y_{HV}$. We can see that for the sample spectrum of portal matter masses in Figure \ref{fig5}, the model might be disallowed for lightest portal matter masses ranging from as light as $\sim 1.5 \; \textrm{TeV}$, or permitted for lightest portal matter masses as large as $\gsim 10 \; \textrm{TeV}$-- this range can be considerably narrowed by finding constraints on the Yukawa couplings $y_{SL}$ and $y_{SE}$ beyond those imposed by perturbative unitarity, however. The more maximal $\Delta a_\mu$ from assuming the portal matter masses are degenerate considered in Figure \ref{fig5-degen} unsurprisingly shift this portal matter mass range somewhat higher, to between $\sim 2 \; \textrm{TeV}$ and $\sim 15 \; \textrm{TeV}$.
Notably, the results in Figure \ref{fig5} suggest that for $O(1)$ portal matter mass splittings and a lightest portal matter mass much above $\sim 10 \; \textrm{TeV}$, it is unlikely for \emph{any} point in parameter space to be able to match the observed $\Delta a_\mu$ without significant fine-tuning-- and hence a muon collider with a maximum center-of-mass energy of $\gsim 20 \; \textrm{TeV}$ could likely probe the entirety of this model's viable parameter space by searching for portal matter pair production and constraining the couplings $y_{SL}$ and $y_{SE}$ via monophoton searches-- this is roughly consistent with the behavior of the more minimal construction in \cite{Wojcik:2022woa}. Being more conservative and instead assuming that $\Delta a_\mu$ is maximized with degenerate portal matter masses, as in Figure \ref{fig5-degen}, we see that the same conclusion can be reached with the upper limit on portal matter mass instead being $\sim 15 \; \textrm{TeV}$, which would potentially lie in the reach of a muon collider with a center-of-mass energy of $\gsim 30 \; \textrm{TeV}$. Of course, we remind the reader that such a high degree of portal matter mass degeneracy should be considered less realistic, and if taken literally, would lead to vanishing kinetic mixing between the photon and the dark photon $A_D$.

\subsection{$g-2$ in Scenario B}\label{sec:g-2-B}
In Scenario B, we identify the muon with the light fermion appearing in the block $M_F^{(1)}$ of the fermion mass matrix in Eq.~(\ref{eq:mf-block-def}), while the $\tau$ is identified with the light fermion appearing in the block $M_F^{(2)}$. The Yukawa couplings $y_H$ and $y_{HS}$ then can be expressed as
\begin{align}
     |y_H| = \frac{\sqrt{2} m_\mu}{v} \sqrt{\frac{({M_L^+}^2+{M_L^-}^2)({M_E^+}^2+{M_E^-}^2)}{(M_L^+ M_E^+ \pm M_L^- M_E^-)^2}}, \;\;\; y_{HS} = \frac{\sqrt{2} m_\tau}{v},
\end{align}
where we adopt the same sign conventions as in Scenario A, namely that $y_{HS}$ is always positive, and $y_H$ could be either positive or negative.

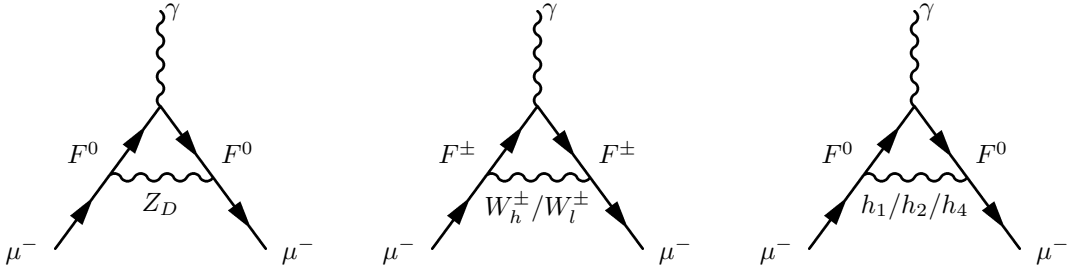
\begin{figure*}[b!]
\centering
\subfloat{
\begin{fmffile}{feyngraph}
  \begin{fmfgraph*}(80,100)
    \fmfbottom{f2,f1}
    \fmftop{g}
    \fmf{boson,t=1.5}{g,v}
    \fmf{fermion}{f2,v1}
    \fmf{fermion}{v1,v}
    \fmf{fermion} {v,v2}
    \fmf{fermion}{v2,f1}
    \fmfv{l.a=-20,l.d=8,l=$\gamma$}{g}
    \fmfv{l.a=-10,l=$\mu^-$}{f1}
    \fmfv{l.a=-170,l=$\mu^-$}{f2}
    \fmfv{l.a=120, d=10,l=$F^0$}{v1}
    \fmfv{l.a=60, d=10,l=$F^0$}{v2}
    \fmffreeze
    \fmf{boson,label=$Z_D$}{v1,v2}
  \end{fmfgraph*}
  \end{fmffile}
}\hspace{15mm}
\subfloat{
\begin{fmffile}{feyngraph1}
  \begin{fmfgraph*}(80,100)
    \fmfbottom{f2,f1}
    \fmftop{g}
    \fmf{boson,t=1.5}{g,v}
    \fmf{fermion}{f2,v1}
    \fmf{fermion}{v1,v}
    \fmf{fermion} {v,v2}
    \fmf{fermion}{v2,f1}
    \fmfv{l.a=-20,l.d=8,l=$\gamma$}{g}
    \fmfv{l.a=-10,l=$\mu^-$}{f1}
    \fmfv{l.a=-170,l=$\mu^-$}{f2}
    \fmfv{l.a=120, d=10,l=$F^\pm$}{v1}
    \fmfv{l.a=60, d=10,l=$F^\pm$}{v2}
    \fmffreeze
    \fmf{boson,label=$W_{h}^\pm/W_{l}^\pm$}{v1,v2}
  \end{fmfgraph*}
  \end{fmffile}
}\hspace{15mm}
\subfloat{
\begin{fmffile}{feyngraph2}
  \begin{fmfgraph*}(80,100)
    \fmfbottom{f2,f1}
    \fmftop{g}
    \fmf{boson,t=1.5}{g,v}
    \fmf{fermion}{f2,v1}
    \fmf{fermion}{v1,v}
    \fmf{fermion} {v,v2}
    \fmf{fermion}{v2,f1}
    \fmfv{l.a=-20,l.d=8,l=$\gamma$}{g}
    \fmfv{l.a=-10,l=$\mu^-$}{f1}
    \fmfv{l.a=-170,l=$\mu^-$}{f2}
    \fmfv{l.a=120, d=10,l=$F^0$}{v1}
    \fmfv{l.a=60, d=10,l=$F^0$}{v2}
    \fmffreeze
    \fmf{dashes,label=$h_1/h_2/h_4$}{v1,v2}
  \end{fmfgraph*}
  \end{fmffile}
}
\caption{ Feynman diagrams contributing to $\Delta a_\mu$ in Scenario B, with $F^0=\{L^0,E^0 \}$, $F^\pm=\{L^\pm,E^\pm \}$. }\label{fig-Bfeyn}
\end{figure*}

In this scenario, the muon's interactions with the new heavy gauge bosons are much stronger, appearing at leading order in $r_\Delta/v_\Phi$. Therefore, the gauge boson contributions and the scalar contributions are comparable and both should be taken into account. Fig~\ref{fig-Bfeyn} shows the Feynman diagrams that have dominant contributions to $\Delta a_\mu$. In Scenario B, it is $y_{HV}$ that is the most relevant coupling, unlike Scenario A in which $y_{SL}y_{SE}$ does the heavy lifting.
Keeping only the chirally enhanced contribution that is numerically significant, and neglecting the term proportional to $\lvert y_H \rvert$, which is now far too small to contribute meaningfully to the magnetic moment anomaly, we have
\begin{align}\label{eq:B-g-2-1}
    a_\mu^{Z_D}&\approx \frac{5 e_D^2 \sigma_\mu \sigma_{E-} m_{HV} m_\mu {M_E^-}{M_E^+}{M_L^-}{M_L^+}}{8\pi^2 s_{2D}^2 {M_L^0}{M_E^0} } \int_0^1 dx \frac{x^2(x-1)}{P_x(M_{Z_D}^2,{M_L^0}^2)P_x(M_{Z_D}^2,{M_E^0}^2)},
\nonumber\\
    a_\mu^{W_h,W_l}&\approx \frac{5e_D^2 \sigma_\mu m_\mu m_{HV}}{16 \pi^2 {M_L^0}{M_E^0} } \left(\sigma_{E-}{M_E^-}{M_L^-}(a_-^{W_h}+a_-^{W_l}) + {M_E^+}{M_L^+}(a_+^{W_h}+a_+^{W_l})\right),
\\
    a_\mu^{h_1,h_2,h_4}&\approx \frac{e_D^2 \sigma_\mu m_\mu m_{HV}}{16\pi^2 {M_L^0}{M_E^0} M_{Z_D}^2 s_{2lh}^2  } \left( 2 M_{h_4}^2 a^{h_4} + ({M_E^-}^2-{M_E^+}^2)({M_L^-}^2-{M_L^+}^2)(M_{h_1}^2 a^{h_1} +M_{h_2}^2 a^{h_2} ) \right),
    \nonumber
\end{align}
where 
\begin{align}\label{eq:B-g-2-2}
  a_-^{W_h}&\equiv \frac{4 K^+_{1}({M_L^+},{M_L^-},{M_E^+},{\sigma_{E-}M_E^-})}{s_{2D}^2} \int_0^1 dx \frac{x^2(x-1)}{P_x(c_{lh}^2 M_{Z_D}^2,{M_E^-}^2)P_x(c_{lh}^2 M_{Z_D}^2,{M_L^-}^2)},
\nonumber\\
 a_-^{W_l}&\equiv K^-_{-1}(\sigma_{E-}{M_E^-},{M_E^+} ,{M_L^-},{M_L^+})  \int_0^1 dx \frac{x^2(x-1)}{P_x(s_{lh}^2 M_{Z_D}^2,{M_E^-}^2)P_x(s_{lh}^2 M_{Z_D}^2,{M_L^-}^2)},
\nonumber\\
  a_+^{W_h}&\equiv \frac{4  K^+_{1}({M_L^-},{M_L^+},{\sigma_{E-}M_E^-},{M_E^+})}{s_{2D}^2} \int_0^1 dx \frac{x^2(x-1)}{P_x(c_{lh}^2 M_{Z_D}^2,{M_E^+}^2)P_x(c_{lh}^2 M_{Z_D}^2,{M_L^+}^2)},
\nonumber\\
   a_+^{W_l}&\equiv K^-_{-1}({M_E^+},\sigma_{E-}{M_E^-} ,{M_L^+},{M_L^-}) \int_0^1 dx \frac{x^2(x-1)}{P_x(s_{lh}^2 M_{Z_D}^2,{M_E^+}^2)P_x(s_{lh}^2 M_{Z_D}^2,{M_L^+}^2)},
\\
    a^{h_4}& \equiv 4 \left({M_L^0}^2- {M_L^+}{M_L^-}s_{2\Phi}\right) \left({M_E^0}^2-\sigma_{E-}{M_E^+}{M_E^-}s_{2\Phi}\right)\int_0^1 dx \frac{x^2(x-1)}{P_x(2M_{h_4}^2,2{M_L^0}^2)P_x(2M_{h_4}^2,2{M_E^0}^2)},
\nonumber\\
 a^{h_1}& \equiv (1-S_h)\int_0^1 dx \frac{x^2(x-1)}{P_x(2M_{h_1}^2,2{M_L^0}^2)P_x(2M_{h_1}^2,2{M_E^0}^2)},
\nonumber\\
a^{h_2}& \equiv (1+ S_h)\int_0^1 dx \frac{x^2(x-1)}{P_x(2M_{h_2}^2,2{M_L^0}^2)P_x(2M_{h_2}^2,2{M_E^0}^2)},
\nonumber
\end{align}
and 
\begin{align}\label{eq:B-g-2-3}
&{M_{L,E}^0}^2=\frac{1}{2}\left( {M_{L,E}^+}^2+ {M_{L,E}^-}^2\right),\quad  P_x(A,B)\equiv (1-x)A + xB, 
\nonumber\\
 &K^\pm_{a}(A,B, C, D) \equiv \frac{1}{4}\left( 2A C s_{D}^{2 a} +2 B D c_{D}^{2 a} + (2A C s_D^{2 a} - 2B D c_D^{2 a} \pm (A D + B C) s_D^a c_D^a t_{2 D} s_{2 \Phi} )\sqrt{\frac{c_{2 D}^2}{c_{2 lh}^2}} \right),
\nonumber    \\
 &S_h \equiv \frac{A \bigg[ 6 B -8 \left(B + A s_{2 \Phi}\right)\frac{s_{2 lh}^2}{s_{2 D}^2} - 4 (M_{h_1}^2 - M_{h_2}^2) s_{2 \Phi}\bigg] + B \bigg(2 B s_{2 \Phi}-(M_{h_1}^2 - M_{h_2}^2)\bigg) \bigg( 2 - \frac{s_{2 D}^2}{s_{2 lh}^2} \bigg)}{(M_{h_1}^2 + M_{h_2}^2)^2+ (M_{h_1}^2 - M_{h_2}^2) \bigg( 2 A - B s_{2 \Phi} \frac{s_{2 D}^2}{s_{2 lh}^2} \bigg)},
\nonumber\\
  &A\equiv \sqrt{-\frac{s_{2D}^2}{s_{2 lh}^2}\left(M_{h_1}^2-2\lambda_1 e_D^{-2}s_{2lh}^2 M_{Z_D}^2  \right)\left(M_{h_2}^2-2\lambda_1 e_D^{-2}s_{2lh}^2 M_{Z_D}^2  \right)},
 \\
 &B \equiv M_{h_1}^2 + M_{h_2}^2 - 4 \lambda_1 e_D^{-2} s_{2 lh}^2 M_{Z_D}^2, \nonumber
\nonumber\\
   & \sigma_{E-}\equiv \text{ sign}[\sin{(\theta_E+\theta_\Phi)}], \quad \sigma_\mu \equiv \text{sign}[y_H ({M_L^+}{M_E^+}\pm {M_L^-}{M_E^-})], \quad \sigma_\Phi\equiv \text{sign}[\sin{2\theta_\Phi}],\quad 
\nonumber\\
 & (s_{nx},c_{nx},t_{nx})=(\sin{n\theta_x},\cos{n\theta_x},\tan{n\theta_x}),\quad s_{2\Phi}=\sigma_\Phi \sqrt{1-\frac{s_{2lh}^2}{s_{2D}^2}},
 \nonumber
\end{align}
Unlike the case of Scenario A, these expressions are not invariant under $M_{L,E}^+ \leftrightarrow M_{L,E}^-$ and $M_{L}^\pm \leftrightarrow M_{E}^\pm$. Nonetheless, following the treatment in Section \ref{sec:g-2-A}, we will now explore the parameter space in which we select $M_{L}^+$ to be the mass of the lightest portal matter field, and leave the exploration of the vector-like leptons' mass asymmetry to the later part of this section.
Among the choices of physical signs of various terms in our model, we find two are particularly dominant. First, the sign of the product $y_{HV} y_{H}$ scales all the chirally enhanced contributions to $\Delta a_\mu$, and therefore in order for the observed anomaly to be generated, the sign of this product must be selected to give a positive contribution to $\Delta a_\mu$. Second, the sign of the term $\sin(\theta_E + \theta_\Phi)$ alters whether certain terms will interfere destructively or constructively with one another: If we select $\sin(\theta_E + \theta_\Phi) < 0$ various contributions to the $y_{HV}$ terms will interfere destructively, severely reducing the magnitude of this contribution relative to the $\sin(\theta_E + \theta_\Phi) > 0$ case.

\begin{figure}[t]
    \centerline{\includegraphics[width=4in]{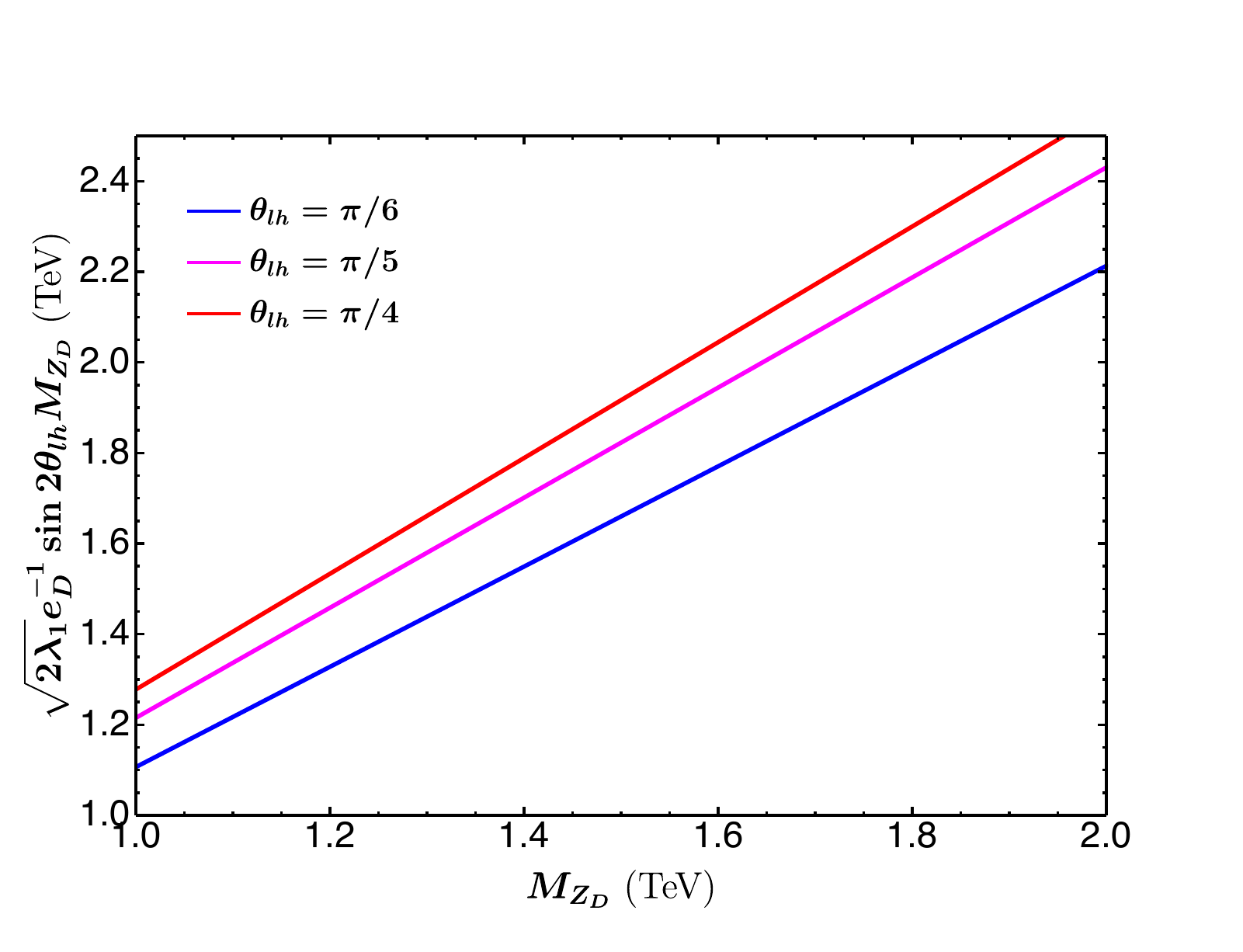}}
    \vspace{-0.5cm}
    \caption{The bound $\sqrt{2\lambda_1}e_D^{-1} \sin{2\theta_{lh}}M_{Z_D} $ of Eq.~(\ref{eq:B-h1h2selectionrule}) as a function of $M_{Z_D}$, for $\lambda_1=0.4$ and $e_D=0.7$. This bound applies in both Scenarios A and B, but is relevant for $g-2$ only in Scenario B.}
    \label{fig-bound}
\end{figure}

Therefore, in Scenario B, the chirally enhanced contributions to $\Delta a_\mu$ will only be dependent on fixed SM parameters, the masses of the portal matter fermions, the masses of the heavy gauge bosons, which are parametrized as $M_{W_h}=\cos{\theta_{lh}} M_{Z_D}$, $M_{W_l}=\sin{\theta_{lh}} M_{Z_D}$, masses of the three heavy scalars $M_{h_{1,2,4}}$, the Yukawa coupling $y_{HV}$, the parameter $\lambda_1$ and $e_D$, and the angle $\theta_D$. We find that this wealth of parameters is highly constrained by certain not immediately apparent physical consistency requirements-- specifically that the mixing angle $\theta_\lambda$, which parameterizes the definition of the $h_1$ and $h_2$ scalar mass eigenstates and is discussed in Eq.~(\ref{eq:scalar-eigenstates}) of Appendix \ref{Sappendix}, must be real. In this case, we need to have the following relation between the gauge boson masses and the scalar masses:
\begin{equation}\label{eq:B-h1h2selectionrule}
    M_{h_2}<\sqrt{2\lambda_1}e_D^{-1} \sin{2\theta_{lh}}M_{Z_D} <  M_{h_1}
\end{equation}
where we note that $M_{h_1}>M_{h_2}$ from the definition of the $h_1$ and $h_2$ masses. This relation heavily constrains the accessible parameter space in our scenario, as 6 different input parameters need to satisfy this relation. In Fig~\ref{fig-bound}, we plot this bound as a function of the mass of $Z_D$, for different $\theta_{lh}$ and fixed $\lambda_1$ and $e_D$. This figure is illustrative for the choice of $M_{h_1}$ and $M_{h_2}$ for our numerical analysis. Here we consider an $O(1)$ splitting for the gauge boson masses, which is a natural assumption since all masses of portal matter fermions and bosons are controlled by the same scale $v_\Phi$.

Before moving on to a direct numerical probe of the $g-2$ corrections in the model, we should highlight one final crucial difference between the construction of Scenario B and that of Scenario A: In Scenario B, the muon magnetic moment correction scales quadratically with the dark gauge coupling $e_D$. However, unlike every other parameter here, $e_D$ is an important parameter not just of the higher-scale, portal matter theory, but also enters into the behavior of the sub-GeV simplified dark matter models which motivate the portal matter framework. As such, the $g-2$ correction (and, as we shall see later on in Section \ref{sec:diboson-B}, other collider probes of the model) can potentially have excellent complementarity with limits from the dark matter sector, although the precise form of the relationship between these constraints is dependent on specifications about the nature of the sub-GeV dark matter candidate that we remain agnostic about in this work.

\begin{figure}[t!]
    \centerline{\includegraphics[width=3.5in]{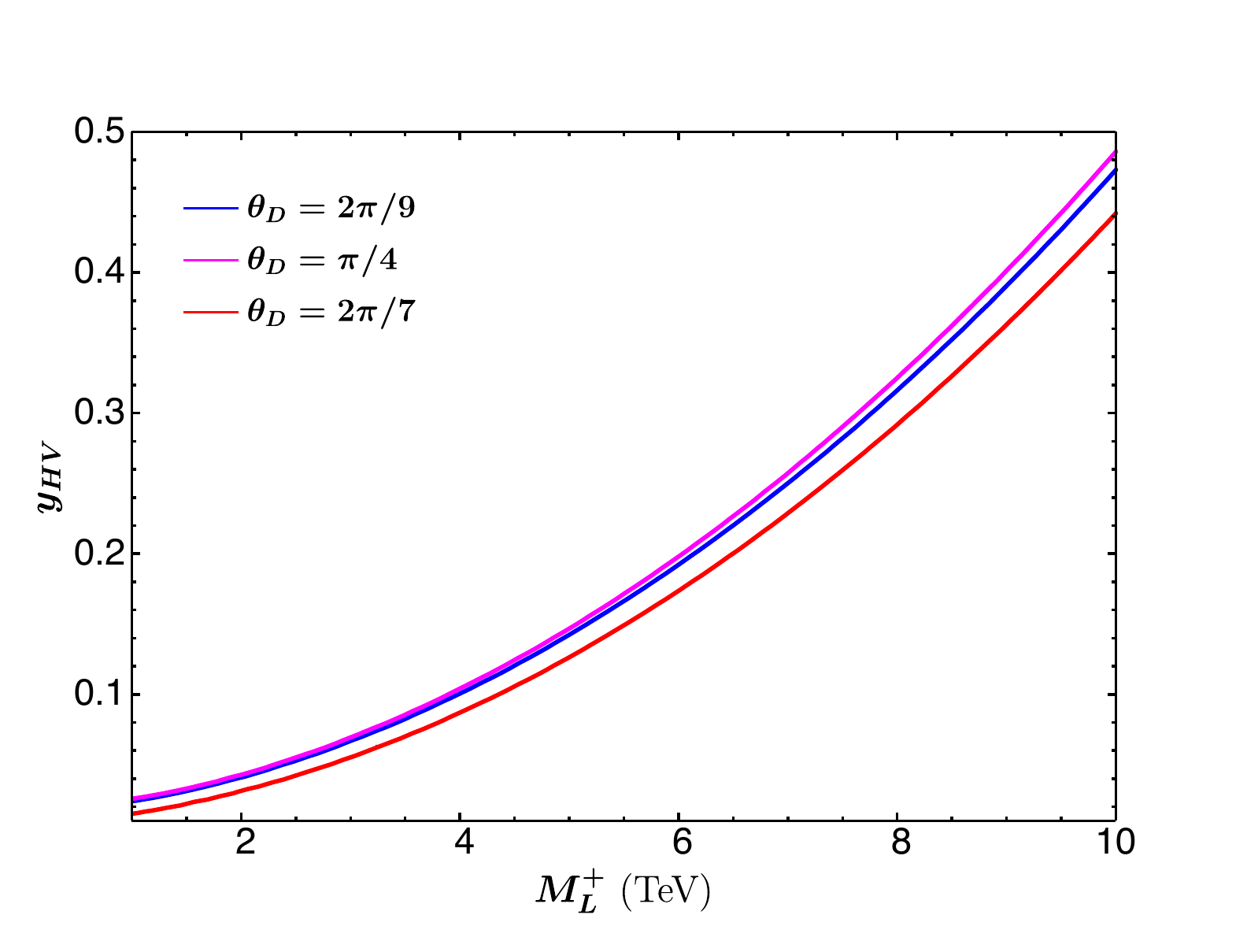}
    \hspace{-2cm}
    \hspace{1cm}\includegraphics[width=3.5in]{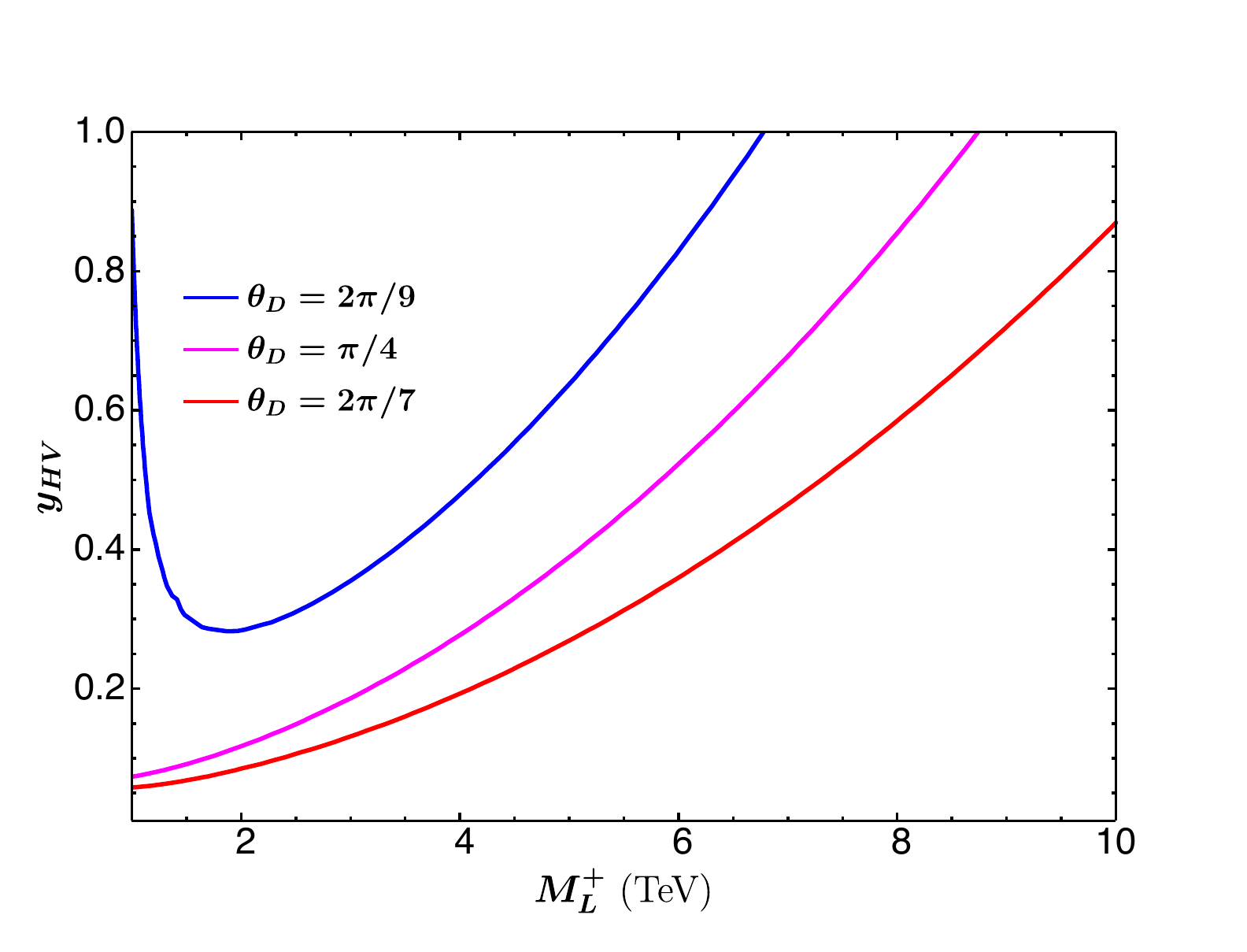}}
    \vspace{-0.5cm}
     \centerline{\includegraphics[width=3.5in]{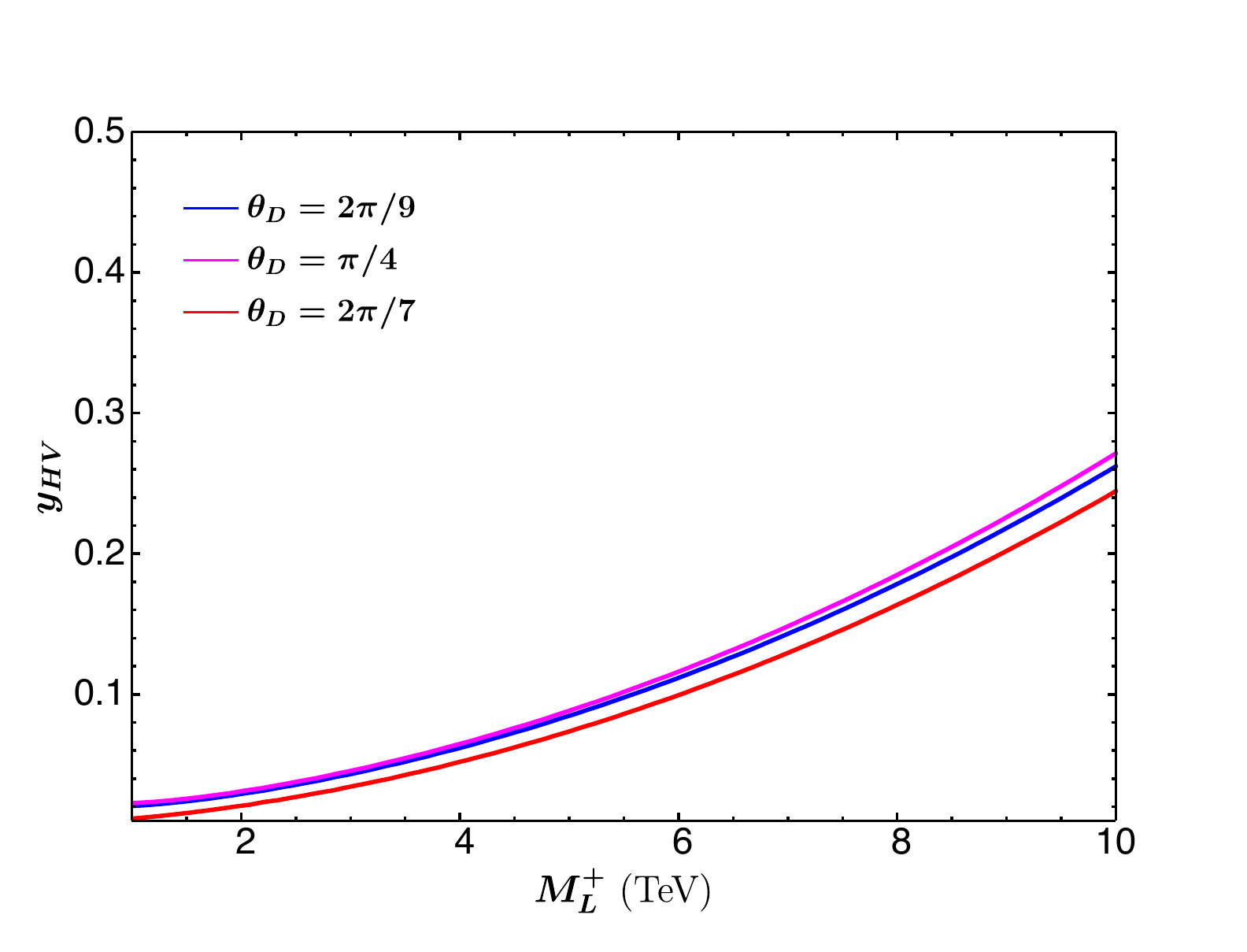}
    \hspace{-2cm}
    \hspace{1cm}\includegraphics[width=3.5in]{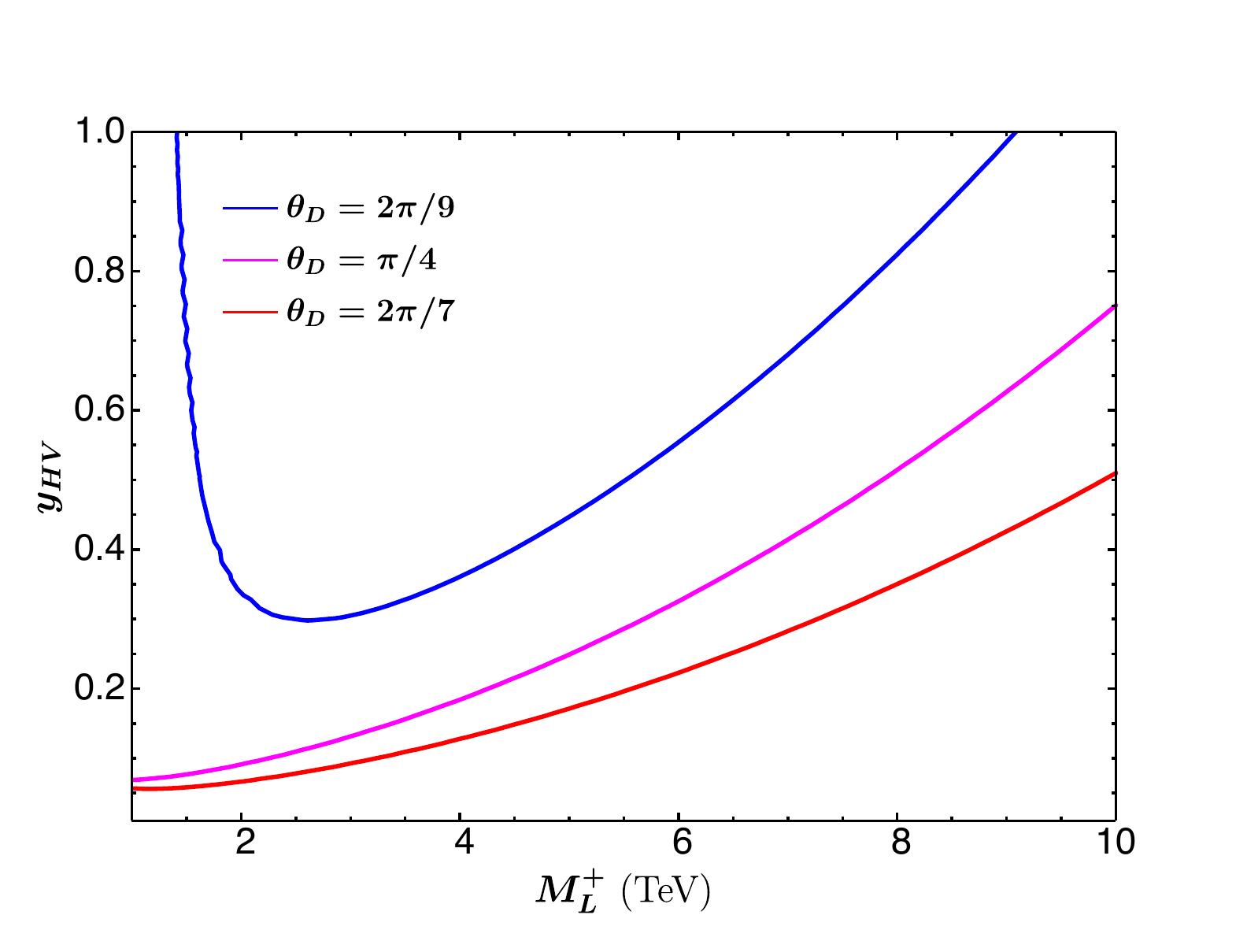}}
    \caption{(Top): The necessary strength of $y_{HV}$ to reproduce $\Delta a_\mu$ in Scenario B as a function of $M_{L}^+$, for $M_{L}^-/M_{L}^+ = 1.3$, $M_{E}^+/M_{L}^+ = 1.5$, $M_{E}^-/M_{L}^+ = 1.8$, $M_{Z_D}=0.74$ TeV, $\theta_{lh}=\arccot{(1.7)}$, $e_D=0.8$, $\lambda_1=0.4$, $M_{h_2}=1.2$ TeV, $M_{h_1}=1.5$ TeV, $M_{h_4}=1$ TeV. 
    (Top left):  $\sin{(\theta_E+\theta_\Phi)}>0$, (top right):  $\sin{(\theta_E+\theta_\Phi)}<0$. 
    (Bottom): The same as the top panels, but for degenerate portal fermion masses.} 
    \label{fig-yHVvsmL+}
\end{figure}

Similar to Scenario A, for our numerical analysis, we will always select $O(1)$ mass splittings between portal matter fermions  $M_{L}^-/M_{L}^+ = 1.3$, $M_{E}^+/M_{L}^+ = 1.5$, and $M_{E}^-/M_{L}^+ = 1.8$ in most cases to get a sense of the numerics of a reasonable point in parameter space, as larger mass splittings between portal matter fermions generally lead to more suppressed loop contributions. Next, we explore the effect of the physical sign of $\sin{(\theta_E+\theta_\Phi)}$. In the top panels of Fig.~\ref{fig-yHVvsmL+}, we depict the necessary strength of the Yukawa coupling $y_{HV}$ in order to achieve the observed $\Delta{a_\mu}$ as a function of the lightest portal matter fermion mass $M_{L}^+$, using the benchmark values for masses of portal matter fermions. Two different physical signs of $\sin{(\theta_E+\theta_\Phi)}$ can give us $\Delta a_\mu$ with the right magnitude and sign at different regions of parameter space. For $\sin{(\theta_E+\theta_\Phi)}>0$ in which various contributions mostly interfere constructively, we just need $y_{HV}$ to be between 0.02 to 0.06 to get the observed anomaly. However, for $\sin{(\theta_E+\theta_\Phi)}<0$ in which various contributions mostly interfere destructively, we need larger $y_{HV}$, which is around $O(0.1)$ to reproduce the observed anomaly. In the bottom panels of Fig.~\ref{fig-yHVvsmL+}, we consider the case in which the portal matter masses are all degenerate, which will maximize the value of $\Delta a_\mu$. For most of the $\theta_D$ values, this indeed lowers the value of $y_{HV}$ required to generate the desired $\Delta a_\mu$. Unlike Scenario A, the value of $y_{HV}$ required to generate the observed $\Delta a_\mu$ is well below the unitarity bound and thus it does not provide a meaningful constraint. However, we might be able to constrain $y_{HV}$ from observations of the branching fractions of some vector-like leptons, as will be discussed in Section \ref{sec:fermion-pheno}.

\begin{figure}[t]
    \centerline{\includegraphics[width=3.5in]{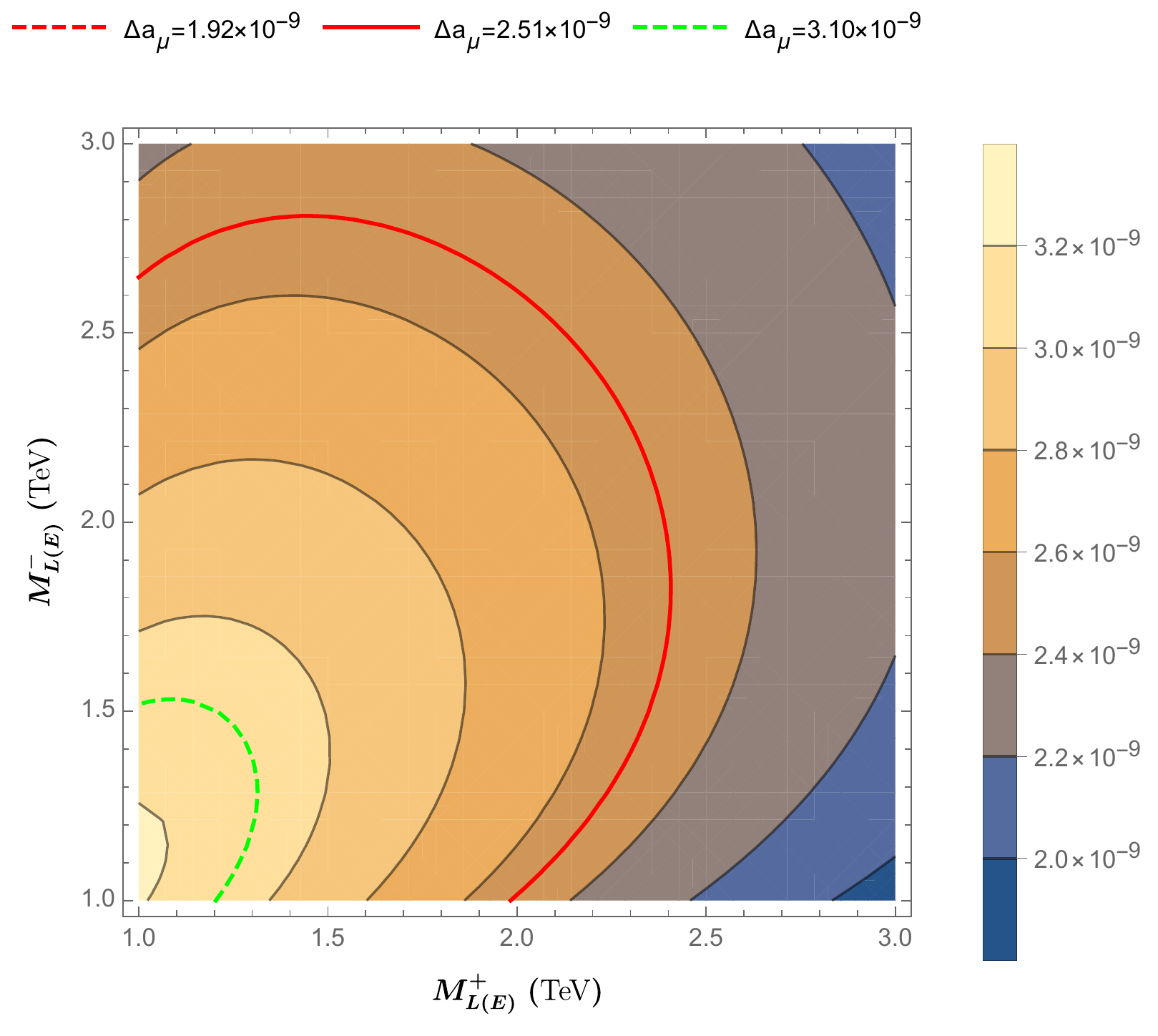}
    \hspace{0.1cm}\includegraphics[width=3.5in]{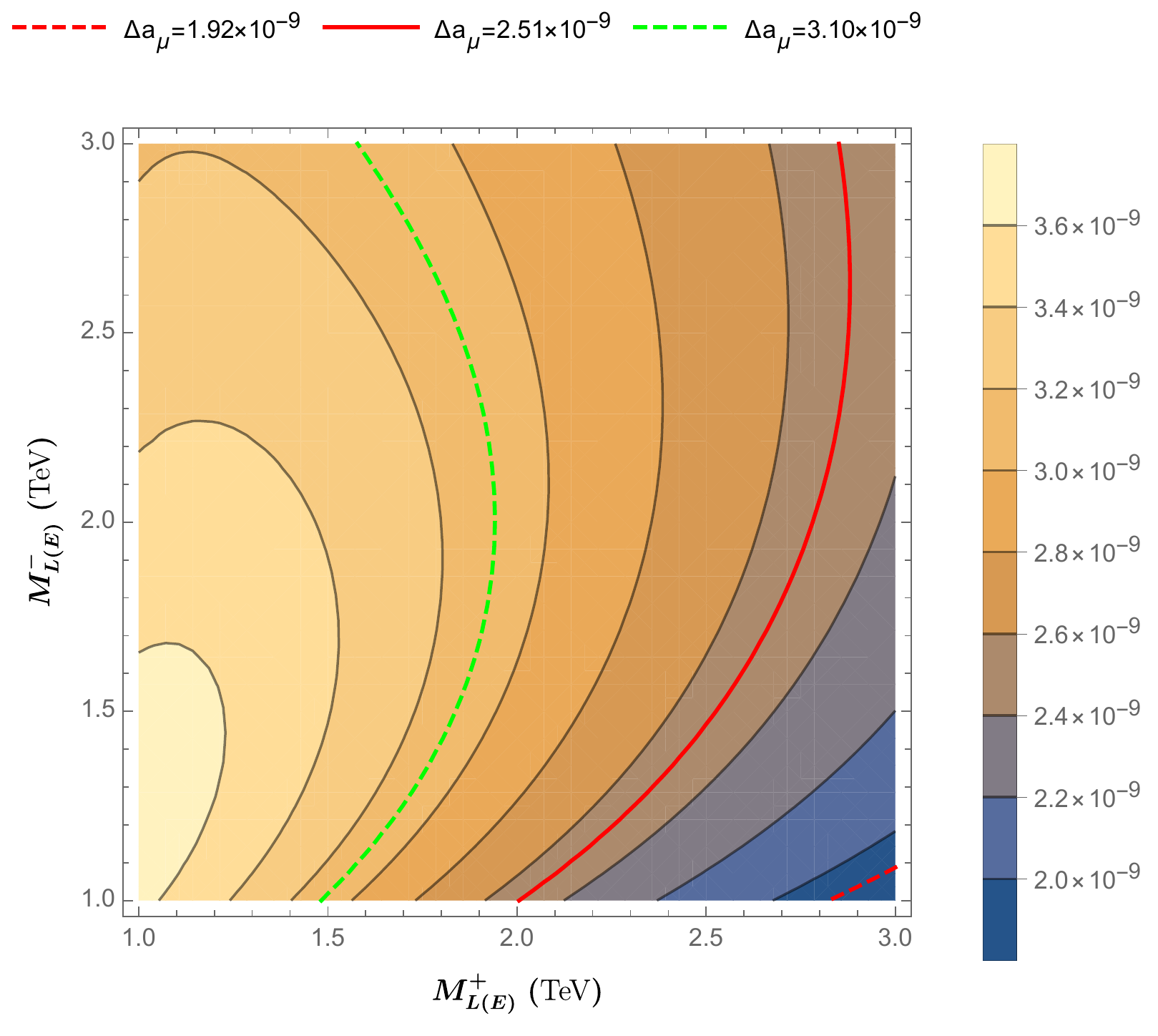}}
    \caption{Contour plots of $\Delta a_\mu$ in Scenario B, for $M_{Z_D}=0.74$ TeV, $\theta_{lh}=\arccot{(1.7)}$, $e_D=0.8$, $\lambda_1=0.4$, $M_{h_2}=1.2$ TeV, $M_{h_1}=1.5$ TeV, $M_{h_4}=1$ TeV, $\theta_D=\pi/4$, $M_{E(L)}^+=1$ TeV, and $M_{E(L)}^- = 1.3$ TeV. 
    (Left): Contour plot as a function of $M_{L(E)}^+$ and $M_{L(E)}^-$, for $y_{HV}=0.03$,  $\sin{(\theta_E+\theta_\Phi)}>0$ . (Right): Contour plot as a function of $M_{L(E)}^+$ and $M_{L(E)}^-$, for $y_{HV}=0.1$, $\sin{(\theta_E+\theta_\Phi)}<0$.
     }
    \label{fig-Btwomasses}
\end{figure}

We now explore the asymmetry of our portal matter fermion masses with respect to $M_{L,E}^+ \leftrightarrow M_{L,E}^-$ and $M_{L}^\pm \leftrightarrow M_{E }^\pm$. Although the expressions in Eq.(~\ref{eq:B-g-2-1}) and Eq.(~\ref{eq:B-g-2-2}) are not manifestly symmetric under $M_{L}^\pm \leftrightarrow M_{E }^\pm$, for the parameter space of interest that is explored, $\Delta a_\mu$ is numerically nearly symmetric under  $M_{L}^\pm \leftrightarrow M_{E }^\pm$, regardless of sign selection of $\sin{(\theta_E+\theta_\Phi)}$. In Fig~\ref{fig-Btwomasses}, we depict contour plots of $\Delta a_\mu$ as a function of different portal matter fermion masses, for various choices of physical signs in the model, assuming the lightest portal matter field has a mass of 1 TeV. In the case that $\sin{(\theta_E+\theta_\Phi)}>0$, we observe a slight asymmetry in $\Delta a_\mu$ value under $M_{L,E}^+ \leftrightarrow M_{L,E}^-$. In the case $\sin{(\theta_E+\theta_\Phi)}<0$, the asymmetry in $\Delta a_\mu$ with respect to $M_{L,E}^+ \leftrightarrow M_{L,E}^-$ is further amplified. In both cases, all other parameters are held constants, except $y_{HV}$ is adjusted to access to the relevant parameter space that gives the desired $\Delta a_\mu$ in each case. However, this difference in $y_{HV}$ does not contribute to the asymmetry of the portal matter masses, since it is a global constant that exists in front of all relevant corrections.

It is also of interest to explore the parameter space dependence on $\theta_D$ and $e_D$. As mentioned earlier, these two parameters significantly constrain the parameter space, given that they appear in the definition of the $h_1,h_2$ mixing angle $\theta_\lambda$ and the bidoublet vev angle $\theta_\Phi$ in a square root form. Therefore, for the physical consistency of $h_1$ and $h_2$ mixing, $\theta_D$ and $e_D$ can only take on certain values, after we have chosen to fix $\theta_{lh}$, $M_{Z_D}$ and $M_{h_{1,2}}$. For $\theta_D$, a different sign selection of $\sin{(\theta_E+\theta_\Phi)}$ does affect the magnitude of $\Delta a_\mu$. However, it does not change the viable region of $\theta_D$, as this sign does not enter the definitions of $\theta_\lambda$ and $\theta_\Phi$. The viable $\theta_D$ range is then around $0.20-0.32 \pi$ or $0.68-0.82 \pi$, under the assumption that $M_{Z_D}=0.74$ TeV, $\theta_{lh}=\arccot{(1.7)}$, $e_D$=0.8, $\lambda_1=0.4$, $M_{h_2}=1.2$ TeV and $M_{h_1}=1.5$ TeV. With the same choice of $\theta_D$ and model parameters as above, we need a larger $y_{HV}$ in the case that $\sin{(\theta_E+\theta_\Phi)}<0$. Turning now to constraints on $e_D$, if we assume $\theta_D=0.7$, $M_{Z_D}=0.74$ TeV, $\theta_{lh}=\arccot{(1.7)}$, $\lambda_1=0.4$, $M_{h_2}=1.1$ TeV and $M_{h_1}=1.9$ TeV, the viable range for $e_D$ is then $e_D=0.6-1.0$. Again for the case $\sin{(\theta_E+\theta_\Phi)}<0$, with the same set of fixed parameter values as in the $\sin{(\theta_E+\theta_\Phi)}>0$ case, $\Delta a_\mu$ always falls short by nearly an order of magnitude, and thus the relevant parameter space in cases with different sign options of   $\sin{(\theta_E+\theta_\Phi)}$ is different. These two examples on the selection of $e_D$ and $\theta_D$ also show that in Scenario B, although there are many  parameters, the allowed range for several of them is severely limited by physical consistency. However, we are still able to realize any $O(1)$ $e_D$ value we need without modifying anything else that affects the sub-GeV dark sector. 

Another interesting feature of the muon $g-2$ anomaly in Scenario B is that the dominant contribution always arises from the contribution of heavy gauge bosons $Z_D$ and $W_{h,l}$, regardless of the choice of signs or of the other parameters. In the case that $\sin{(\theta_E+\theta_\Phi)}>0$, $W_{h,l}$ and $Z_D$ both contribute constructively to $\Delta a_\mu$, whereas $h_{1,2,4}$ contribute destructively, with a contribution which is an order of magnitude smaller than that of the heavy gauge bosons. In the case that $\sin{(\theta_E+\theta_\Phi)}<0$, the contribution of $W_{h}$ and $W_{l}$ is opposite in sign, causing the overall contribution from $W_{h,l}$ to be small but constructive, comparable to the size of the constructive contributions from $h_{1,2,4}$. The contribution of $Z_D$ is dominant in this case, which is about an order of magnitude larger. This relative contributions from heavy gauge bosons and scalars hold true in all regions of parameter space explored.

\section{Other Model Phenomenology: Constraints and Experimental Prospects}\label{sec:pheno}

Having explored the valid parameter regimes for which this construction might address the observed muon magnetic moment anomaly, we must now consider what other observational consequences this model might lead to in both Scenario A and Scenario B, and how these compare to the more minimal construction in \cite{Wojcik:2022woa}.

\subsection{Fermion Collider Production}\label{sec:fermion-pheno}
We will begin with a discussion of the segment of the theory which most resembles our minimal construction in \cite{Wojcik:2022woa}, and constitute the signals which are the most characteristic of portal matter models: The direct production of the vector-like fermions. Both Scenario A and Scenario B will have extremely similar phenomenology in this sector, so we shall address both cases together. These fields are all vector-like leptons with SM quantum numbers, and so they may be pair-produced in abundance at current and future colliders through SM interactions. A distinctive characteristic of this model is the spectrum of the vector-like leptons-- because the squared mass of the electroweak doublet (singlet) $U(1)_D$-neutral portal matter field is equal to the average of the squared masses of the $U(1)_D$-charged isospin doublet (singlet) portal fields, we will have 
\begin{align}
    M_{L,E}^\pm \leq M_{L,E}^0 \leq M_{L,E}^\mp,
\end{align}
that is, the weak doublet (singlet) portal matter field content consists of two $U(1)_D$-charged weak doublet (singlet) states and a $U(1)_D$-neutral weak doublet (singlet) case with a mass between them. Since a $U(1)_D$-charged state will therefore represent the lightest (and hence most likely to be kinematically accessible) new vector-like fermion, it behooves us to consider collider signatures of these states first. For brevity, when computing partial decay widths in this Section, we shall present our results in Scenario A, reminding the reader of this by applying an ``A'' superscript to all of our width expressions here. The corresponding results for Scenario B can be obtained by swapping the muons and $\tau$'s in all expressions.

The $U(1)_D$-charged portal matter fields will exhibit similar phenomenology to that of minimal portal matter discussed in \cite{Rizzo:2018vlb}-- due to their nontrivial $U(1)_D$ charge, their decay widths to SM fermions through the emission of electroweak gauge bosons are highly suppressed. These states always have far larger partial widths for decay into a muon or muon neutrino (in Scenario A) or a $\tau$ or $\tau$ neutrino (Scenario B) associated with the emission of $A_D$ or $h_D$; these widths are given by
\begin{align}\label{eq:PMtoADSM}
    \Gamma^A_{L_\pm,E_\pm \rightarrow \mu^- A_D} = \Gamma^A_{L_\pm,E_\pm \rightarrow \mu^- h_D} = \frac{M_{L,E}^\pm y_{SL, SE}^2 \cos^2(\theta_\Delta)}{128 \pi}.
\end{align}
For the lightest portal matter state, these final states will have a branching fraction of close to $100\%$ if there are no other kinematically accessible two-body channels. In this case, the collider signature of the portal matter is simply the familiar case discussed in \cite{Rizzo:2018vlb}: A highly boosted lepton (in this case a muon or $\tau$, depending on the scenario we have chosen) accompanied by missing energy (or a lepton-jet, if the dark photon decays visibly within the detector). For Scenario A, in which the portal matter decays into muons, the case in which both the dark photon and the dark Higgs are invisible is explored in the study of \cite{Guedes:2021oqx} by rescaling an existing ATLAS slepton search \cite{ATLAS:2018ojr}. Assuming that the lightest portal matter state decays with a branching fraction of 1 through the processes outlined in Eq.~(\ref{eq:PMtoADSM}), then if the electroweak singlet portal matter is the lightest state, $M_E^\pm \gsim 895 \; \textrm{GeV}$, while if the lightest state is an electroweak doublet, the constraint becomes $M_L^\pm \gsim 1050 \; \textrm{GeV}$ (the electroweak doublet constraints are marginally stronger due to larger production cross sections \cite{Rizzo:2022qan,OsmanAcar:2021plv}). These are considerably stronger than the constraints on a vector-like lepton decaying dominantly through traditional channels via the emission of electroweak gauge bosons, which limit the mass of an electroweak singlet vector-like muon partner to $\gsim 420 \; \textrm{GeV}$ \cite{Guedes:2021oqx}.\footnote{In the case of electroweak doublet vector-like fermions, constraints from such a particle decaying via electroweak gauge boson emissions can be considerably strengthened (to $\gsim 720 \; \textrm{GeV}$ \cite{Guedes:2021oqx}) by including SM $W^\pm$-mediated production of a charged vector-like lepton associated with a neutral one. An analogous analysis for portal matter, in which the process $\overline{q} q \rightarrow W^\pm \rightarrow L^\pm \overline{N^\pm} \rightarrow \mu^- + \textrm{MET}$ will produce mono-lepton events, has not been performed, but might be conjectured to yield marginally stronger constraints than the $\gsim 1050 \; \textrm{GeV}$ limit quoted for pair production of charged electroweak doublet portal matter.} 

We can now consider the corresponding case in Scenario B, in which the lightest portal matter field will decay dominantly into $\tau$'s instead of muons. Unfortunately, a detailed study of such exotically decaying $\tau$ partners in the manner of \cite{Guedes:2021oqx} does not yet exist, but we can generally anticipate that the LHC constraints on the lightest portal matter mass will be considerably weaker than they are in the case of muons, due to the greater difficulty involved in $\tau$ reconstruction. We can most saliently see this by comparing the results of slepton searches for muon and $\tau$ partners, which provide the closest analogy to portal matter searches dominantly decaying through dark photon and dark Higgs emission: In \cite{CMS:2022rqk}, CMS finds a $95\%$ CL lower mass bound on electroweak singlet $\tau$ sleptons of $M_{\tilde{\tau}} \gsim 240 \; \textrm{GeV}$ from 138 $\textrm{fb}^{-1}$ of data, while the ATLAS study in \cite{ATLAS:2019lff} finds a constraint of $M_{\tilde{\mu}} \gsim 450 \; \textrm{GeV}$ in the corresponding scenario for muons. In both Scenario A and B, it should be noted that a multi-TeV muon collider will generally have a far better reach than current or future proton-proton colliders, and likely be able to constrain portal matter masses up to just below half of its center-of-mass energy.

\begin{figure}
    \centerline{\includegraphics[width=3.5in]{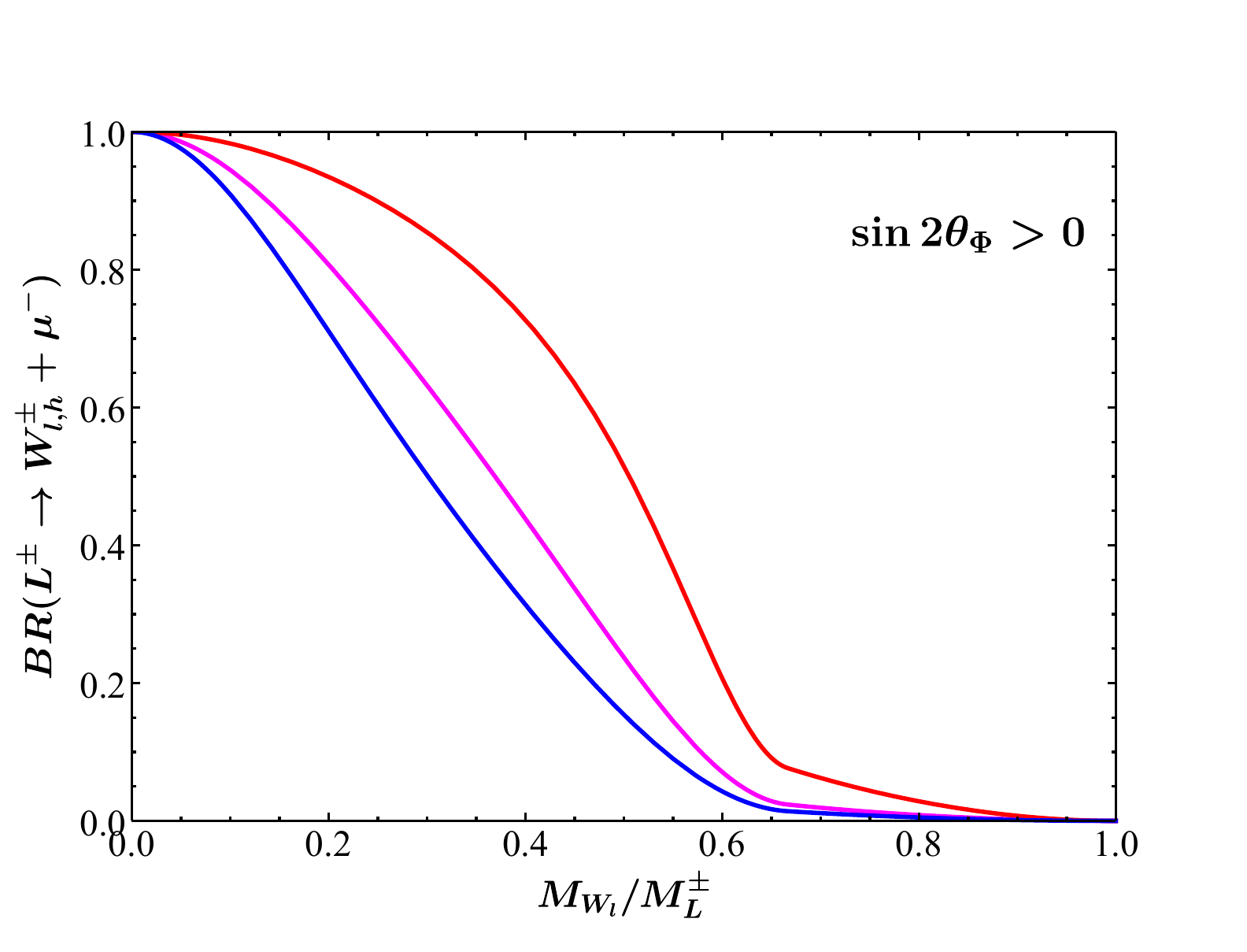}
    \hspace{-0.75cm}
    \includegraphics[width=3.5in]{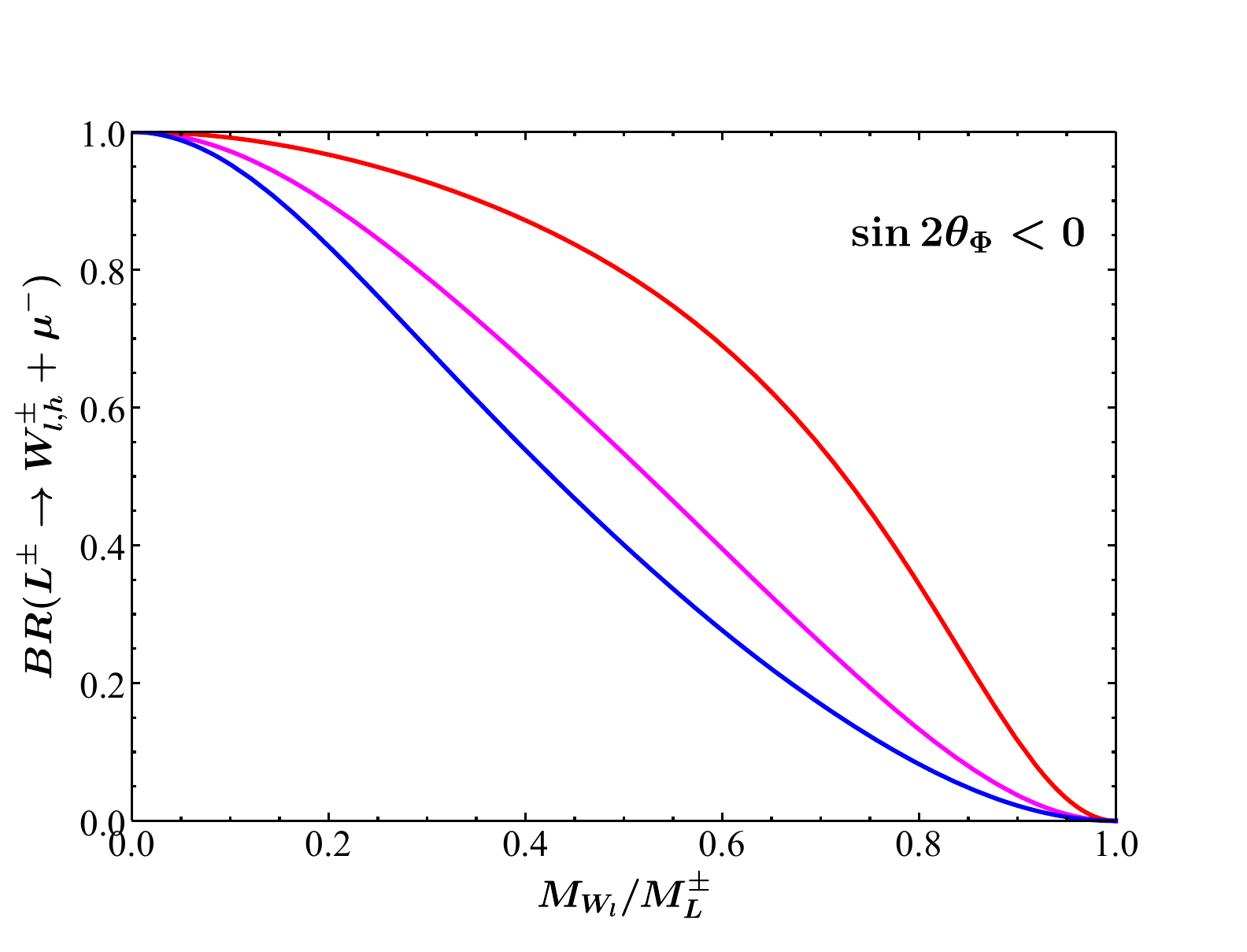}}
    \vspace*{-0.75cm}
    \centering
    \includegraphics[width=3.5in]{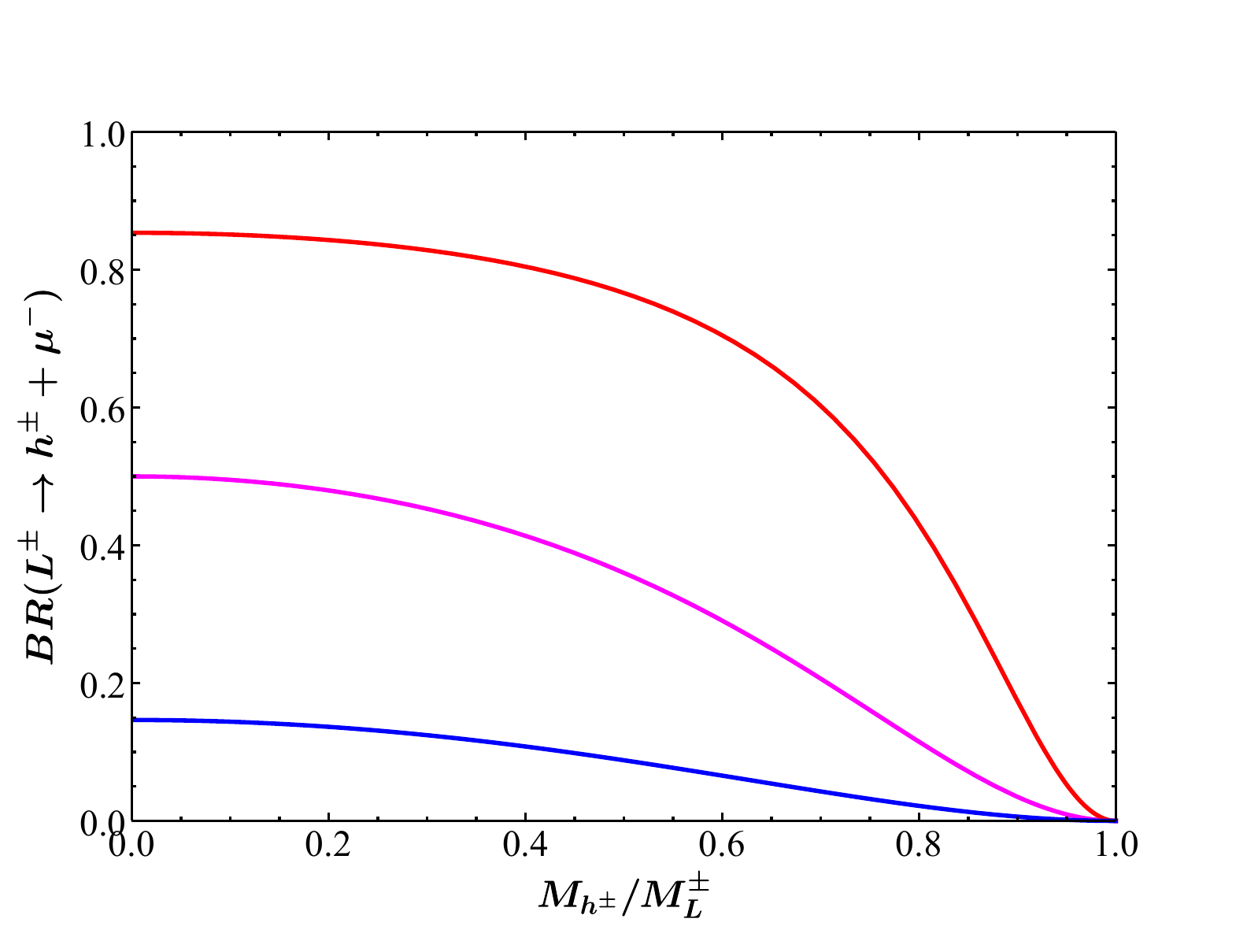}
    \caption{(Top): The branching fraction $BR(L_\pm \rightarrow \mu^- + W_{h,l}^\pm)$, assuming that the only significant kinematically accessible decay channels are $L^\pm \rightarrow \mu^- + W_{h,l}^\pm$ and the familiar portal matter channels $L^\pm \rightarrow \mu^- + h_3,A_D$. We have taken $M_{W_h} = 1.5 M_{W_l}$, $M_L^- = 1.3 M_L^+$, $y_{SL} = 1$, $e_D = \sqrt{4 \pi \alpha_{em}}$, and $\cos 2 \theta_D = (\cos{2 \theta_{hl}})/2$ and $\theta_\Delta = \pi/8$ (Blue), $\pi/4$ (Magenta), and $3 \pi/8$ (Red). The sign of $\sin{2 \theta_\Phi}$ is taken to be positive (left) and negative (right). (Bottom): The branching fraction $BR(L^\pm \rightarrow \mu^- + h^\pm)$, assuming that the only significant kinematically accessible decay channels are $L_\pm \rightarrow \mu^- + h^\pm$ and the familiar portal matter channels $L^\pm \rightarrow \mu^- + h_3, A_D$. Selections for $\theta_\Delta$, the only free parameter in this quantity other than the mass $M_{h^\pm}/M_L^\pm$, follow the same convention as the top figure. These charts are computed for Scenario A, but apply equally well for Scenario B by substituting all muons for $\tau$ leptons.}
    \label{fig:fermion-br}
\end{figure}

Our considerations thus far are entirely consistent with a minimal construction of portal matter of the form discussed in \cite{Rizzo:2018vlb}-- we should note that any constraints from fermion production will be further modified by the fact that in our model, the dark gauge and scalar sectors are no longer minimal and include a significant number of new heavy scalar and gauge bosons. Specifically, these new heavy bosons open up more decay channels for the lightest portal matter state. In particular, the $U(1)_D$-charged portal matter states may now decay into SM fields via the emission of a scalar $h_\pm$ boson, as well as a $W_h$ and $W_l$ bosons. For emission of a $W_h$ boson in Scenario A, the partial width is given by
\begin{align}\label{eq:PMtoWhSM}
    &\Gamma^A_{{L,E}_\pm \rightarrow \tau^- W^\pm_h} = \frac{e_D^2 M_{L,E}^\pm }{64 \pi \cos(2 \theta_{lh}) \cos^2 \theta_{lh}} \mathcal{G}\bigg( \frac{M_{L,E}^\pm}{M_{Z_D}}\bigg) \mathcal{F} \bigg( \frac{M_{L,E}^\pm}{M_{Z_D}}, \frac{M_{L,E}^\mp}{M_{Z_D}},\theta_{lh}, \theta_D, \sigma_\Phi \bigg),\\
    &\mathcal{F} ( x_A, x_B, \theta_{lh}, \theta_D, \sigma_\Phi) \equiv \bigg[ \bigg( 1- \frac{\sin^2 \theta_{lh}}{\cos^2 \theta_D}\bigg)x_A^2 + \bigg( 1- \frac{\sin^2 \theta_{lh}}{\sin^2 \theta_D}\bigg)x_B^2 + 2 x_A x_B \sqrt{1-\frac{\sin^2 2 \theta_{lh}}{\sin^2 2 \theta_D}} \sigma_\Phi  \bigg],\nonumber\\
    &\mathcal{G}(x) \equiv 1 + x^{-2} - 2 x^{-4},\;\;\; \sigma_\Phi \equiv \textrm{sign} [\sin 2 \theta_\Phi],\nonumber
\end{align}
where $\theta_{lh}$ and $\theta_D$ are defined in Eq.~(\ref{eq:gauge-sector-defs}). The corresponding width for the emission of a $W_l$ boson is given by making the substitutions, $\theta_{lh, D} \rightarrow \theta_{lh, D} + \pi/2$.
The partial widths for the corresponding neutrino-like portal matter decays into neutrinos and $W_{h,l}$ bosons are identical to those appearing in Eq.~(\ref{eq:PMtoWhSM}). Meanwhile, for emission of the $h_\pm$ scalar, the partial widths are
\begin{align}\label{eq:PMtohPSM}
    \Gamma^A_{L,E_\pm \rightarrow \mu^- h_\pm} = \frac{M_{L,E}^\pm y_{SL,SE}^2}{64 \pi} \bigg(1-\frac{M_{h^\pm}^2}{{M_{L,E}^\pm}^2}\bigg)^2 \sin^2 \theta_\Delta.
\end{align}
When either $h^\pm$ or $W_{h,l}$ are too heavy to be produced on-shell via portal matter decays, channels with virtual $W^\pm_{l,h}$ and $h_\pm$ are unlikely to have significant branching fractions. In such a scenario, the lightest portal matter field will overwhelmingly decay through the emission of $A_D$ or $h_D$, as in the minimal portal matter construction of \cite{Rizzo:2018vlb}. If, however, any of the particles $W_{h,l}$ or $h_\pm$ are lighter than a given portal matter particle, the partial widths of Eqs.(\ref{eq:PMtoWhSM}) and (\ref{eq:PMtohPSM}) can easily yield significant branching fractions into these states. These additional decay channels will in turn diminish the branching fraction of the lightest portal matter state to the final state considered in \cite{Guedes:2021oqx}. To get a sense for the scale of this effect, we plot the branching fractions of the portal matter field $L_\pm$ to $h_\pm$ and $W_{h,l}^\pm$ in Figure \ref{fig:fermion-br}.

Notably, the results of Figure \ref{fig:fermion-br} demonstrate that feasible points in parameter space allow for substantial branching fractions of even the lightest portal matter field into final states other than the minimal model's dominant channels to an SM lepton and a dark photon or dark Higgs. As a result, a considerable reduction of the predicted signal in the analysis of \cite{Guedes:2021oqx}, and therefore a significant weakening of constraints-- if in Scenario A, $BR(E_\pm \rightarrow \mu^- + A_D,h_D ) \approx 0.45$, for example, then \cite{Guedes:2021oqx} finds that current LHC constraints require only that $M_E^\pm \gsim 600 \; \textrm{GeV}$, rather than the constraint $M_E^\pm \gsim 895 \; \textrm{GeV}$ if $BR(E_\pm \rightarrow \mu^- + A_D,h_D ) = 1$. 

Of course, this diminished sensitivity could be ameliorated by including analyses with the new heavy boson final states. While a full analysis of this type is beyond the scope of this paper, we can consider what sorts of signals might be of interest in such searches. The exact final states corresponding to the new heavy boson channels depends on the decay channel of the on-shell $W_{h,l}$ or $h_\pm$. In many regions of parameter space visible $W_{h,l}$ decays will be dominated by two-body decays featuring an SM $\tau$ (in Scenario A) or $\mu$ (in Scenario B) and a $U(1)_D$-charged portal matter field, for example  $W^\pm_{h,l} \rightarrow \tau^- + \overline{L}_\mp$, if such a channel is kinematically accessible, or if no portal matter fields are lighter than $W_{h,l}$, through a three-body decay such as $W^\pm_{h,l} \rightarrow \tau^- + \mu^+ A_D,h_D$ with a virtual internal portal matter field.\footnote{In the event that $W_{h,l}$ decays through on-shell portal matter emissions are kinematically inaccessible, it is also possible for the $W_{h,l}$ to decay via a two-body flavor violating decay such as $W_{h,l}^\pm \rightarrow \mu^+ \tau^-$, however this partial width will be suppressed by $O(v^2_\Delta/v_\Phi^2) \sim 10^{-6}$, and so is unlikely to exceed the partial width of the three-body channel from a virtual portal matter field, which only suffers $(2 \pi)^{-3} \sim O(10)^{-3}$ relative suppression to a two-body decay from the three-body phase space.} Meanwhile, $h_\pm$ will dominantly decay into a two-body muon plus portal matter state or via bosonic interactions into a $Z_D$ plus an $h_D$, if any such final states are kinematically accessible. If no two-body decays are kinematically accessible to the $h_\pm$, possible dominant three-body decays include muon-antimuon pairs from virtual portal matter $h_\pm \mu^+ \mu^- +  A_D, h_D$, or somewhat more exotic channels such as $h_\pm \rightarrow 3 h_D$ and $h_\pm \rightarrow h_D + Z_D^* \rightarrow \tau^+ \tau^- + h_D$. In any case, these final states are considerably more complicated than the canonical $\mu + h_D,A_D$ final states of the minimal model and invariably feature significant missing energy. In the case of pair production of the lightest portal matter fermion, the two-body decays of the on-shell gauge and scalar bosons will invariably be kinematically forbidden, and so we see that signals for these events will generally involve two nearly back-to-back lepton-jets (collimated clusters of charged particles) plus missing energy (because of the heavy on-shell intermediate particles in these processes, it is likely that the clusters of charged leptons will have some exotic event geometry which may help in tagging the signal). Such signals are likely discoverable, however it should be noted that because of the larger masses of the $W_{h,l}$ and $h_\pm$ bosons compared to the dark photon and dark Higgs, the individual leptons (in particular for pair production of the lightest portal matter fermions) will in general be significantly softer in cases with portal matter decay through the heavy boson emission than in cases where the portal matter decays through the more familiar dark photon/dark Higgs channel, making it less likely that events of this type will pass kinematic cuts \cite{Guedes:2021oqx}.

In addition to the $U(1)_D$-charged portal matter states, one of which must constitute the lightest vector-like fermion in the model, we can also consider how the collider signatures for the $U(1)_D$-neutral vector-like leptons which appear in the model. The absence of a $U(1)_D$ charge for these fields permits these fermions to decay via the emission of electroweak bosons into SM fermions, in this case the SM $\tau$ (for Scenario A) or muon (for Scenario B), in a manner completely analogous to that of a conventional $\tau$ or $\mu$-philic vector-like lepton. For the isodoublet, the decay widths for these processes are given by
\begin{align}
    \Gamma^A_{L_0 \rightarrow \tau^- h} = \Gamma^A_{L_0 \rightarrow \tau^- Z} = \frac{1}{2}\Gamma^A_{L_0 \rightarrow \nu_\tau W^-} = \frac{M_L^0}{32 \pi} \bigg(\frac{M_L^+ M_E^- - \sigma_{E-} M_L^+ M_E^-}{ M_L^+ M_E^+ + \sigma_{E-} M_L^- M_E^-}\bigg)^2 \frac{m_\tau^2}{v^2},\\
    \sigma_{E-} = \textrm{sign}[\sin(\theta_E + \theta_\Phi)],\nonumber
\end{align}
with corresponding results for the electroweak singlet portal matter, albeit without channels featuring the SM $W^\pm$ boson. Depending on the mass of the $W_{h,l}$ gauge bosons, an additional two-body decay channel may be open to the $U(1)_D$-neutral vector-like leptons of the type $L_0 \rightarrow W_{h,l}^{\mp} L_\pm$. Because the electroweak decay partial widths suffer substantial $m_\tau^2/v^2 \sim 10^{-5}$ suppression due to the small Yukawa coupling of the $\tau$, the decays from $W_{h,l}^{\mp}$ emission will be overwhelmingly favored if they are kinematically accessible, however because $W_{h,l}$ itself tends to have a suppressed decay width (as discussed above), these channels quickly become negligible once $W_{h,l}$ becomes massive enough to kinematically forbid their on-shell emission. Therefore, there are two conceivable qualitative scenarios for the collider signature of the lightest $U(1)_D$-uncharged vector-like lepton: It may either decay like a traditional, non-portal matter vector-like lepton via emission of electroweak bosons, or it may decay dominantly into $U(1)_D$-charged portal matter and $W_{h,l}$ bosons.

In the former case, searches for these vector-like leptons will directly correspond to searches for any predominantly $\tau$-mixed (in Scenario A) or $\mu$-mixed (Scenario B) vector-like lepton. In Scenario A, constraints from present searches, such as that of \cite{CMS:2019hsm} are considerably weaker than constraints on the portal matter masses-- even the analysis of \cite{CMS:2019hsm}, which considers an electroweak doublet vector-like lepton, leverages the production of both charged lepton pairs and production of a single charged lepton with a vector-like neutrino, and benefits from a statistical deficit in the measured signal region events at CMS, only constrains $M_L^0 \gsim 790 \; \textrm{GeV}$, considerably worse than the $M_L^\pm \gsim 1050 \; \textrm{GeV}$ constraint achieved by rescaling the constraint of \cite{Guedes:2021oqx} to limit electroweak doublet, rather than singlet, portal matter. In Scenario B, the weakened constraints on portal matter may make constraints on the $U(1)_D$-neutral fermions more important. Quantitatively, assuming that the model's heavy gauge boson and scalar content are massive enough that the $U(1)_D$-neutral fermions decay dominantly via electroweak boson emission, the analysis of \cite{Guedes:2021oqx} finds an expected limit of $\geq 420(720) \; \textrm{GeV}$ on the isospin singlet(doublet) $U(1)_D$-neutral fermion.

The collider signature for the case in which the $U(1)_D$-neutral vector-like fermions decay dominantly via $W_{h,l}$ has been less studied. However, we can comment briefly that the preferred decay channels for the $W_{h,l}$ bosons and the $U(1)_D$-charged portal matter suggest that these channels will involve large numbers of collimated leptons (lepton-jets) and significant missing energy, and likely yield constraints similar to an analysis of the $W_{h,l}$ emission channels in the case of $U(1)_D$-charged portal matter, as mentioned earlier.

The analysis thus far has restricted our attention to the least massive portal matter fermions of that are $U(1)_D$-charged and $U(1)_D$-uncharged, but the results for the heavier portal matter fields are qualitatively quite similar, if somewhat complicated by the presence of additional possible decay channels from intermediate vector-like fermions. Notably, in both the $U(1)_D$-charged and $U(1)_D$-uncharged cases, these permit the electroweak doublet vector-like leptons to decay into the electroweak singlets (or vice versa, depending on the relative masses of the particles in question) of the same $U(1)_D$ charge via the emission of SM electroweak bosons. The partial widths of these decay channels will scale with the Yukawa couplings $y_H \sim m_{\tau,\mu}/v$ and $y_{HV}$, as $O(m_{\tau,\mu}^2/v^2, y_{HV} m_{\tau,\mu}/v, y_{HV}^2)$-- because $m_{\tau,\mu}/v \ll 1$, this suppression will render these channels largely irrelevant for the $U(1)_D$-charged fermions unless $y_{HV} \sim 1$, but in the case of the $U(1)_D$-neutral vector-like fermions these channels can compete with the $O(m_{\tau,\mu}^2/v^2)$-suppressed decay width to SM fermions, or even dominate them when $y_{HV} \gg m_{\tau,\mu}/v$, leading to atypical signatures. In the event that the complete spectrum of vector-like fermions is discovered (likely at a multi-TeV muon collider where such a search has significantly greater reach), it may then be feasible to place limits on or even measure the coupling $y_{HV}$ from the branching fractions of the vector-like fermions-- given the central role this coupling constant can play in the generation of $\Delta a_\mu$ in both Scenario A and Scenario B, such a measurement would play a key role in determining if the model remains a viable explanation of the muon magnetic moment anomaly.

\subsection{Boson Sector}\label{sec:boson-pheno}
Having discussed the fermion sector in some detail, we can now move on to the boson sector. Because the new gauge and scalar bosons are leptophilic (up to small kinetic mixing), significant production of these states (apart from in association with vector-like fermions, as discussed in Section \ref{sec:fermion-pheno}) is not especially feasible at the LHC, where the leptons which couple to the dark sector bosons must be produced from quark and gluon collisions. However, in both Scenario A and Scenario B, various bosons in the dark sector may be produced copiously at a multi-TeV muon collider, through $t$-channel exchanges of heavy vector-like leptons. The heavy gauge bosons and scalars will have visible decay channels featuring final-state charged leptons in both scenarios, while if we assume dark photons and dark Higgses decay invisibly, their production may still be measured via searches for monophoton events. Furthermore, because many of these diagrams are essentially the same as the one-loop diagrams that generate the new physics contributions to the muon magnetic moment (albeit without an external photon and with the internal scalar/gauge boson line cut), the production cross section of these bosons will necessarily be related to the same parameters which govern the muon magnetic moment correction. It is therefore of interest to consider the rate at which we might expect the dark sector bosons to be produced at a muon collider, and discuss the role that these probes might play in constraining the model's ability to address the muon $g-2$ anomaly. Because of the preserved $Z_2'$ parity, it should be noted that there is no overlap between the bosons that can be produced at a muon collider in Scenario A and those which can be produced in Scenario B. As a result, we shall address the two scenarios separately here.

\subsubsection{Diboson Production (Scenario A)}\label{sec:diboson-A}

In Scenario A, the exotic bosons which may be produced at leading order in $r_\Delta/v_\Phi$ are the dark photon $A_D$, the dark Higgs $h_3$, and the heavier scalars $h_5$, $h_6$, and $h_\pm$. Perhaps the most intriguing signatures here will be those from the production of solely $A_D$ and $h_3$ states, since these both  have sub-GeV masses and will always be kinematically accessible at a multi-TeV muon collider. In the event that these particles decay invisibly, it should be possible to constrain the production rate for these particles via a simple monophoton search, such as what has been considered for a WIMP dark matter model in \cite{Han:2020uak,Black:2022qlg}.\footnote{The production of dark photons and dark Higgses from muon collisions in our model is also analogous to the corresponding case in the model of \cite{Rueter:2019wdf}, in which quark portal matter mediated similar processes at the LHC and from which the authors found monojet constraints on their parameter space.} As these cross sections scale quartically with $y_{SL}$ and $y_{SE}$, such a measurement can be used to constrain or observe these couplings, which from Eqs.~(\ref{eq:amu-1}-\ref{eq:amu-3}) play a crucial role in generating the anomalous muon magnetic moment. The precise reach of these searches remains unclear, pending further work on the beamline background in muon colliders and, of course, knowledge of the beam energy, detector coverage, and integrated luminosity of the hypothetical future machine, but we can make a rudimentary assessment by simply computing the production cross sections of the monophoton processes $\mu^+ \mu^- \rightarrow \gamma A_D A_D$, $\mu^+ \mu^- \rightarrow \gamma A_D h_3$, and $\mu^+ \mu^- \rightarrow \gamma h_3 h_3$, and estimating their significance against the dominant background $\mu^+ \mu^- \rightarrow \gamma \nu \overline{\nu}$.\footnote{This estimate ignores the monophoton signature from electroweak production of the portal matter neutrino partners, which should be degenerate in mass with the charged isospin doublet portal matter, the dominant process of which is $\mu^+ \mu^- \rightarrow \gamma N^\pm \overline{N}^\pm$. When the beam energy is sufficient to render these processes kinematically accessible, they produce comparable cross sections to those of the signal channels we do analyze when $y_{SL,SE} = 1$, but do not scale with these couplings. As such, generally for $y_{SL,SE}<1$ this signal would somewhat enhance our BSM signal (and hence improve the reach of any given collider experiment), but is generally a subdominant signal contribution when $y_{SL,SE}>1$ due to the quartic scaling of the other BSM monophoton processes with these couplings.} Analytically, we find that in the limit that the muon, $h_D$, and $A_D$ masses are negligible, all three LO cross sections for the signal processes here can be written as\footnote{As a $2 \rightarrow 3$ process, the monophoton cross sections are lengthy and not especially enlightening, so we do not present them in full here.}
\begin{align}
    \sigma_{XY} = (y_{SL}^4 \sigma^L_{XY} + y_{SE}^4 \sigma^E_{XY})\cos^4 \theta_\Delta ,
\end{align}
where $X$ and $Y$ can denote $h_D$ or $A_D$, depending on the final state for the given signal process, and $\sigma^{L,E}_{XY}$ depend solely on SM parameters and the masses of the portal matter fermions. In turn, we can estimate the significance of the monophoton signal here for a muon collider with integrated luminosity $\mathcal{L}$ as
\begin{align}
    N_{SD} = \frac{\sqrt{\mathcal{L}} \cos^4 \theta_\Delta }{\sqrt{\sigma_{SM}}} (y_{SL}^4 \sigma^L + y_{SE}^4 \sigma^E), \;\;\; \sigma^{L,E} \equiv \sum_{XY} \sigma^{L,E}_{XY}, \;\;\; XY = h_D h_D, A_D h_D, A_D A_D,
\end{align}
where $\sigma_{SM}$ is the production cross section for the dominant SM background process $\mu^+ \mu^- \rightarrow \gamma \nu \overline{\nu}$. For the purposes of constraining the parameter space of this model that can address the muon $g-2$ anomaly, we are primarily concerned with the value of the product $y_{SL} y_{SE}$, which scales the chirally enhanced contribution of the portal matter to $\Delta a_\mu$, rather than the two Yukawa couplings' individual values. Given a spectrum of portal matter states, any given value of the product $y_{SL} y_{SE}$, will in turn have a corresponding minimum possible value for $\sum_{XY} ( y_{SL}^4 \sigma^L + y_{SE}^4 \sigma^E)$, and hence $N_{SD}$. Using this minimum, we can estimate that for a given product $y_{SL} y_{SE}$, we shall achieve a significance of at least
\begin{align}\label{eq:n-sd-product}
   N_{SD} > 2 y_{SL}^2 y_{SE}^2 \cos^4 \theta_\Delta \sqrt{\mathcal{L}} \sqrt{\frac{\sigma^L \sigma^E}{\sigma_{SM}}}.
\end{align}
We can then use Eq.~(\ref{eq:n-sd-product}) to estimate the $2 \sigma$ exclusion and $5 \sigma$ discovery reach for the product $y_{SL} y_{SE}$, given hypothetical muon collider experiments with various center-of-mass energies and integrated luminosities.
We compute the signal and background cross sections using MadGraph \cite{Alwall:2014hca}, with our model implemented with Feynrules \cite{Alloul:2013bka}. We impose kinematic cuts such that
\begin{align}
    E_\gamma > 50 \; \textrm{GeV}, \;\;\; m_{\textrm{miss}}^2 \equiv (p_{\mu^+} + p_{\mu^-} - p_\gamma)^2 > (200 \; \textrm{GeV})^2, \;\;\; |\eta_\gamma|<2.5,
\end{align}
where $E_\gamma$ is the energy of the final-state photon, $m_{\textrm{miss}}^2$ is the invariant mass of the combined four-momentum of the invisible final-state particles, and $\eta_\gamma$ is the pseudorapidity of the photon. For the benchmark center-of-mass energies and corresponding luminosities we consider in this study, we have chosen the values displayed in Table \ref{tab:mu-collider-stats}, following those of the similar study in \cite{Han:2020uak}.

\begin{table}[]
    \centering
    \begin{tabular}{| c | c |}
        \hline
        $\sqrt{s}$ ($\textrm{TeV}$) & $\mathcal{L}$ ($\textrm{ab}^{-1}$) \\
        \hline
        3 & 1\\
        6 & 4\\
        10 & 10\\
        14 & 20\\
        30 & 90\\
        \hline
    \end{tabular}
    \caption{\footnotesize The benchmark center-of-mass energies and corresponding integrated luminosities for the study of monophoton constraints on the product  $y_{SL} y_{SE}$.}
    \label{tab:mu-collider-stats}
\end{table}

\begin{figure}
    \centerline{\includegraphics[width=3.5in]{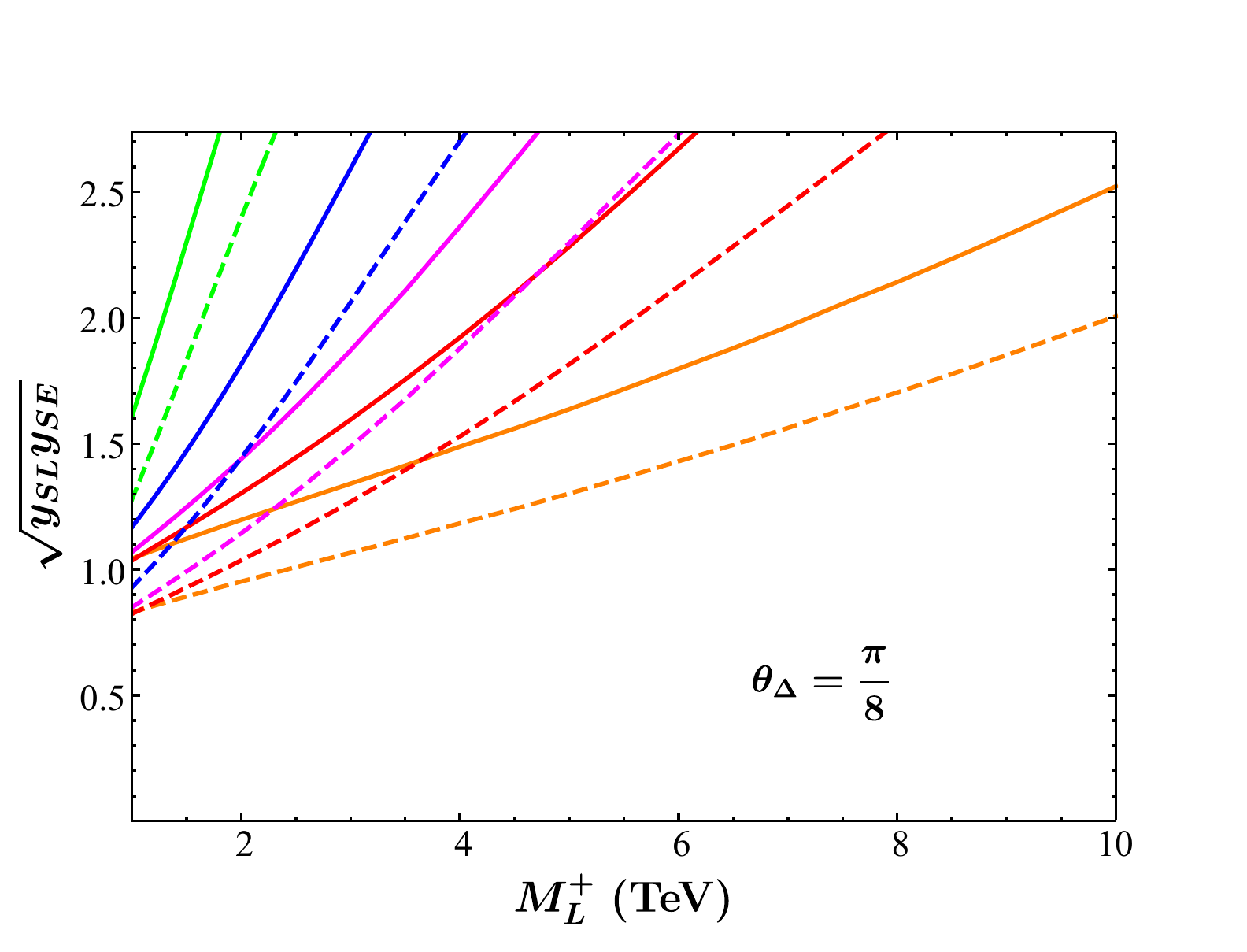}
    \hspace{-0.75cm}
    \includegraphics[width=3.5in]{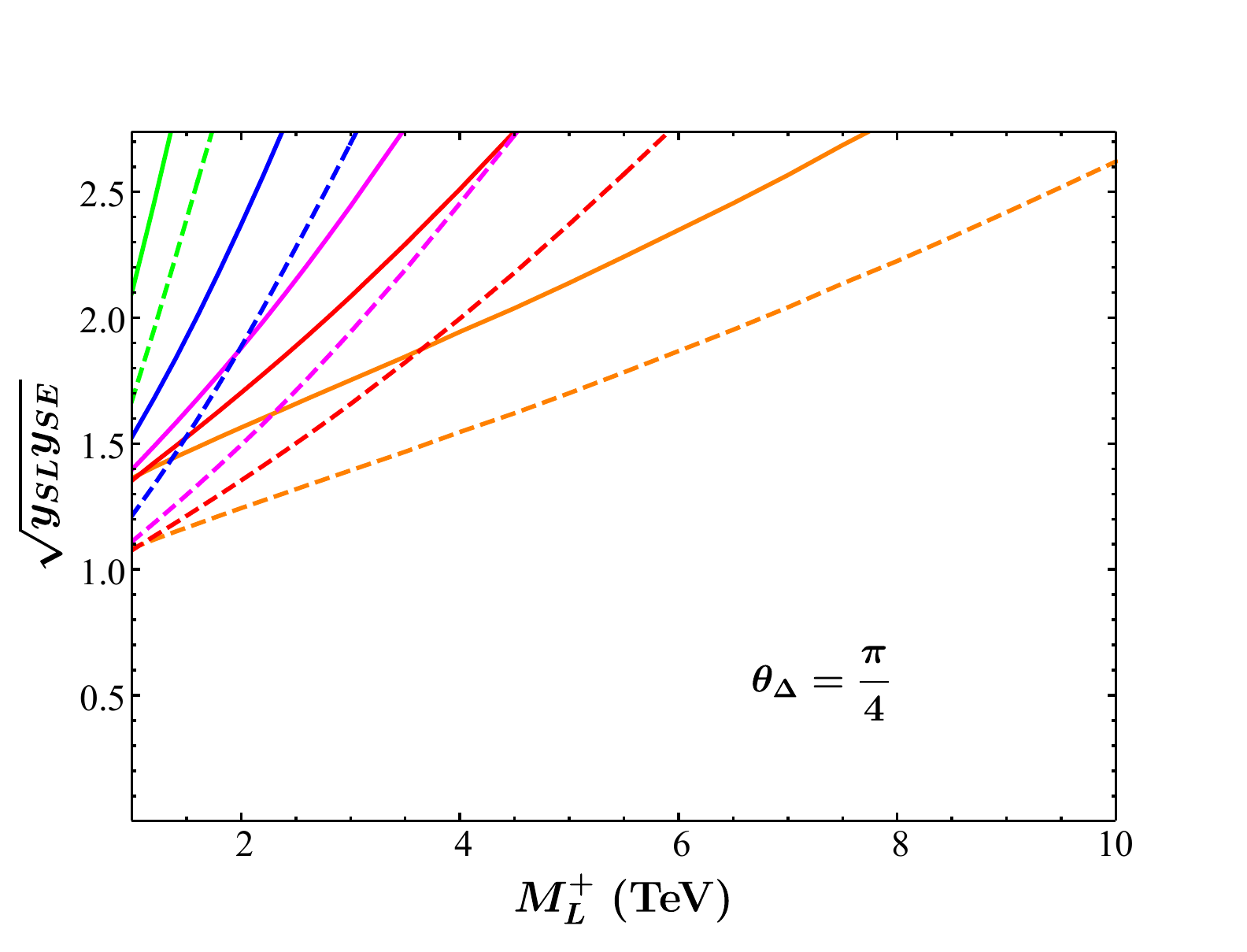}}
    \vspace*{-0.75cm}
    \centering
    \includegraphics[width=3.5in]{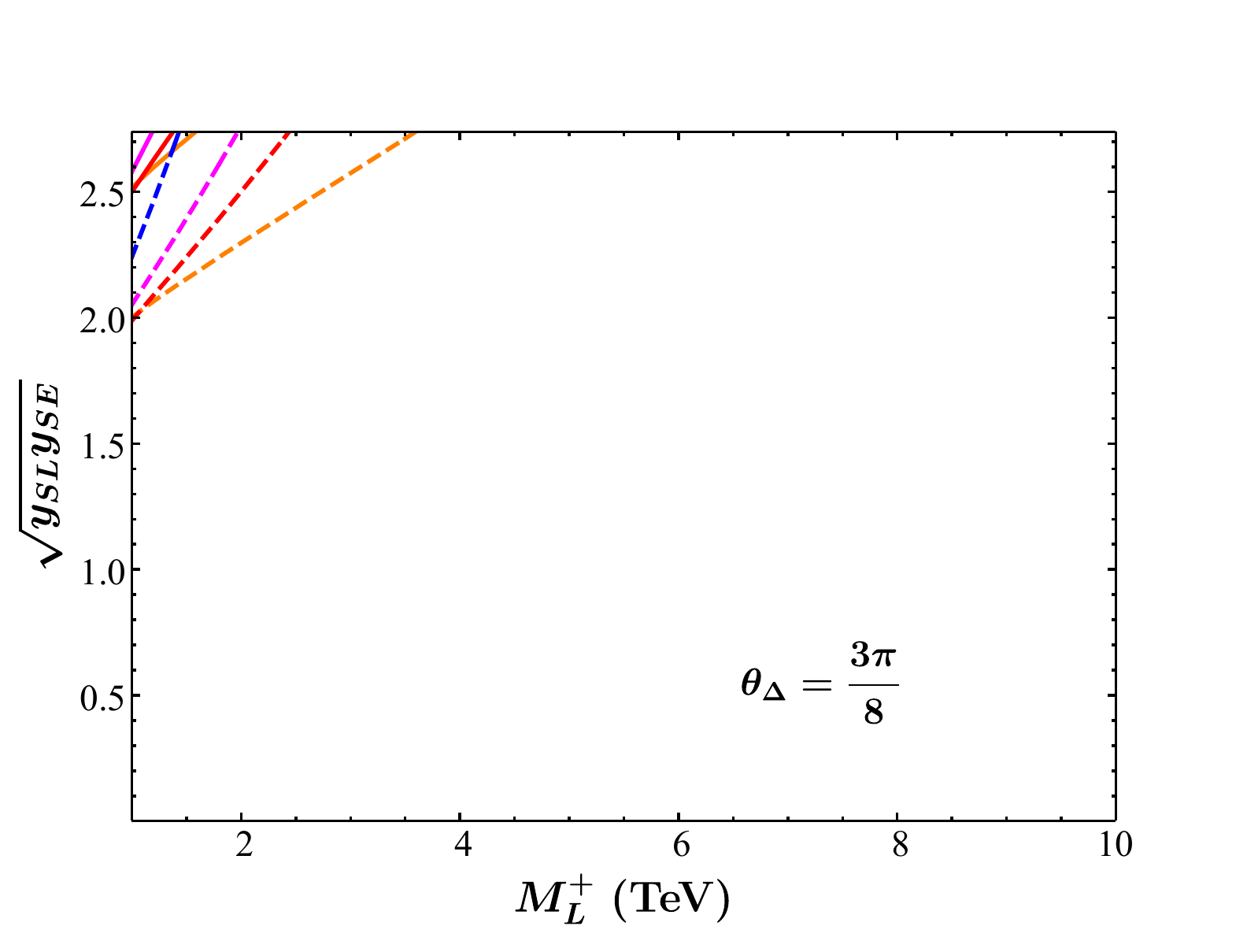}
    \caption{The estimated $5\sigma$ discovery (solid) and the $2\sigma$ exclusion (dashed) reach in Scenario A for $\sqrt{y_{SL} y_{SE}}$ at a potential future muon collider with center-of-mass energies $\sqrt{s} = 3$ TeV (Green), 6 TeV (Blue), 10 TeV (Magenta), 14 TeV (Red), and 30 TeV (Orange), with corresponding integrated luminosities given in Table \ref{tab:mu-collider-stats}. We have assumed that $M_L^- = 1.3 M_L^+$, $M_E^+ = 1.5 M_L^+$, $M_E^- = 1.8 M_L^+$, and taken the $\theta_\Delta$ values displayed on each chart. The $y$-axis at each chart terminates at the upper bound on the product $\sqrt{y_{SL} y_{SE}}$ from partial wave unitarity, given in Eq.~(\ref{eq:yL-yE-unitarity}), $\sqrt{y_{SL} y_{SE}} \lesssim 2.74$.}
    \label{fig-mono-photon}
\end{figure}

In Figure \ref{fig-mono-photon}, we depict the $2 \sigma$ exclusion and $5 \sigma$ discovery reach for these benchmark muon collider experiments, assuming that the spectrum of portal matter states follows the canonical benchmark we have used thus far, namely that $M_L^-=1.3 M_L^+$, $M_E^+= 1.5 M_L^+$, and $M_E^-=1.8 M_L^+$, at various values of the angle $\theta_\Delta$ in order to get a sense for the robustness of these constraints. Because the cross sections scale quartically with the couplings $y_{SL}$ and $y_{SE}$, we see that even though the signal-to-background ratio in these searches is quite low (generally between $10^{-2}$ and $10^{-3}$), the monophoton search can still present a powerful probe of the couplings, often significantly outperforming the bound on the product $y_{SL} y_{SE}$ from partial wave unitarity given in Eq.~(\ref{eq:yL-yE-unitarity}). Given that we have found in Section \ref{sec:g-2-A} that $\sqrt{y_{SL} y_{SE}} \gsim O(1)$ is generally necessary in order to achieve the desired value of $\Delta a_\mu$, especially in the case where $y_{HV} \ll y_{H}$, it is clear that this monophoton search can potentially significantly constrain the viable parameter space in which the model can address the muon magnetic moment anomaly. Given that a multi-TeV muon collider is likely to have a discovery reach for the portal matter fermions up to half of its center-of-mass energy, it is even feasible that such a muon collider could discover all portal matter fermions in the model, and hence have their complete spectrum known in a monophoton search.The monophoton search's constraints on the product $\sqrt{y_{SL} y_{SE}}$ are potentially exceptionally powerful.

A similarly quantitative evaluation of the discriminating power of searches for other diboson processes, namely $\mu^+ \mu^- \rightarrow h_{5,6}, h_{5,6}$, $\mu^+ \mu^- \rightarrow h^+ h^-$, $\mu^+ \mu^- \rightarrow h^+ h_3$, and $\mu^+ \mu^- \rightarrow h^+ h_D$, is considerably more complicated, and because a detailed investigation of the branching fractions of the heavy scalars in the model is highly nontrivial and dependent on the extremely complicated scalar potential terms, we will not attempt it. However, we can comment on the fermionic decay channels for these bosons which will presumably represent a sizable portion of their total decay width. The $U(1)_D$-charged scalar $h_\pm$ will invariably have open decay channels into an SM muon plus $U(1)_D$-charged portal matter fields, leading to a final state featuring a muon-antimuon pair and a dark photon or dark Higgs. Similarly, $h_5$ and $h_6$ will have open channels into an SM muon plus a virtual or on-shell $U(1)_D$-neutral vector-like lepton, $L_0$ or $E_0$, which will lead to signals featuring muons plus some electroweak boson (if the electroweak boson decays visibly, and all decay products remain hard enough to be observed by the detector, it may in fact be possible to entirely reconstruct the mass peak of the $h_5$ or $h_6$, although a detailed study of whether this is in fact feasible is not within the purview of the current work). Hence, it is feasible to suspect that in wide regions of parameter space (perhaps depending on values of scalar potential parameters, which might influence the branching fractions of the heavy scalars to other scalar states), a significant fraction of the heavy scalars will produce events with visible final state particles.
Assuming that is the case, it is then useful to then get a feel for the number of events with heavy scalars that might be produced at a multi-TeV muon collider experiment. We note that for the process $\mu^+ \mu^- \rightarrow X Y$, where $X$ and $Y$ are some final-state scalars (or the dark photon), we will have a cross section given by
\begin{align}
    \bigg( \frac{d\sigma}{d \cos\theta} \bigg)_{\mu^+ \mu^- \rightarrow X Y} = \frac{1}{32 \pi} \frac{1}{s} \sqrt{\bigg( 1 - \frac{m_X^2 + m_Y^2}{s}\bigg)^2 - 4 \frac{m_X^2 m_Y^2}{s^2}} | \mathcal{M}_{XY}|^2,
\end{align}
where $s$ is the usual Mandelstam variable, the final-state particles $X$ and $Y$ have masses $m_X$ and $m_Y$, respectively, and $|\mathcal{M}_{XY}|^2$ is the squared amplitude (with symmetry factors accounted for and averaged over initial spins), which for the processes involving the production of heavy scalars will be
\begin{align}
    |\mathcal{M}_{h_5 h_5}|^2 &= \frac{1}{8} \bigg(1 - \frac{\cos(2 \theta_\Delta)}{\cos(2 \theta_M)} \bigg)^2 \bigg( y_{SL}^4 \frac{(t-u)^2 (t u - M_{h^\pm}^4 \cos^4 \theta_M )}{(2t - {M_L^+}^2 - {M_L^-}^2)^2(2u - {M_L^+}^2 - {M_L^-}^2)^2} + (L \rightarrow E)\bigg), \nonumber \\
    |\mathcal{M}_{h_6 h_6}|^2 &= \frac{1}{8} \bigg(1 + \frac{\cos(2 \theta_\Delta)}{\cos(2 \theta_M)} \bigg)^2 \bigg( y_{SL}^4 \frac{(t-u)^2 (t u - M_{h^\pm}^4 \sin^4 \theta_M )}{(2t - {M_L^+}^2 - {M_L^-}^2)^2(2u - {M_L^+}^2 - {M_L^-}^2)^2} + (L \rightarrow E)\bigg), \nonumber \\
    |\mathcal{M}_{h_5 h_6}|^2 &= \frac{1}{4} \bigg(1 - \frac{\cos^2 (2 \theta_\Delta)}{\cos^2 (2 \theta_M)} \bigg) \bigg( y_{SL}^4 \frac{(t-u)^2 (t u - M_{h^\pm}^4 \sin^2 \theta_M \cos^2 \theta_M )}{(2t - {M_L^+}^2 - {M_L^-}^2)^2(2u - {M_L^+}^2 - {M_L^-}^2)^2} + (L \rightarrow E)\bigg), \nonumber \\
    |\mathcal{M}_{h_+ h_-}|^2 &= \frac{\sin^4 \theta_\Delta}{16} \bigg( y_{SL}^4 \frac{({M_L^-}^2 - {M_L^+}^2 + t - u)^2 (t u - M_{h^+}^4)}{(t-{M_L^+}^2)^2(u- {M_L^-}^2)^2} + (L \rightarrow E ) \bigg),\\
    |\mathcal{M}_{h_D h_+}|^2 &= |\mathcal{M}_{A_D h_+}|^2 = \frac{\sin^2(2 \theta_\Delta)}{8} \bigg( y_{SL}^4 \frac{({M_L^-}^2 - {M_L^+}^2 + t - u)^2 t u}{(t-{M_L^+}^2)^2(u- {M_L^-}^2)^2} + (L \rightarrow E ) \bigg), \nonumber \\
    |\mathcal{M}_{h_D h_-}|^2 &= |\mathcal{M}_{A_D h_+}|^2 = \frac{\sin^2(2 \theta_\Delta)}{8} \bigg( y_{SL}^4 \frac{({M_L^-}^2 - {M_L^+}^2 + u - t)^2 t u}{(u-{M_L^+}^2)^2(t- {M_L^-}^2)^2} + (L \rightarrow E ) \bigg). \nonumber
\end{align}
Notice that the requirement that $|\cos(2 \theta_\Delta)|< |\cos(2 \theta_M)|$ (see Appendix \ref{Sappendix}) is manifested in the squared amplitude for $\mu^+ \mu^- \rightarrow h_5 h_6$, where if it is not satisfied the cross section will become negative. 

\begin{figure}
    \centerline{\includegraphics[width=3.5in]{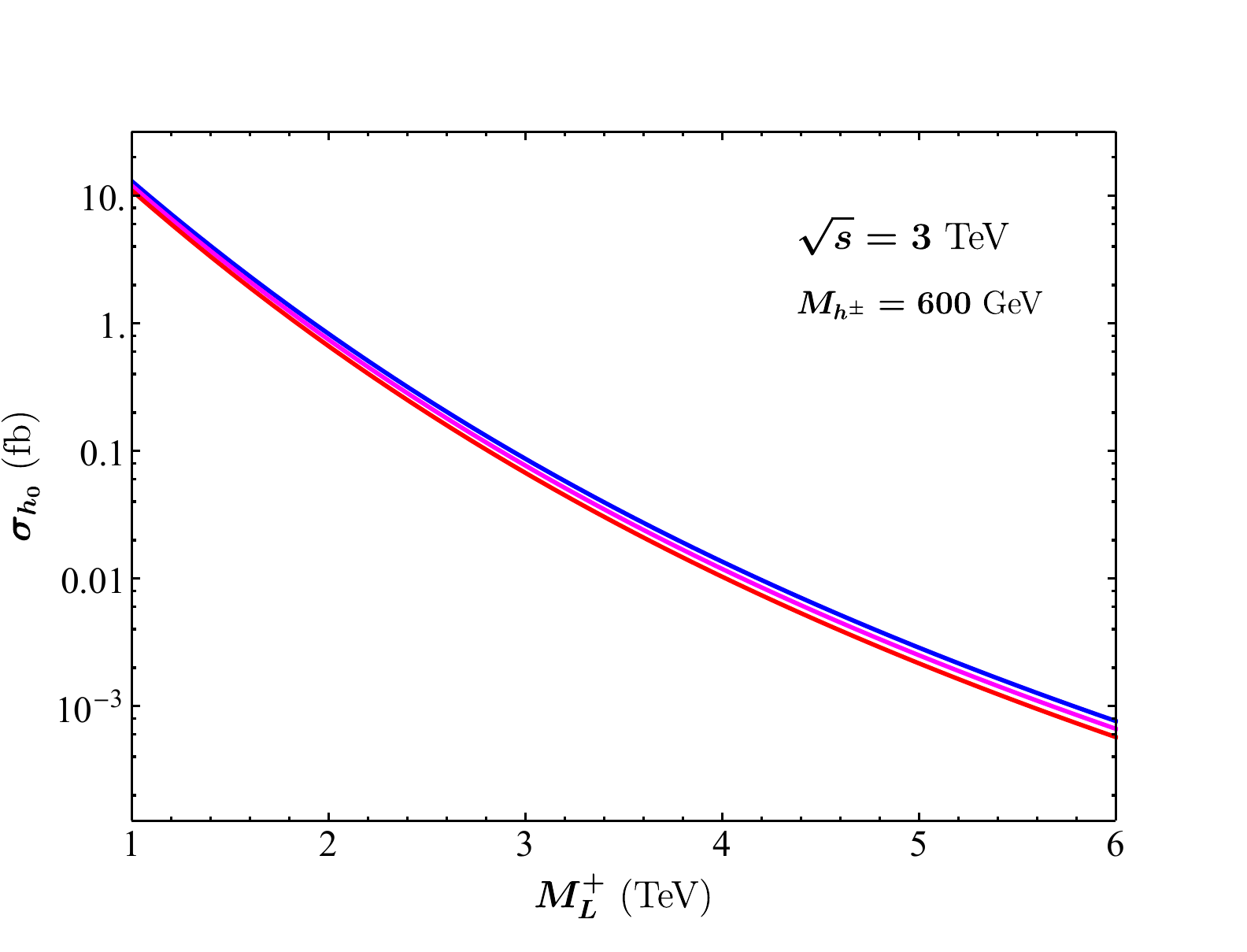}
    \hspace{-0.75cm}
    \includegraphics[width=3.5in]{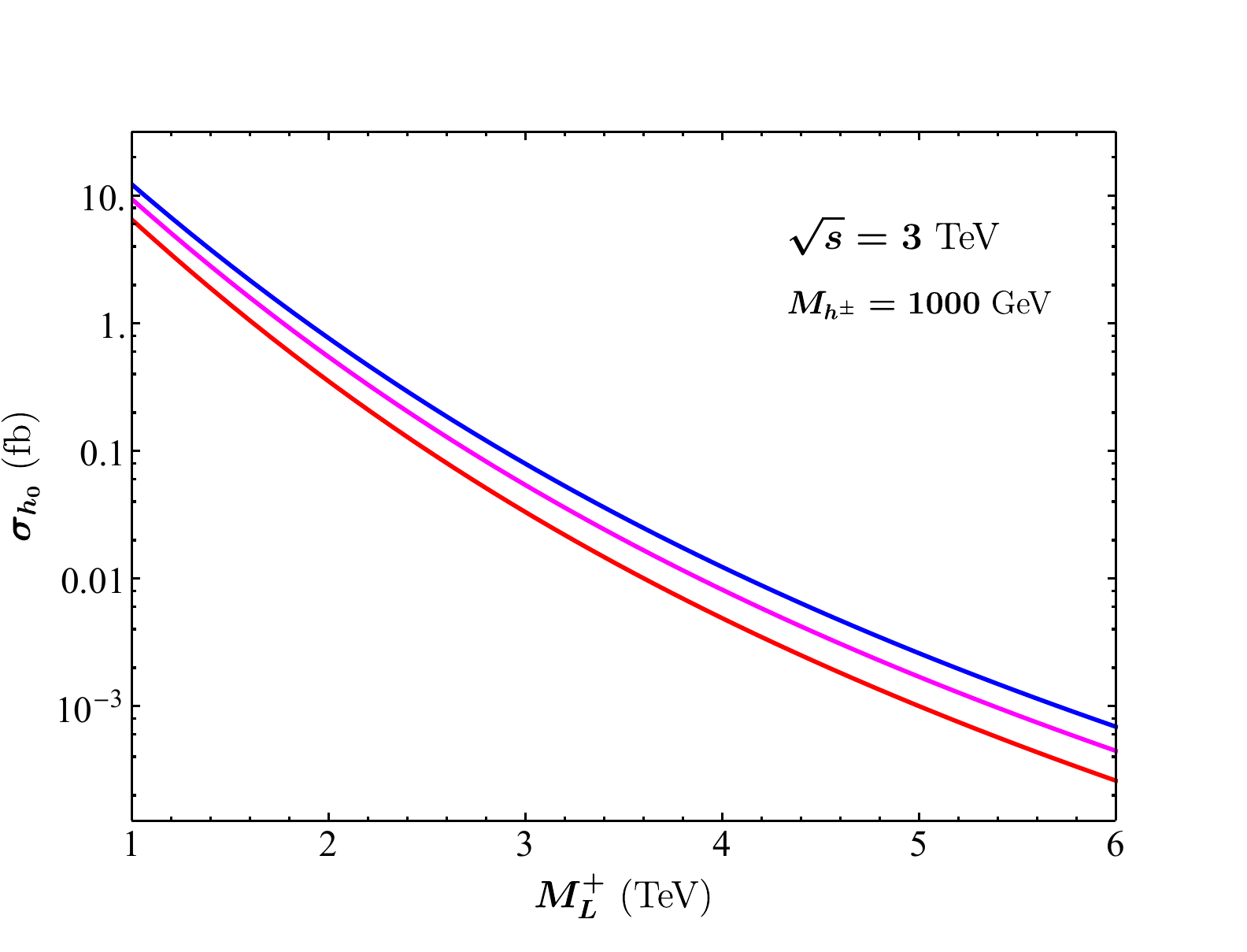}}
    \vspace*{-0.75cm}
    \centering
    \includegraphics[width=3.5in]{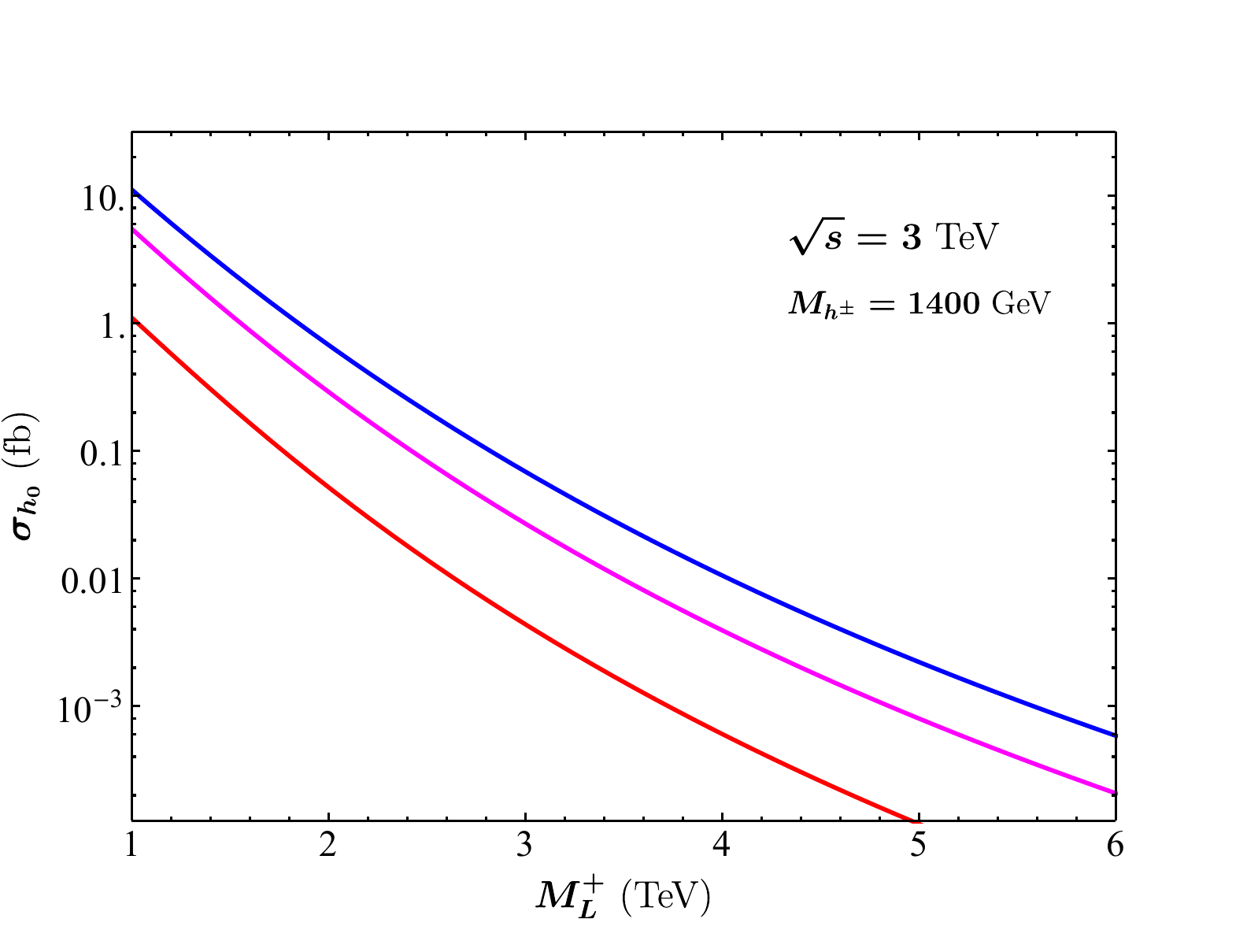}
    \caption{The combined total production cross section $\sigma_{h_0} \sigma_{\mu^+ \mu^- \rightarrow h_5 h_5} + \sigma_{\mu^+ \mu^- \rightarrow h_5 h_6} + \sigma_{\mu^+ \mu^- \rightarrow h_6 h_6}$ in Scenario A as a function of the portal matter mass $M_L^+$ at a $\sqrt{s} = 3 \; \textrm{TeV}$ muon collider for $\theta_\Delta = \theta_M = \pi/8$ (Blue), $\theta_\Delta = \theta_M = \pi/4$ (Magenta), and $\theta_\Delta = \pi/8$, $\theta_M = 3 \pi/8$ (Red). We have taken $y_{SL} = y_{SE} = 1$, and assumed a benchmark portal matter mass spectrum $M_L^- = 1.3 M_L^+$, $M_E^+ = 1.5 M_L^+$, and $M_E^- = 1.8 M_L^+$.}
    \label{fig-h5h6-xsections}
\end{figure}

\begin{figure}
    \centerline{\includegraphics[width=3.5in]{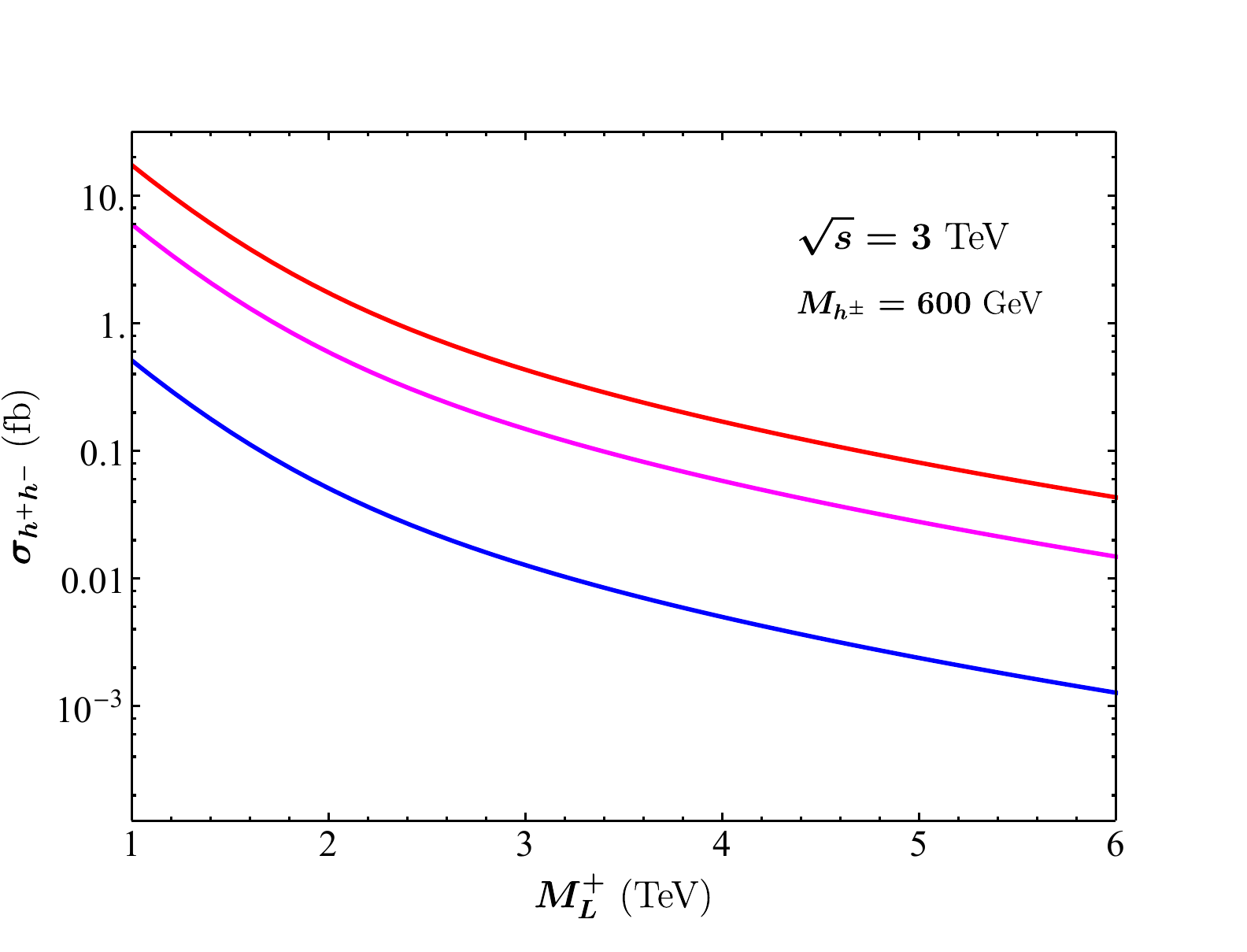}
    \hspace{-0.75cm}
    \includegraphics[width=3.5in]{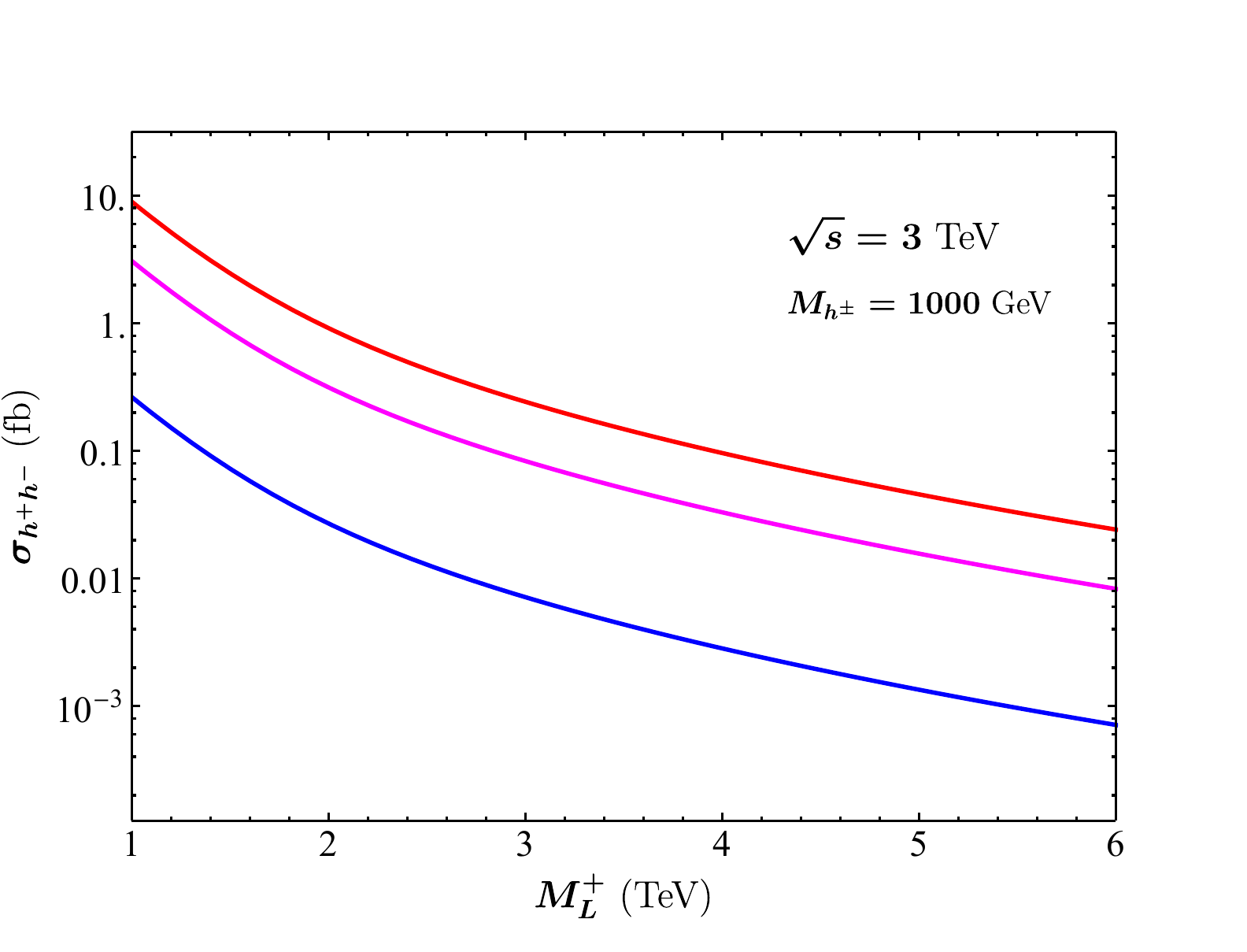}}
    \vspace*{-0.75cm}
    \centering
    \includegraphics[width=3.5in]{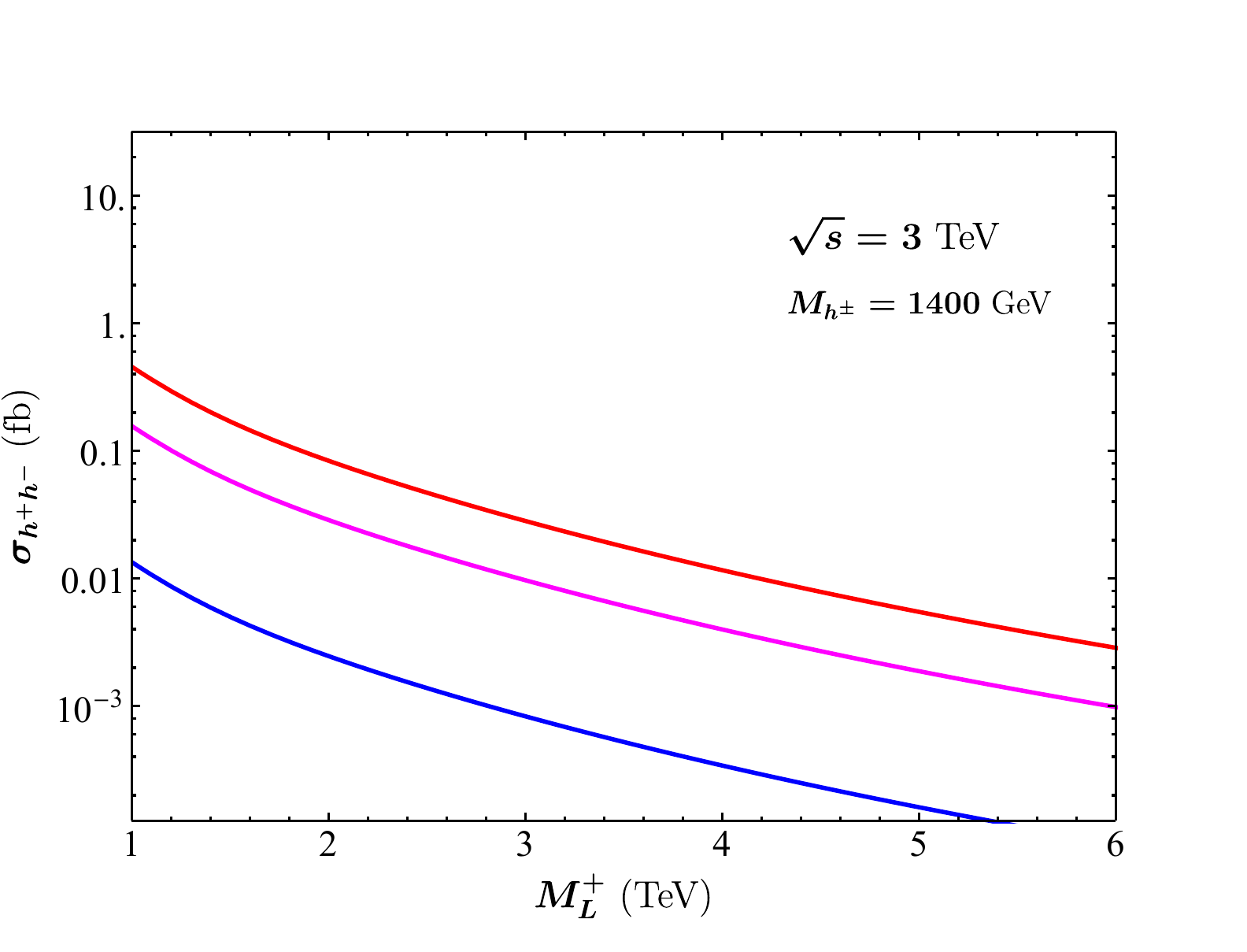}
    \caption{The total production cross section $\sigma_{\mu^+ \mu^- \rightarrow h_+ h_-}$ in Scenario A as a function of the portal matter mass $M_L^+$ at a $\sqrt{s} = 3 \; \textrm{TeV}$ muon collider for $\theta_\Delta = \pi/8$ (Blue), $\theta_\Delta = \pi/4$ (Magenta), and $\theta_\Delta = 3\pi/8$ (Red).We have taken $y_{SL} = y_{SE} = 1$, and assumed a benchmark portal matter mass spectrum $M_L^- = 1.3 M_L^+$, $M_E^+ = 1.5 M_L^+$, and $M_E^- = 1.8 M_L^+$.}
    \label{fig-hPhM-xsections}
\end{figure}

\begin{figure}
    \centerline{\includegraphics[width=3.5in]{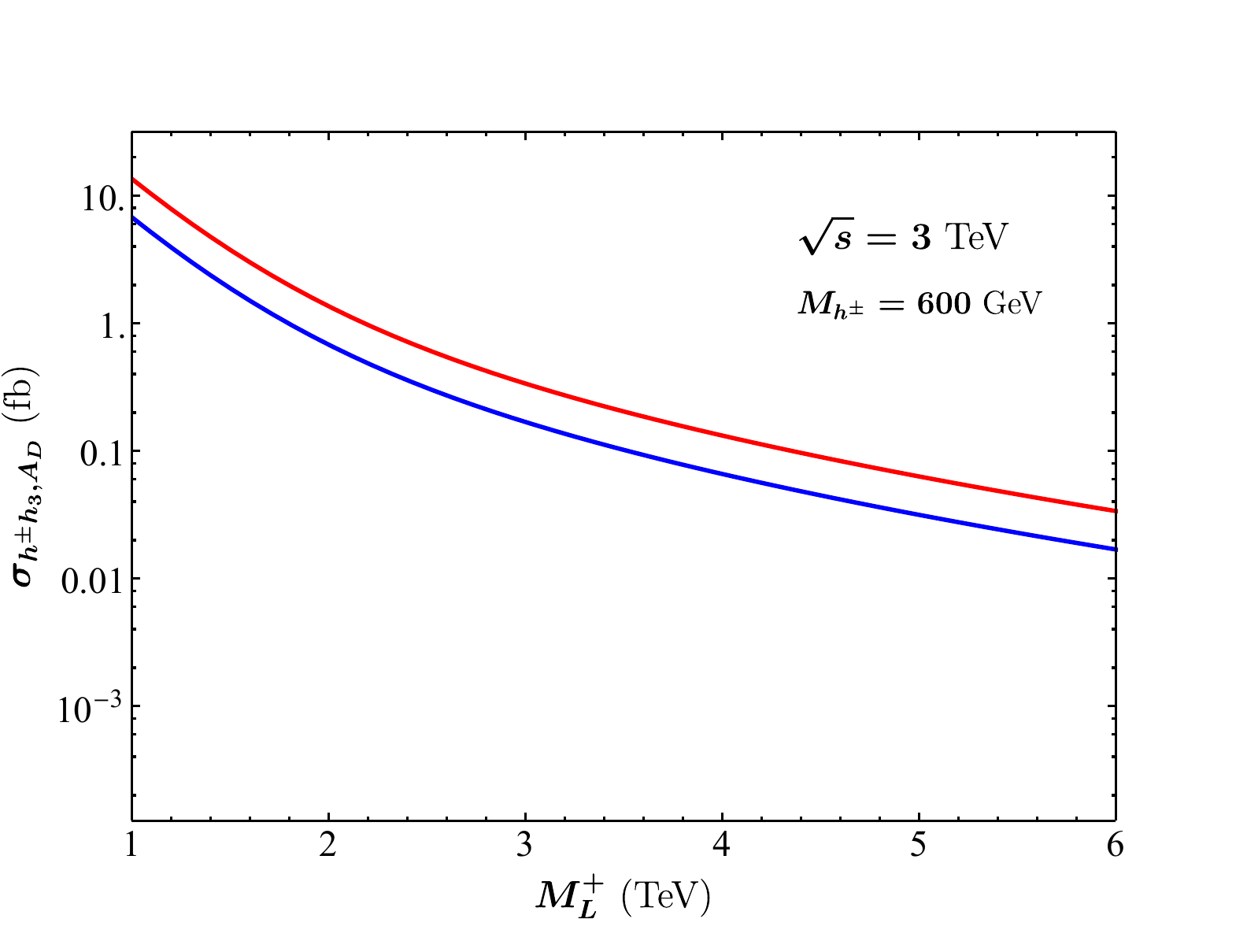}
    \hspace{-0.75cm}
    \includegraphics[width=3.5in]{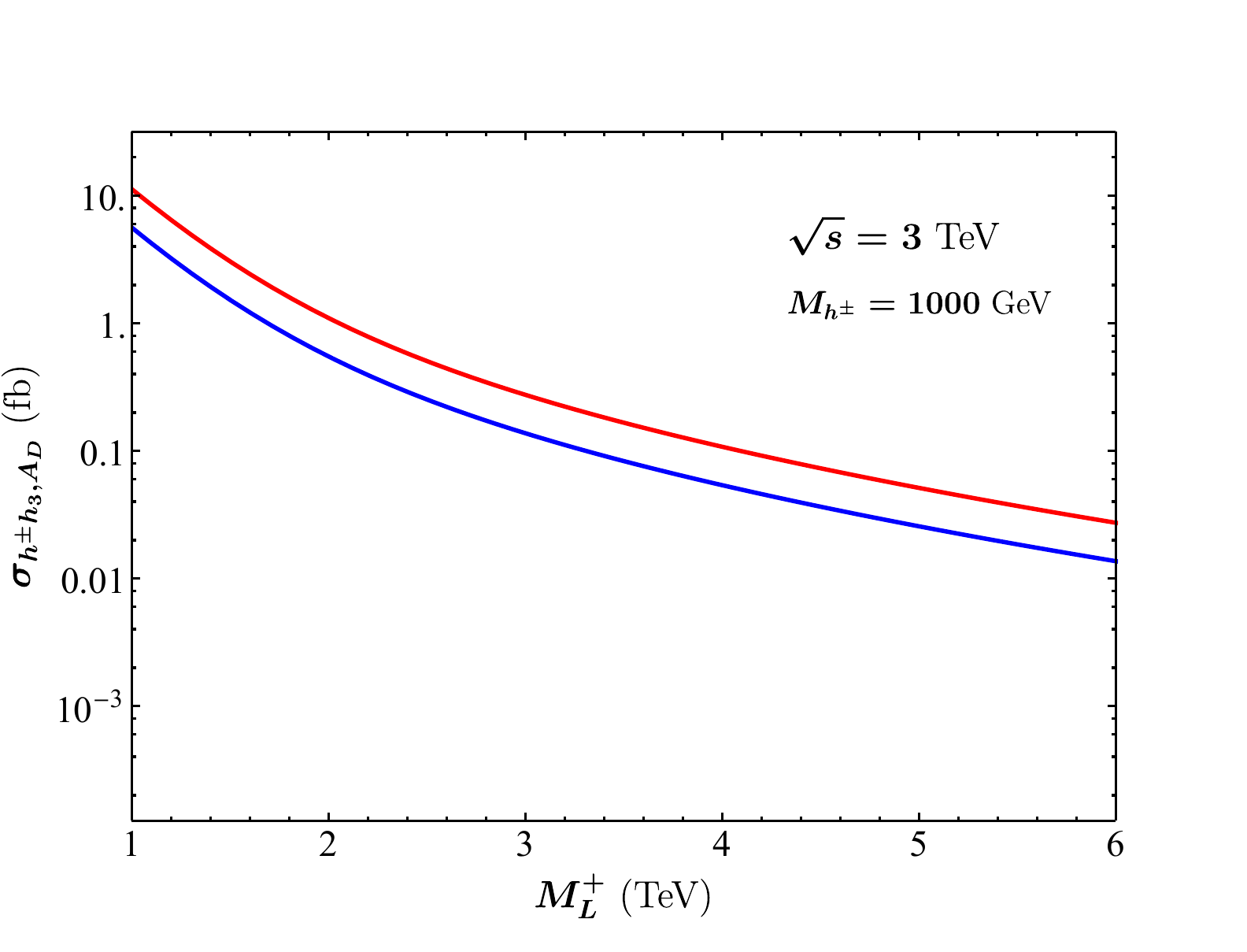}}
    \vspace*{-0.75cm}
    \centerline{\includegraphics[width=3.5in]{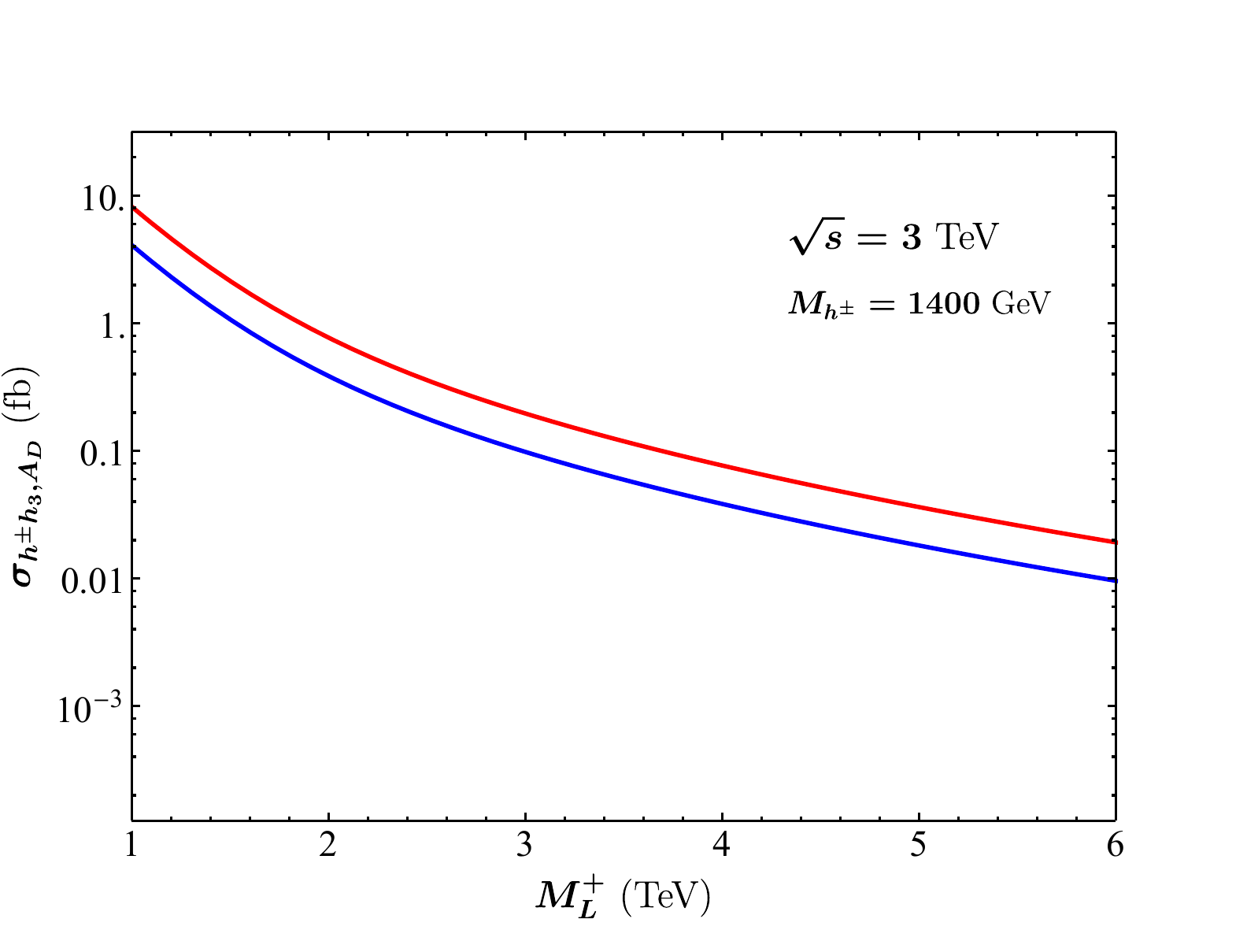}
    \hspace{-0.75cm}
    \includegraphics[width=3.5in]{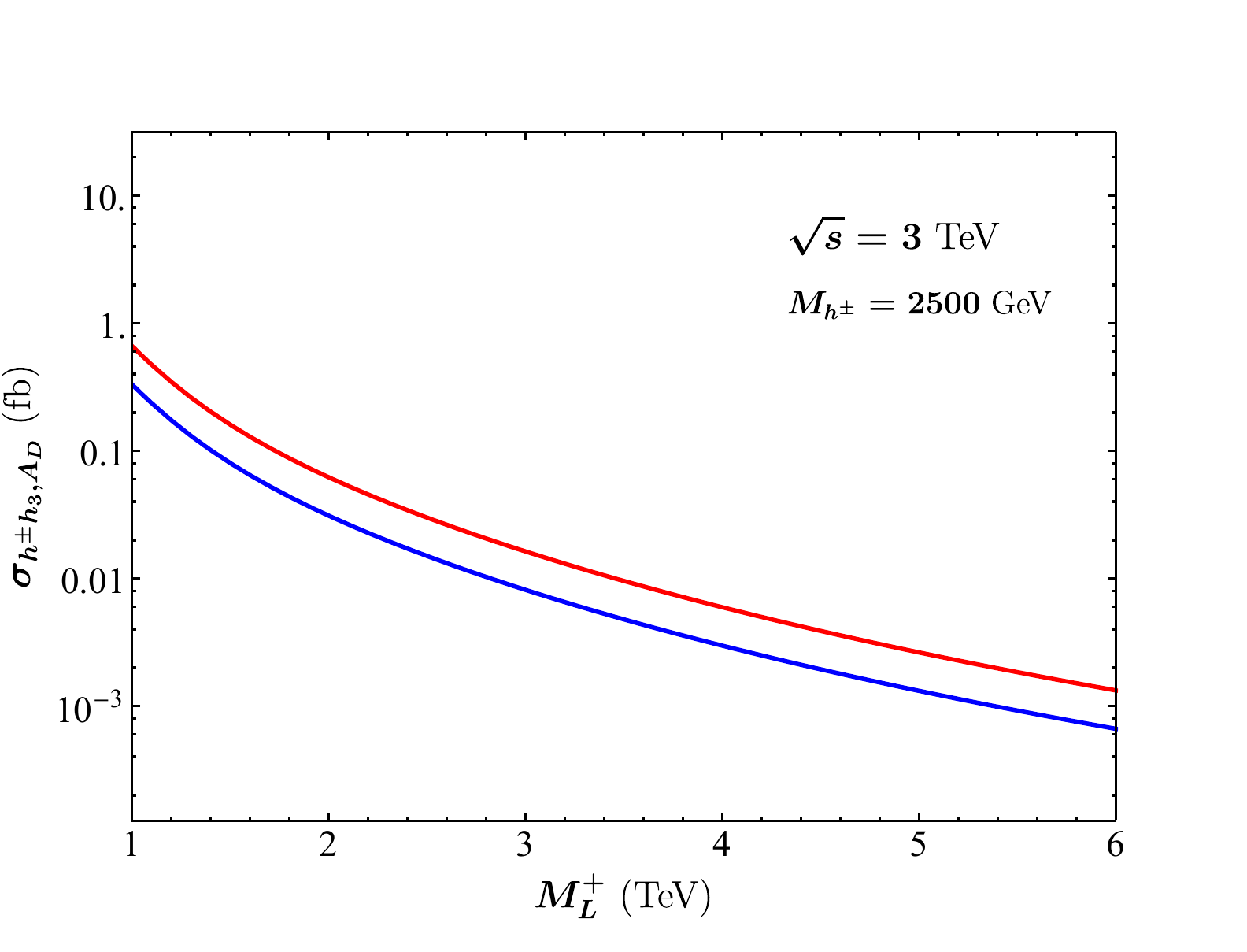}}
    \caption{The combined total production cross section $\sigma_{h_\pm h_3,A_D} = \sigma_{\mu^+ \mu^- \rightarrow h_+ h_3} + \sigma_{\mu^+ \mu^- \rightarrow h_- h_3} + \sigma_{\mu^+ \mu^- \rightarrow h_+ A_D} + \sigma_{\mu^+ \mu^- \rightarrow h_- A_D}$ in Scenario A as a function of the portal matter mass $M_L^+$ at a $\sqrt{s} = 3 \; \textrm{TeV}$ muon collider for $\theta_\Delta = \pi/8$ (Blue) and  $\theta_\Delta = \pi/4$ (Red) (note that these cross sections are invariant under $\theta_\Delta \rightarrow \pi/2 - \theta_\Delta$). We have taken $y_{SL} = y_{SE} = 1$, and assumed a benchmark portal matter mass spectrum $M_L^- = 1.3 M_L^+$, $M_E^+ = 1.5 M_L^+$, and $M_E^- = 1.8 M_L^+$.}
    \label{fig-hPhD-xsections}
\end{figure}

In Figures \ref{fig-h5h6-xsections}, \ref{fig-hPhM-xsections}, and \ref{fig-hPhD-xsections}, we depict the heavy scalar production cross sections at a $\sqrt{s} = 3 \; \textrm{TeV}$ muon collider as functions of the portal matter mass $M_L^+$, again assuming our benchmark values $M_L^- = 1.3 M_L^+$, $M_E^+ = 1.5 M_L^+$, and $M_E^- = 1.8 M_L^+$, for different values of the scalar mass scale $M_{h^\pm}$ and different choices of the angles $\theta_\Delta$ and $\theta_M$. For convenience, In Figures \ref{fig-h5h6-xsections} and \ref{fig-hPhD-xsections}, we have combined several final states that will likely have similar or identical final-state signals, such as $h_5$ and $h_6$, $h_D$ and $A_D$, and $h^+$ and $h^-$. Assuming a luminosity of $O(\textrm{ab}^{-1})$ for a multi-TeV muon collider, we then see that if the final states with heavy scalars are kinematically accessible, they can be produced copiously at such a machine-- for $M_L^+$ near the current bound of $\sim 1 \; \textrm{TeV}$ we can expect thousands of heavy scalar production events. We should also note that these processes, like the monophoton processes we inspected earlier, will scale quartically with the Yukawa couplings $y_{SL}$ and $y_{SE}$, introducing possibly significant constraints on these couplings which play a key role in the new physics contribution to the anomalous muon magnetic moment. 

\subsubsection{Diboson Production (Scenario B)}\label{sec:diboson-B}

In Scenario B, only heavy scalars and gauge bosons might be produced via the $t$-channel exchange of vector-like fermions here, since the dark photon and the dark Higgs fields both lack appreciable fermionic couplings featuring the muon. The heavy boson production, however, might be constrained through their visible decay channels to vector-like and/or SM leptons, as with the heavy scalars in Scenario A, discussed in Section \ref{sec:diboson-A}. In Scenario B, the exotic bosons that might be produced are $W_{h,l}^\pm$, $Z_D$, $h_1$, $h_2$, and $h_4$-- as noted earlier, precisely those fields which may \emph{not} be appreciably produced at a muon collider in Scenario A-- final states with $Z_D$ are likely to be of particular phenomenological interest, since it is feasible that they have an appreciable branching fraction to $\mu^+-\mu^-$ pairs, and can therefore be easily reconstructed if their width is sufficiently narrow. The symbolic forms of the cross sections in this scenario are  quite lengthy, and we do not display them here (although for those who are curious about the specific forms, they are given in Appendix \ref{Bappendix}. For our purposes, the salient points to be aware of are that all cross sections scale quartically with the quantity $e_D/M_{Z_D}$, and are otherwise dependent on the same parameters which govern the new physics correction to the muon magnetic moment here: the spectrum of vector-like fermion masses, $\lambda_1$, $\theta_D$, $\theta_{lh}$, the masses of the heavy scalars $h_1$, $h_2$, and $h_4$, and that of the gauge boson $Z_D$. As discussed in Section \ref{sec:g-2-B}, physical consistency of the various parameters leads to a rather narrow allowed space of selections of these parameters-- as in our $g-2$ analysis, then, although our space appears to have a significant number of free parameters, in practice our selections will be necessarily limited.

 \begin{figure}[h!]
   \centerline{ \includegraphics[width=3.5in]{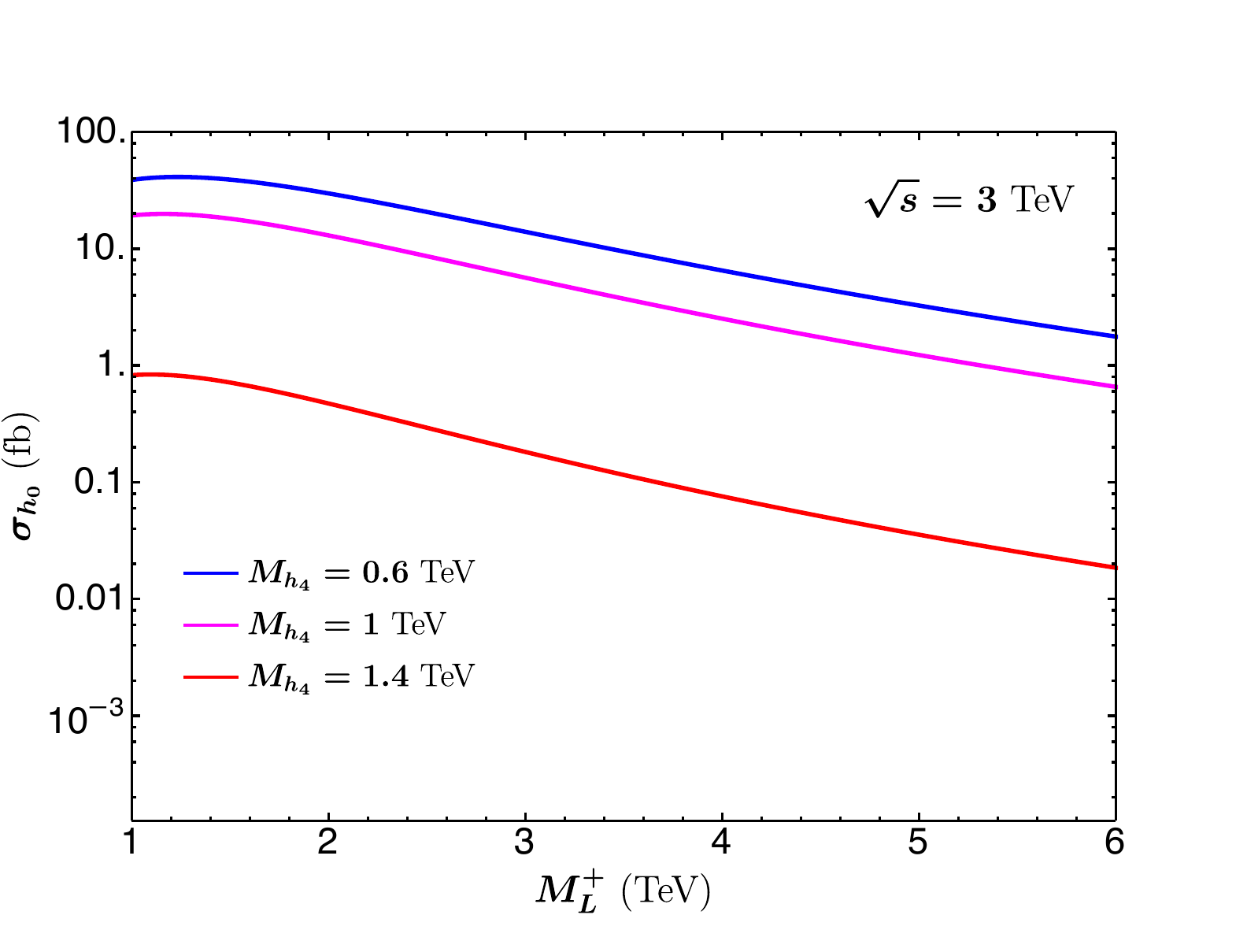}
    \hspace{-0.75cm}
    \includegraphics[width=3.5in]{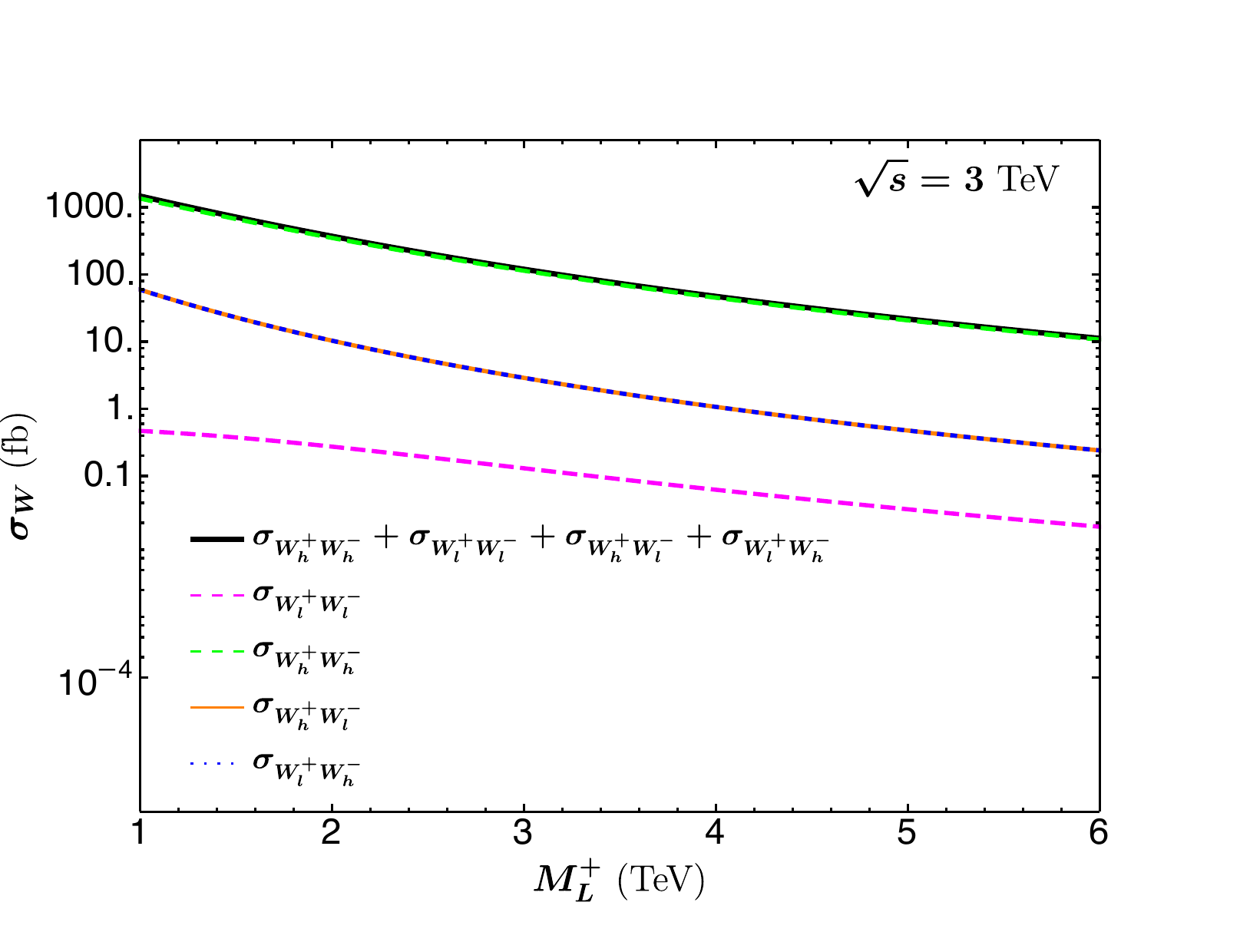}}
    \vspace*{-0.5cm}
    \centering
    \caption{(Left): The combined total production cross section $\sigma_{h_0 } = \sigma_{\mu^+ \mu^- \rightarrow h_1 h_1} + \sigma_{\mu^+ \mu^- \rightarrow h_2 h_2} + \sigma_{\mu^+ \mu^- \rightarrow h_1 h_2} + \sigma_{\mu^+ \mu^- \rightarrow h_4 h_4}$ as a function of the portal matter mass $M_{L}^+$ at a $\sqrt{s} = 3 \; \textrm{TeV}$ muon collider for Scenario B, for 
    $M_{L}^-/M_{L}^+ = 1.3 $, $M_{E}^+/M_{L}^+ = 1.5 $, $M_{E}^-/M_{L}^+ = 1.8$, 
    $M_{h_1}=1.5$ TeV and $M_{h_2}=1.2$ TeV.
    (Right): The combined total and individual production cross sections $\sigma_{W } = \sigma_{\mu^+ \mu^- \rightarrow W_l^+W_l^-} + \sigma_{\mu^+ \mu^- \rightarrow  W_h^+W_h^-}+ \sigma_{\mu^+ \mu^- \rightarrow  W_h^+W_l^-}+ \sigma_{\mu^+ \mu^- \rightarrow  W_l^+W_h^-}$ in Scenario B, for the same portal masses, but with 
     $ s_{lh} M_{Z_D}= 0.75\,\text{TeV} ,\; c_{lh} M_{Z_D}= 1.3\,\text{TeV}$, and $ \theta_D=\pi/4$. 
    }
    \label{fig-B-h0-xsections}
\end{figure}

\begin{figure}[h!]
   \centerline{ \includegraphics[width=3.5in]{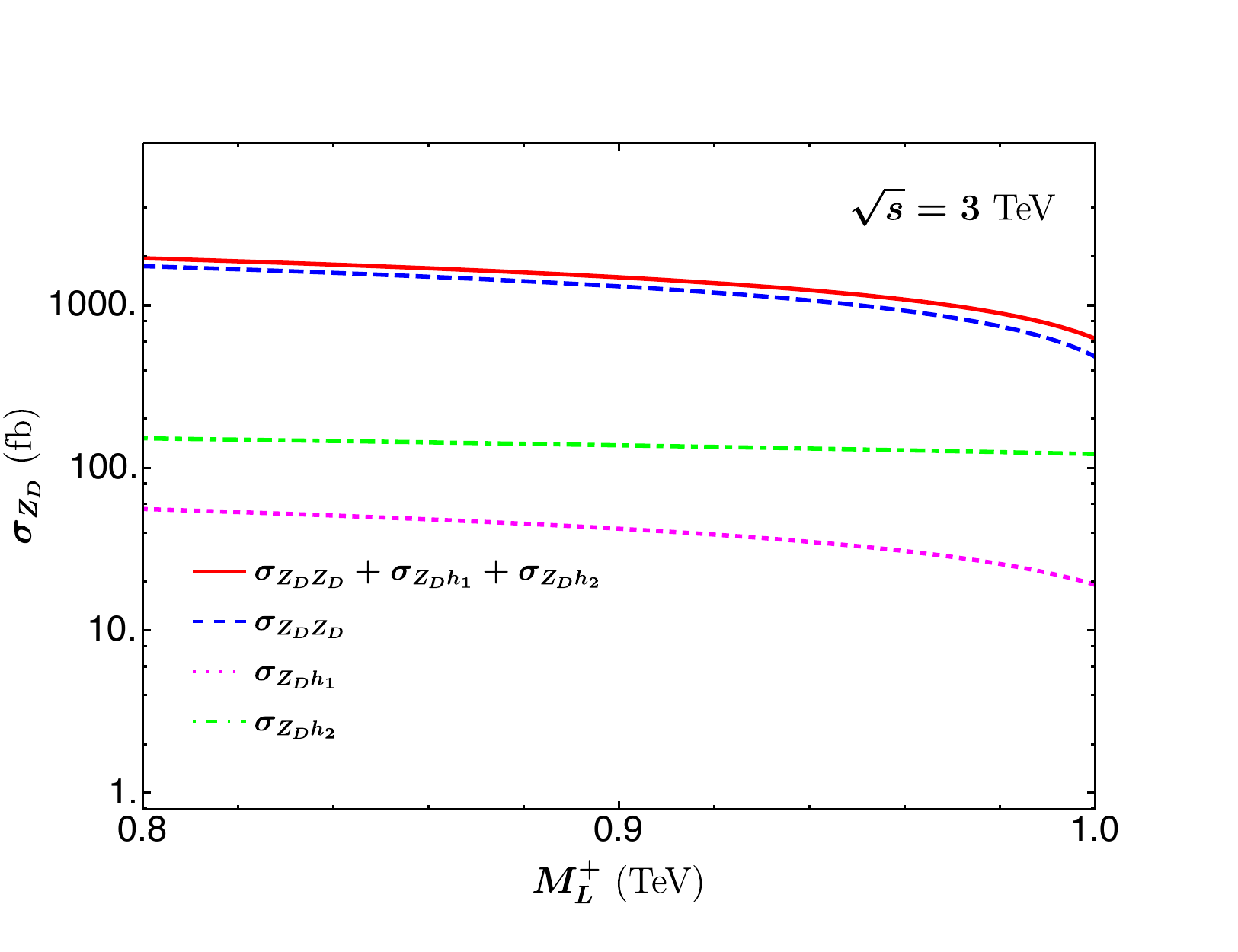}
    \hspace{-0.75cm}
    \includegraphics[width=3.5in]{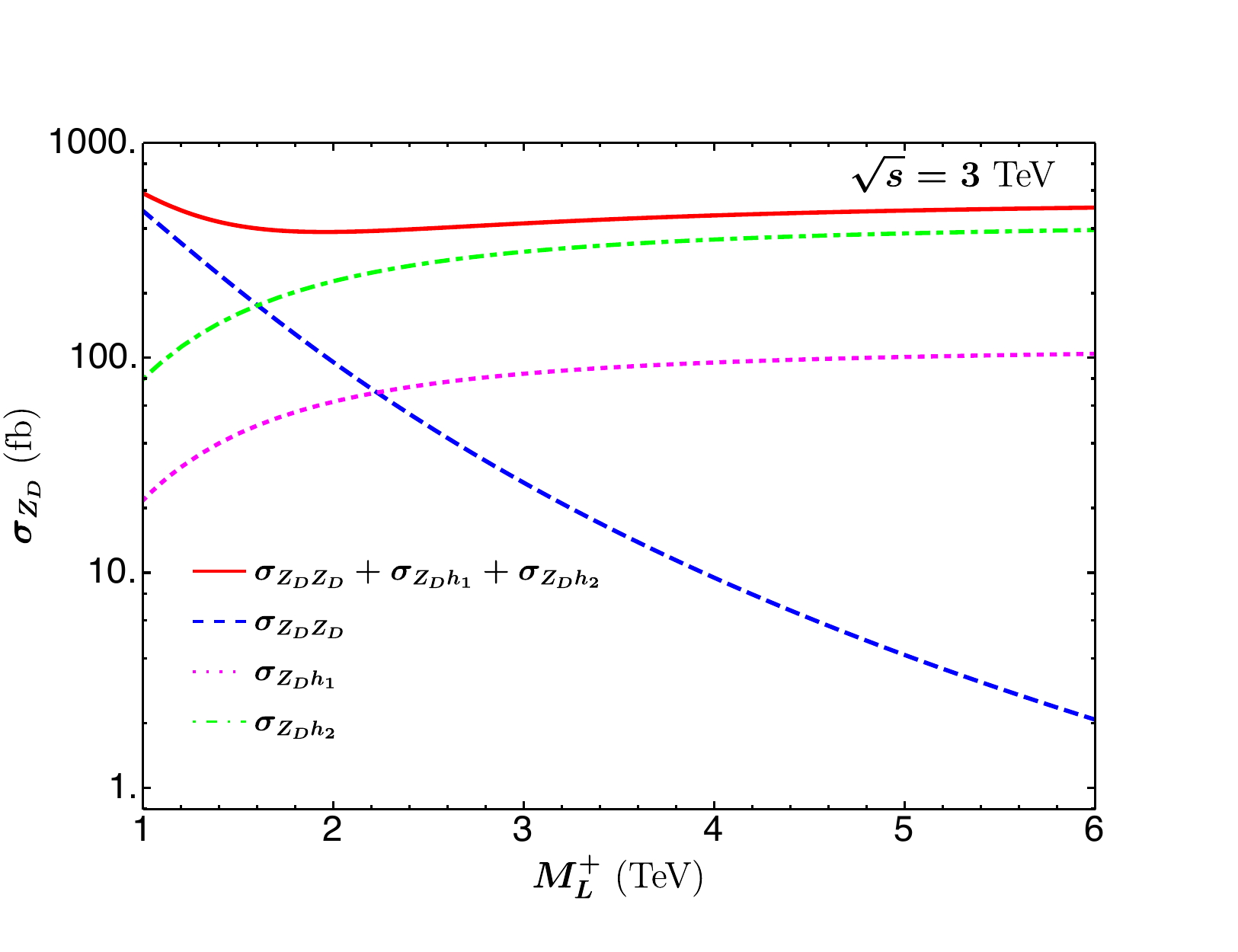}}
    \vspace*{-0.5cm}
    \centering
    \caption{The combined total and individual production cross sections of $\sigma_{Z_D } = \sigma_{\mu^+ \mu^- \rightarrow Z_D Z_D} + \sigma_{\mu^+ \mu^- \rightarrow  Z_D h_1}+ \sigma_{\mu^+ \mu^- \rightarrow  Z_D h_2}$  as a function of $M_{L}^+$ at a $\sqrt{s} = 3 \; \textrm{TeV}$ muon collider for Scenario B, assuming that (Left): $M_{L}^-/M_{L}^+ = 1.3 $, $M_{E}^+/M_{L}^+ = 1.5 $, $M_{E}^-/M_{L}^+ = 1.8$, $ s_{lh} M_{Z_D}/M_{L}^+ = 0.75,\, c_{lh} M_{Z_D}/M_{L}^+ = 1.3,\, M_{h_2}/M_{L}^+ =0.8,\, M_{h_1}/M_{L}^+ =1.5,\;\theta_D=\pi/4  $, (Right): $M_{L}^-/M_{L}^+ = 1.3 $, $M_{E}^+/M_{L}^+ = 1.5 $, $M_{E}^-/M_{L}^+ = 1.8$, $ s_{lh} M_{Z_D}= 0.75\,\text{TeV} ,\, c_{lh} M_{Z_D}= 1.3\,\text{TeV},\, M_{h_2}=1.2\,\text{TeV},\, M_{h_1}=1.5 \,\text{TeV}$, and $\theta_D=\pi/4$. 
    }
    \label{fig-B-Z-xsections}
\end{figure}

\begin{figure}[t!]
\hspace{-0.3cm}
\centerline{
    \includegraphics[width=3.5in]{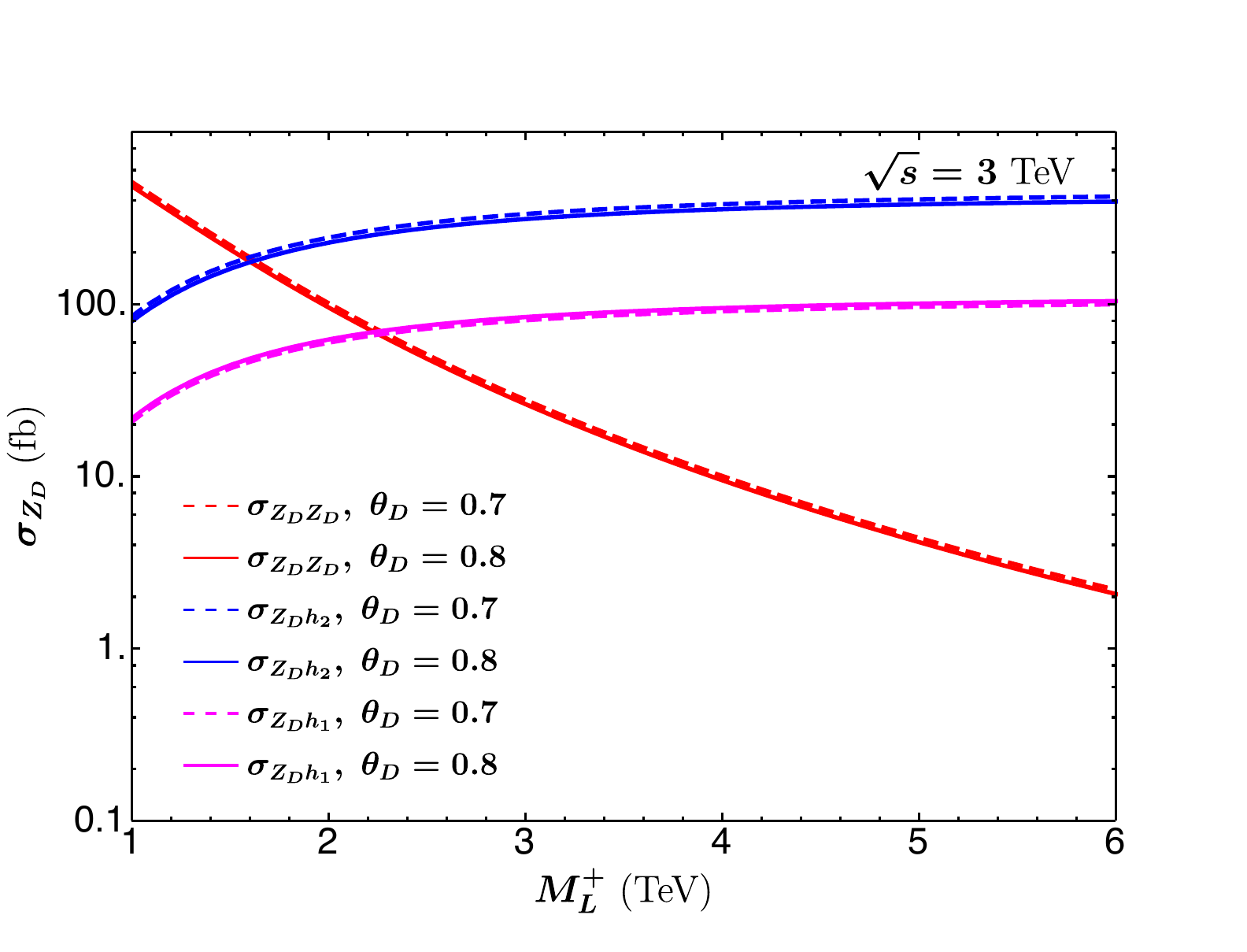}
     \hspace{-0.75cm}
    \includegraphics[width=3.5in]{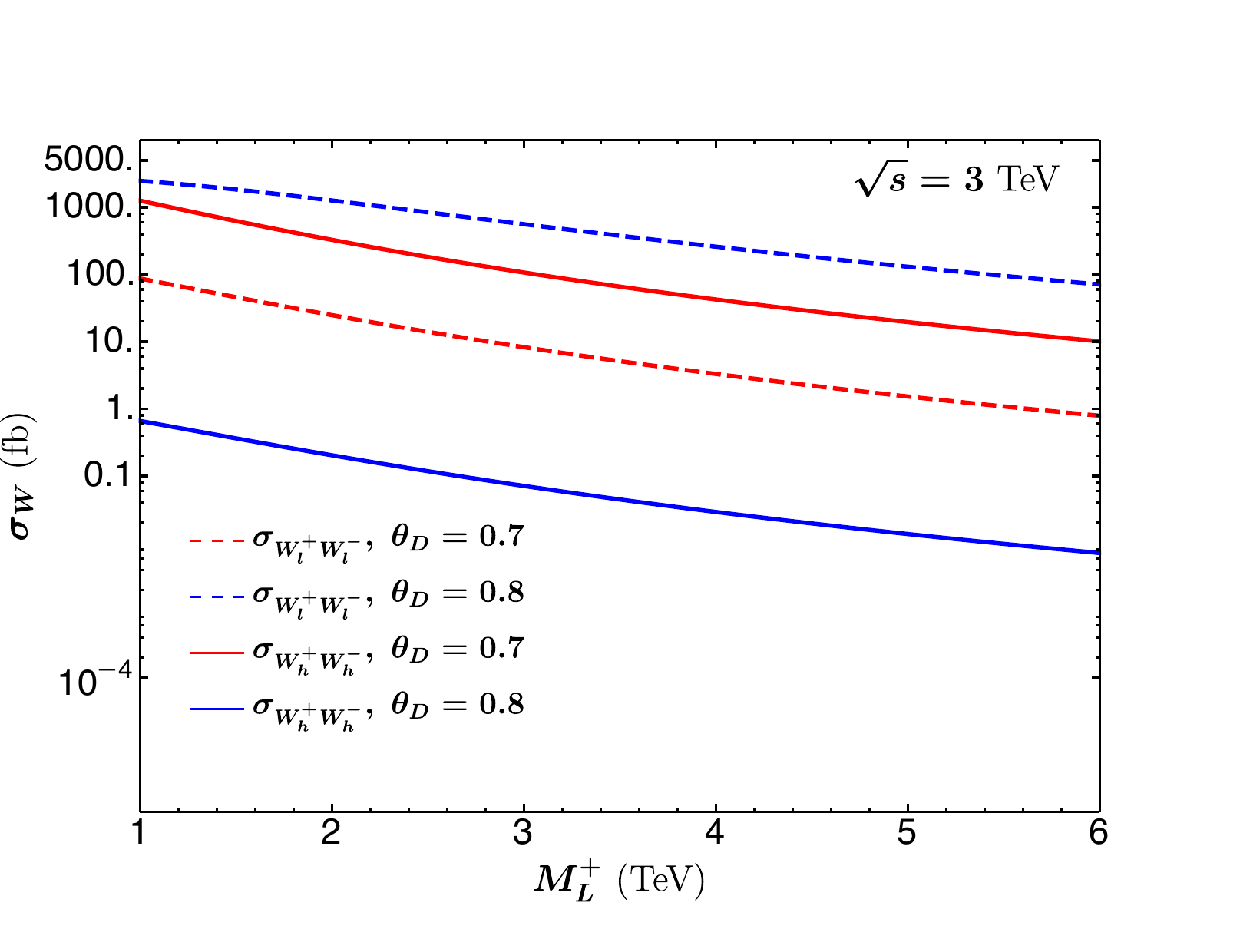}
   }
    \vspace*{-0.5cm}
    \centering
    \caption{The effect of $\theta_D$ on the individual cross sections for heavy gauge bosons in Scenario B, assuming $M_{L}^-/M_{L}^+ = 1.3 $, $M_{E}^+/M_{L}^+ = 1.5 $, $M_{E}^-/M_{L}^+ = 1.8$, $ s_{lh} M_{Z_D}= 0.75\,\text{TeV} ,\, c_{lh} M_{Z_D}= 1.3\,\text{TeV},\, M_{h_2}=1.2\,\text{TeV},\, M_{h_1}=1.5 \,\text{TeV},\; e_D=0.8  $. (Left): Different channels for $Z_D$ production are shown for $\theta_D=0.7$ and $\theta_D=0.8$.  
    (Right): $\sigma_{W_l^+,W_l^-}$ are plotted in dashed curves and $\sigma_{W_h^+,W_h^-}$ are shown for $\theta_D=0.7$ and $\theta_D=0.8$. 
    Note that for different values of $\theta_D$, the main contributing channels are different.}
    \label{fig-B-WZ-xsections-thetaD}
\end{figure}

On the left panel of Fig~\ref{fig-B-h0-xsections}, we plot the combined heavy scalar production cross sections  $\sigma_{h_0 } = \sigma_{\mu^+ \mu^- \rightarrow h_1 h_1} + \sigma_{\mu^+ \mu^- \rightarrow h_2 h_2} + \sigma_{\mu^+ \mu^- \rightarrow h_1 h_2} + \sigma_{\mu^+ \mu^- \rightarrow h_4 h_4}$ in Scenario B at a $\sqrt{s}=$ 3 TeV muon collider as functions of the portal matter mass $M_{L}^+$, with our canonical selection of benchmark values  $M_{L}^- = 1.3 M_{L}^+$, $M_{E}^+ = 1.5 M_{L}^+$, $M_{E}^- = 1.8 M_{L}^+$. Here we select $M_{h_2}=1.2$ TeV and $M_{h_1}=1.5$ TeV following the selection rule in Eq.(~\ref{eq:B-h1h2selectionrule}) to ensure that the selection in parameter space is physical, and select $\theta_D=\pi/4$, which, combined with the above-mentioned bound, also ensures that the $\Phi$ vev angle $\theta_\Phi$ appearing in Eq.~(\ref{eq:vevs}) is real. The available parameter range for $e_D$ will then also be governed by the same bound.
The main channel for this process is $ \sigma_{\mu^+ \mu^- \rightarrow h_4 h_4}$. We thus plot the combined heavy scalar production cross section $\sigma_{h0}$ as a function of $M_L^+$ with different choice of $M_{h_4}$. There is a decrease in $\sigma_{h_0}$ when we scale up $M_{h_4}$, but overall these heavy scalars can be copiously produced in a multi-TeV muon collider if they are kinematically allowed. On the right panel of Fig~\ref{fig-B-h0-xsections}, we plot the combined total and individual production cross section $\sigma_{W } = \sigma_{\mu^+ \mu^- \rightarrow W_l^+W_l^-} + \sigma_{\mu^+ \mu^- \rightarrow  W_h^+W_h^-}+ \sigma_{\mu^+ \mu^- \rightarrow  W_h^+W_l^-}+ \sigma_{\mu^+ \mu^- \rightarrow  W_l^+W_h^-}$, with the benchmark values for heavy vector-like leptons and constant heavy gauge bosons and scalars masses. The main contribution comes from $\sigma_{\mu^+ \mu^- \rightarrow  W_h^+W_h^-}$, due to the choice of $\theta_D=\pi/4$. Fig~\ref{fig-B-WZ-xsections-thetaD} shows that the main contributing channel for $\sigma_W$ actually changes with respect to $\theta_D$. For $\theta_D=0.7$, $\sigma_{W_h^+,W_h^-}$ dominates the total $\sigma_W$ whereas for $\theta_D=0.8$, $\sigma_{W_l^+,W_l^-}$ dominates the total $\sigma_W$. The contributions of $\sigma_{\mu^+ \mu^- \rightarrow  W_l^+W_h^-}$ and $ \sigma_{\mu^+ \mu^- \rightarrow  W_h^+W_l^-}$ are identical, emerging in the figure as two overlapping curves, since their squared amplitudes differ by only $t\leftrightarrow u$.

In Fig~\ref{fig-B-Z-xsections}, we show the combined total and individual production cross sections of $Z_D$, wih $\sigma_{Z_D } = \sigma_{\mu^+ \mu^- \rightarrow Z_D Z_D} + \sigma_{\mu^+ \mu^- \rightarrow  Z_D h_1}+ \sigma_{\mu^+ \mu^- \rightarrow  Z_D h_2}$, as a function of $M_{L}^+$, again assuming benchmark values for masses of other heavy vector-like leptons. For the masses of heavy gauge bosons and scalars, we make two different assumptions. In the right panel of Fig~\ref{fig-B-Z-xsections}, we set the values of the heavy gauge bosons and scalars masses to be constants, and observe an increase in $ \sigma_{\mu^+ \mu^- \rightarrow  Z_D h_1}$ and $ \sigma_{\mu^+ \mu^- \rightarrow  Z_D h_2}$  as a function of $M_{L}^+$. This is because the masses of all heavy particles (scalars, vector-like leptons and gauge bosons) scale with $v_\Phi$, the $O(\textrm{TeV})$ vev of the bidoublet $\Phi$. When we define our input parameters, $v_\Phi$ is replaced with $M_{Z_D}$ and other proportional constants and angles. Therefore, without assuming masses of heavy scalars scale with $M_{L}^+$, we are actually assuming that the Yukawa couplings $y_L$ and $y_E$, which govern the strength of the couplings of the muon to $h_1$ and $h_2$, increase when $M_{L}^+$ increases, following the relations in Eq.(~\ref{eq:m-plus-minus}). This running of Yukawa couplings, though resulting in unexpected increase of $ \sigma_{\mu^+ \mu^- \rightarrow  Z_D h_1}$ and $ \sigma_{\mu^+ \mu^- \rightarrow  Z_D h_2}$ as $M_L^+$ increases, still respects the perturbative bound for this mass range of $M_{L}^+$. On the left panel of Fig~\ref{fig-B-Z-xsections}, we fix the ratio of various heavy gauge bosons and scalars masses to $M_{L}^+$. Now, as expected, all channels of $\sigma_{Z_D}$ decrease with respect to the increase of $M_{L}^+$. However, we can not show $\sigma_{Z_D}$ over a larger range of $M_{L}^+$, as the masses of heavy gauge bosons and scalars become too heavy very easily, and will quickly violate the bound of Eq.(~\ref{eq:B-h1h2selectionrule}). $\sigma_{Z_D}$ also depends non-trivially on the choice of $\theta_D$. In Fig~\ref{fig-B-WZ-xsections-thetaD}, we plot individual cross sections with different choice of $\theta_D$, assuming benchmark values for vector-like leptons masses, and constant heavy gauge bosons and scalars masses. Numerically, we find that larger $\theta_D$ results in larger $\sigma_{Z_D,Z_D}$ and $\sigma_{Z_D,h_2}$, and vice versa for $\sigma_{Z_D,h_1}$-- there is unfortunately no immediately satisfactory intuitive argument that this relation should be so, but the relationship can be ultimately observed analytically by inspection of the cross sections in Appendix \ref{Bappendix}.

Ultimately, we see that a TeV-scale muon collider with $O(\textrm{ab}^{-1})$ integrated luminosity will readily produce thousands, if not hundreds of thousands, of potentially visible exotic diboson pairs in this model if the relevant particles are kinematically accessible. In particular, because the results scale quartically with the dark coupling $e_D/M_{Z_D}$, which enters into the $g-2$ calculation, searches for these particles have the potential to significantly constrain the model's ability to address the magnetic moment anomaly. Furthermore, if $M_{Z_D}$ can be directly measured, possibly in its dimuon decay channel, constraints on the coupling $e_D$ from physical consistency requirements with other model parameters can be determined. Because $e_D$ directly affects the coupling strength of our conjectured dark matter candidate to the dark photon mediator, such constraints have the potential to have excellent complementarity with dark sector constraints from the relic abundance and direct and indirect detection.

\subsection{Precision Constraints}\label{sec:precision-constraints}

In this section, we discuss the precision constraints on this model. We first note that precision electroweak observables do not provide significant constraints on this model, similar to the case of the minimal portal matter construction of \cite{Rizzo:2018vlb}. As the new fermions are all vector-like with respect to the SM group, their loop-level effects on $S$ and $T$ are heavily suppressed. The only other potential source of corrections to the oblique parameters in this model is the kinetic mixing between the $Z$ boson and the dark sector gauge bosons, $A_D$ and $Z_D$.\footnote{The $Z$ boson can also mix with the $W_{h,l}$ bosons, however due to approximate $U(1)_D$ charge conservation this mixing is highly suppressed. If we were to relax our assumption that the SM Higgs sector is not coupled to the dark scalar sector, nontrivial mass mixing between the $Z$ boson and the scalar sector would emerge from these mixed terms in the scalar potential. However, because these couplings are under harsh phenomenological constraints, the effects of these couplings will be quite similar to those of the kinetic mixing-induced effect.} Oblique corrections stemming from kinetic mixing scale quadratically with the kinetic mixing coefficient, however, and as such generally only constrain kinetic mixing of $O(10^{-2})$ \cite{Zhang:2022nnh,Curtin:2014cca}. We can readily compute the kinetic mixing coefficients of both the $A_D$ and the $Z_D$ with $Z$, which due to the fact that the dark group is semisimple, will be finite and calculable. At the $Z$ pole (the relevant scale for the precision electroweak measurements), we arrive at kinetic mixing coefficients of
\begin{align}
    \epsilon_{Z-A_D} &= \frac{e_D e}{6 \pi^2} \frac{s_w}{c_w} \log \bigg( \frac{M_L^+ M_E^+}{M_L^- M_E^-} \bigg),\\
    \epsilon_{Z-Z_D} &= \frac{e_D e}{12 \pi^2 \sin(2 \theta_D)} \frac{s_w}{c_w} \bigg[\frac{{M_L^+}^2 - {M_L^-}^2}{{M_L^+}^2 + {M_L^-}^2} \bigg( \frac{5}{6}+ \log \frac{M_L^0}{m_Z} \bigg) + (1 - 2 \cos(2 \theta_D)) \log \frac{M_L^+}{M_L^-} + (L \rightarrow E) \bigg], \nonumber
\end{align}
to leading order in $m_Z^2/{M_{L,E}^\pm}^2$ and $m_{\tau,\mu}^2/m_{Z}^2$. Following the analysis of \cite{Bauer:2022nwt}, because $Z_D$ interacts at leading order with SM fields (the $\tau$ or $\mu$, depending on whether we consider Scenario A or Scenario B, and its corresponding neutrino), we have included the effects of both the kinetic mixing of the $Z_D$ with the $U(1)_Y$ \emph{and} $SU(2)_L$ gauge fields. Given that the coefficient $e s_w/(12 \pi^2 c_w) \approx 1.4 \times 10^{-3}$, it is trivial to see that both kinetic mixing coefficients trivially satisfy precision electroweak constraints from, \eg, \cite{Curtin:2014cca}, as long as $M_{Z_D} - m_Z \gsim 10 \; \textrm{GeV}$. Since it is unlikely that $Z_D$ would be so light while the portal matter (which presumably has masses of the same magnitude) would remain undetected, we can reasonably conclude that precision electroweak observables play no serious role in constraining the parameter space.

\begin{figure}[t!]
    \includegraphics[width=4in]{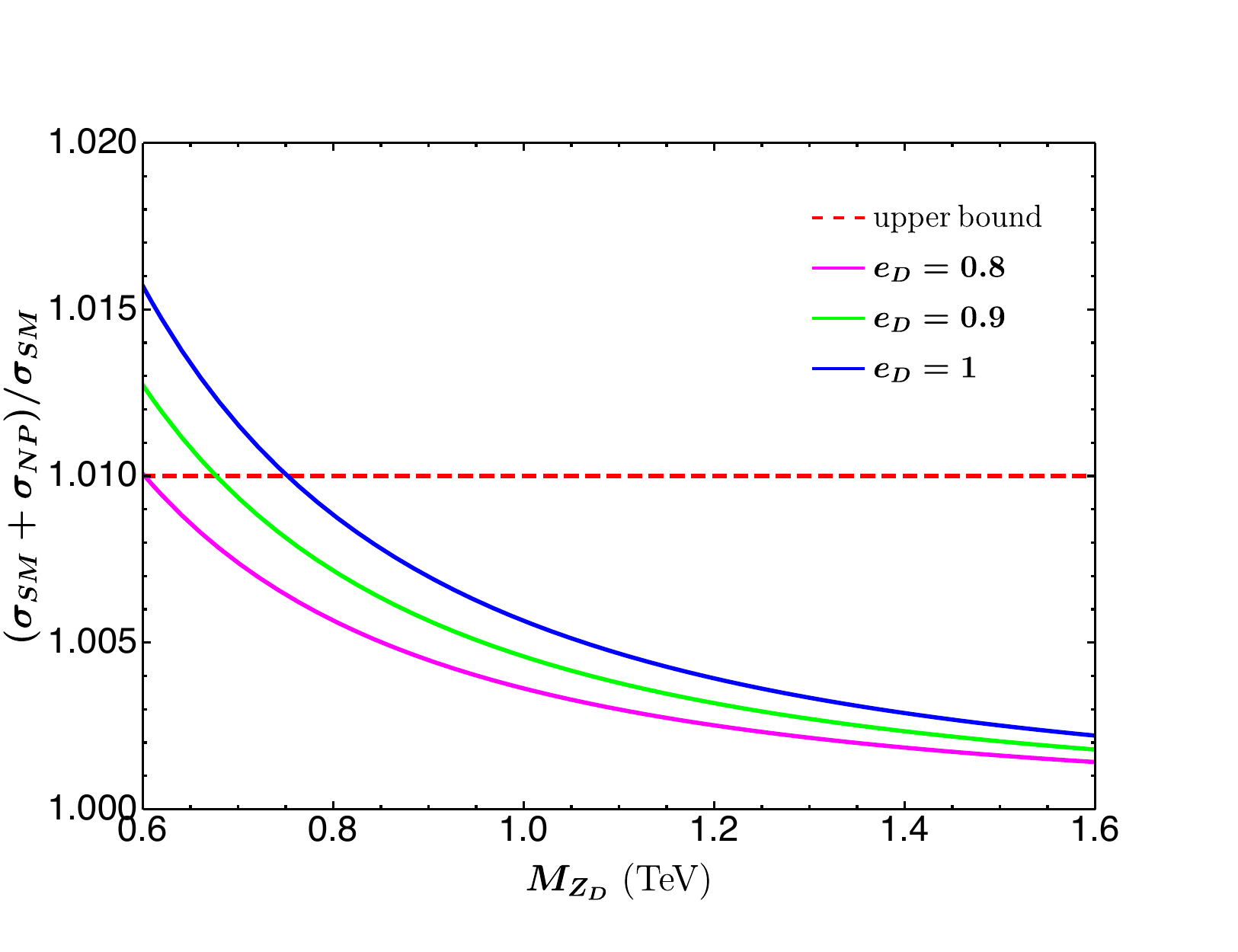}
    \centering
    \caption{The ratio $ \sigma_{SM+NP}/\sigma_{SM}$ that emerges in Scenario B from neutrino trident production. The experimental upper bound is obtained from averaging CHARM-II, CCFR and NuTEV.
    }
    \label{fig-B-neutrino-trident}
\end{figure}

In Scenario B, the fact that $Z_D$ couples to muons, rather than $\tau$'s, requires us to consider further precision constraints, emerging from four-lepton operators featuring muons and muon neutrinos. The strongest of these constraints comes from the neutrino trident production process $\nu_\mu N \rightarrow \nu_\mu \mu^+ \mu^- N$. Following the results of \cite{Buras:2021btx} for chiral $Z_D$ couplings, we have
\begin{equation}
    \frac{\sigma_{SM+NP}}{\sigma_{SM}}=1+ 8 \frac{(1+4s_W^2)\frac{g^L_{\mu\mu} (g^L_{\mu\mu}+g^R_{\mu\mu}) }{g_2^2} \frac{m_W^2}{M_{Z_D}^2}- \frac{g^L_{\mu\mu} (g^L_{\mu\mu}-g^R_{\mu\mu}) }{g_2^2} \frac{m_W^2}{M_{Z_D}^2}}{1+(1+4s_W^2)^2},
\end{equation}
where $s_W$, $m_W$ and $g_2$ are the usual Standard Model parameters, and $g^{L,R}_{\mu\mu}$ describe left and right $\Bar{\mu}\mu Z_D$ couplings. In Scenario B, they take the forms
\begin{align}
    g^L_{\mu\mu}=\frac{e_D ({M_{L}^-}^2-{M_{L}^+}^2) }{\sin{2\theta_D}({M_{L}^-}^2+{M_{L}^+}^2))}, \qquad
    g^R_{\mu\mu}=\frac{e_D ({M_{E}^-}^2-{M_{E}^+}^2) }{\sin{2\theta_D}({M_{E}^-}^2+{M_{E}^+}^2)}, 
\end{align}
Average results from CHARM-II \cite{CHARM-II:1990dvf}, CCFR \cite{CCFR:1991lpl} and NuTEV \cite{nutevcollaboration1998neutrino} places a bound on this ratio
\begin{equation}
     \frac{\sigma_{SM}\sigma_{NP}}{\sigma_{SM}}=0.83\pm0.18
\end{equation}
We plot this ratio as a function of $Z_D$ mass in Fig~\ref{fig-B-neutrino-trident}, for 3 different values of $e_D$. We find that for reasonable ranges of the choice of $e_D$, for $M_{Z_D}>0.8 \,\text{TeV}$, this ratio is always well bounded. For smaller values of $e_D$, the lower bound on $M_{Z_D}$ could be further pushed down to around 0.6 TeV.

Turning now to bounds from flavor violation, we first note that charged lepton flavor constraints also do not provide significant constraints on this model. More precisely, because the dark gauge symmetry breaking preserves a residual $Z_2'$, as discussed in Section \ref{sec:model}, the muon and $\tau$ lepton flavors are to excellent approximation isolated from one another, and charged lepton flavor-violating couplings are so suppressed that constraints, from, \eg, flavor-violating $\tau$ decays are trivially satisfied in both Scenario A and Scenario B. Flavor-violating couplings between SM fermions mediated by the $W_h$ and $W_l$ gauge bosons do exist, but they are heavily suppressed. Note also that since the electron is taken to be a singlet with respect to $G_D$ and $Z_2$ (and thus even with respect to $Z_2^\prime$), charged lepton flavor violation involving the first generation would necessarily be absent at tree level in this model.

That said, we note that if the model is taken literally in the neutrino sector, the preserved $Z_2'$, which  conveniently insulates the model from large flavor-changing neutral currents in the charged lepton sector, also prevents mixing among the $\nu_\tau$ and $\nu_\mu$, in conflict with current best-fits for the parameters of the PMNS matrix (see \eg \cite{ParticleDataGroup:2022pth} and references therein). 
Hence, the $Z_2'$ must not be an exact symmetry, which ultimately means that this model needs to be extended. The $Z_2'$ breaking might be introduced solely into the neutrino sector via, for example, sterile neutrinos with $Z_2'$-violating Majorana mass terms, or the underlying global $Z_2$ we originally introduced into the model might be removed or softly broken in the scalar sector, leading to additional non-trivial constraints on the model from charged lepton flavor-violation searches. As a detailed exploration of the neutrino flavor physics within this general framework is beyond the scope of this paper, we shall defer such an analysis to future work.

\section{Conclusions and Outlook}\label{sec:conclusions}
   
In this work, we have presented an extension to a minimal framework of leptonic portal matter inspired by the model presented in \cite{Wojcik:2022woa} to address the muon magnetic moment anomaly. By augmenting the SM gauge group with a semisimple symmetry that contains a dark $U(1)_D$, in our case $SU(2)^2$ augmented by a global $Z_2$, we are able to accommodate both portal and SM matter as constituents of single multiplets such that any kinetic mixing between Abelian factors of the new dark gauge group and SM gauge fields are finite and calculable. Compared to the minimal construction, this extended dark gauge group presents a far richer parameter space in which the observed correction to the muon magnetic moment can be recreated, including a number of nontrivial physical phase differences (or, in our simplifying assumption that the parameters are real, signs) between parameters, contributions to the magnetic moment involving an extended scalar and gauge sector, and even an implementation of the model in which the muon does not directly mix with the portal matter fields, but still experiences a sizable correction to $g-2$ arising from loops involving both the portal matter fields and new $U(1)_D$-neutral vector-like leptons. 

We have explored other possible experimental signatures of the model, noting that the model presents rich collider phenomenology. In the fermion sector, we found that the extended dark gauge group gives rise to the possibility of non-trivial additional decay channels for the portal matter fields \emph{and} the $U(1)_D$-neutral vector-like leptons, which can lead to atypical experimental signatures for these fields. We have also seen that a multi-TeV muon collider offers excellent prospects to discover the model's new vector-like leptons as well as the TeV-scale scalar and vector boson content arising from the extended dark gauge group. Additionally, the similarity between the Feynman diagrams contributing to the production of the model's new scalars and gauge bosons at such a machine and those diagrams which contribute to the anomalous magnetic moment of the muon make searches for these particles a potentially useful method of constraining or measuring various critical parameters governing the magnitude of the $g-2$ correction.

There are several immediate directions in which this work might be extended. As mentioned in Section \ref{sec:precision-constraints}, implementing a phenomenologically realistic model for neutrinos within this framework presents challenges, including relaxing the $Z_2$ symmetry which offers flavor protection in the charged lepton sector. As a result, a detailed exploration of the feasibility of preserving this $Z_2$ for the charged leptons and violating it in the neutrino sector, or the degree to which recreating realistic neutrino mixing would require explicit $Z_2$ breaking in the charged sector, would be of interest. Additionally, inspired by the idea of interrelating the muon $g-2$ contribution in this framework to the $\tau$ lepton Yukawa coupling and avoiding experimental constraints on new physics coupled to electrons, we have limited the discussion of this lepton flavor symmetry to involve only the second and third generations. Because both the electron and muon anomalous magnetic moments have been measured, a study of the kind given in, \eg \cite{CarcamoHernandez:2019ydc,Li:2021koa} of an analogous framework to the model presented here, featuring a flavor symmetry of the first two lepton generations instead of the last two could be an enlightening additional regime of the model to consider. More broadly, the model we have presented here represents a reasonable ``proof of concept'' in incorporating the muon $g-2$ paradigm of \cite{Wojcik:2022woa} into a more UV complete portal matter construction, and in the process has demonstrated a rich nontrivial phenomenology descending from the extended dark gauge group. In principle a broad number of similar frameworks, perhaps associated with an extended dark gauge group containing other phenomenologically significant BSM symmetries, can be developed.

\section*{Acknowledgements}
This work was supported by the U.S. Department of Energy under the contract number DE-SC0017647.

\appendix
\section{Dark Scalar Masses and Eigenstates}
\label{Sappendix}
In this Appendix, we provide the detailed symmetry breaking analysis for the potential, Eq.~(\ref{eq:scalarpot}). We see that to completely specify this potential, we need $17$ inputs, the $4$ $\mu$ parameters and the 13 $\lambda$ parameters. We will express the dark scalar bidoublet $\Phi$ and the two real dark scalar triplets $\Delta_{A,B}$ as
\begin{equation}
    \Phi=\frac{1}{2}\left(\Re \Phi_0+i \Im \Phi_0 \right)+\sum_{a=x,y,z}\tau_a  \left(\Re \Phi_a + i \Im \Phi_a \right), \qquad \Delta_{A,B}=\sum_{a=x,y,z}\tau_a  \Delta^{a}_{A,B},
\end{equation}
in which $\tau_a\equiv \sigma_a/2$, where $\sigma_a$ are the usual Pauli matrices. 

We will organize the $14$ independent variables in $\Phi$ and $\Delta_{A,B}$ as $ X_{S}=(X_{S}^{B1},X_{S}^{B2},X_{S}^{B3},X_{S}^{B4})$, with
\begin{align}
    X_{S}^{B1}&=(\Re \Phi_0,\Re \Phi_z,\Delta^x_B,\Delta^x_A), \qquad 
    X_{S}^{B2}=(\Im \Phi_0,\Im \Phi_z,\Delta^y_B,\Delta^y_A), \\ X_{S}^{B3}&=(\Delta^z_B,\Delta^z_A,\Re \Phi_x,\Im \Phi_y),
    \qquad 
    X_{S}^{B4}=(\Re \Phi_y,\Im \Phi_x). \nonumber
\end{align}
The system of equations generated upon extremizing the potential can be simplified by noticing that we can use the $6$ rotations in $SU(2)_A \otimes SU(2)_B$ to eliminate $6$ out of the $14$ degrees of freedom. In particular, using this gauge freedom, we can eliminate $2$ complex off-diagonal terms and the imaginary parts of the diagonal of the bi-doublet $\Phi$. This allows us to write the vevs as 
\begin{align}
    \expval{X_{S}^{B1}}&=(\expval{\Re \Phi_0},\expval{\Re \Phi_z},\expval{\Delta^x_B},\expval{\Delta^x_A)}, \qquad 
    \expval{X_{S}^{B2}}=(0,0,\expval{\Delta^y_B},\expval{\Delta^y_A}), \\ \expval{X_{S}^{B3}}&=(\expval{\Delta^z_B},\expval{\Delta^z_A},0,0),
    \qquad 
    \expval{X_{S}^{B4}}=(0,0). \nonumber
\end{align}
We will write the nonzero components of the bi-doublet's vev as
\begin{equation}
\expval{\Re\Phi_0}=v_\Phi(\cos\theta_\Phi+\sin \theta_\Phi)~ \quad \text{and} \quad \expval{\Re\Phi_z}=v_\Phi(\cos\theta_\Phi-\sin\theta_\Phi),
\end{equation}
and the components of the triplets as
\begin{align}
~\expval{\Delta^x_{A}}=r_{\Delta}v_{\Phi}s_{\theta_\Delta}c_{\phi_{A}}s_{\theta_{A}},~\expval{\Delta^y_{A}}=r_{\Delta}v_{\Phi}s_{\theta_\Delta}s_{\phi_{A}}s_{\theta_{A}},~\expval{\Delta^z_{A}}=r_{\Delta}v_{\Phi}s_{\theta_\Delta}c_{\theta_{A}},\\
~\expval{\Delta^x_{B}}=r_{\Delta}v_{\Phi}c_{\theta_\Delta}c_{\phi_{B}}s_{\theta_{B}},~\expval{\Delta^y_{B}}=r_{\Delta}v_{\Phi}c_{\theta_\Delta}s_{\phi_{B}}s_{\theta_{B}},~\expval{\Delta^z_{B}}=r_{\Delta}v_{\Phi}c_{\theta_\Delta}c_{\theta_{B}},\nonumber
\end{align}
in which $0 \leq \theta_\Delta \leq \pi/2$, $v_\Phi,r_{\Delta} \geq 0$, $0 \leq \theta_{\Phi,A,B} \leq \pi$, and $0 \leq \phi_{A,B} \leq 2\pi$. 
Extremizing the potential yields three inequivalent families of vacua, all of which are CP preserving:
\begin{itemize}
    \item Family (a): $r_{\Delta}=0$ (preserved dark subgroup $U(1)_D \otimes Z_2$).
    \item Family (b): $r_{\Delta} \neq 0$, $0 < \theta_\Delta <\pi/2$, $\theta_{A,B}=0$  (preserved dark subgroup $U(1)_D$).
    \item Family (c): $r_{\Delta} \neq 0$,  $0 < \theta_\Delta <\pi/2$, $\theta_{A,B}=\pi/2$ 
    (preserved dark subgroup $Z'_2$). This case is  true for an arbitrary $\phi_{A}=\phi_{B}$. We can then, for simplicity, take $\phi_A=\phi_B=0$.
\end{itemize}
In the above, we note that $U(1)_D$ is the group formed by $z$-rotations in the gauge space $G_{D}$, i.e., the group formed by the transformations
\begin{equation}
    D_{A}(\hat{z},\phi)\otimes D_{B}(\hat{z},\phi),
\end{equation}
and $Z'_{2}$ is the group generated by the transformation
\begin{equation}
    D_{A}(\hat{z},\pi)\otimes D_{B}(\hat{z},\pi) \otimes Z_2.
\end{equation}
To break the dark group $G_D$ completely, we now see the need for the two triplets $\Delta_{A,B}$: if either one is zero, vacuum (a) is the only option and there will always be a preserved $U(1)_D$ subgroup. 

As described earlier in this work, the symmetry breaking pattern of interest is composed of two steps: The bi-doublet $\Phi$ acquires a vev, breaking $G_D$ to $U(1)_D$, which is  subsequently broken by the (smaller) vevs of the triplets $\Delta_{A,B}$. To achieve this symmetry-breaking pattern, we need (i) $V(a)>V(b)$ and (ii) $V(b) >V(c)$.
Condition (i) can be satisfied by imposing
\begin{equation}
     (\mu^2_{3}(b)+\mu^2_{4}(b))<(\mu^2_{3}(b)-\mu^2_{4}(b))c_{2\theta_{\Delta}}.
\end{equation}
Furthermore, for the vacua families (b) and (c) we can also exchange the  $\mu_{3},\mu_{4}$ for $\theta_{\Delta},r_{\Delta}$. In that case, we see that
\begin{equation}
     V(b)-V(a)=\frac{v_{\Phi}^2}{8}r^2_{\Delta}(\lambda_8-\lambda_9)s_{2\theta_\Delta}(s_{2\theta_\Phi}-1),
\end{equation}
and hence condition (ii) can be easily satisfied by requiring $\theta_\Phi \neq \pi/4$ and $\lambda_8>\lambda_9$.

Once the desired symmetry-breaking pattern, which corresponds to vacuum family (c), has been obtained, we can analyze issues of vacuum stability.
The extremum condition yields the following relations:
\begin{align}
    \mu_1^2v_{\Phi}^2&=2\lambda_1+\lambda_3 \sin(2\theta_\Phi)+\frac{r_\Delta^2}{4}\left[(\lambda_5+\lambda_{11})+(\lambda_5-\lambda_{11})c_{2\theta_\Delta}+\frac{\lambda_9}{2}\sin(2\theta_\Delta) \right], \nonumber\\
    \mu_2^2 v_{\Phi}^2&=\lambda_3+2(\lambda_2+\lambda_7)\sin(2\theta_\Phi)+\frac{r_\Delta^2}{4}\left[(\lambda_6+\lambda_{12})+(\lambda_6-\lambda_{12})\cos(2\theta_ \Delta)+\frac{\lambda_8}{2}\sin(2\theta_\Delta) \right], \\
    \mu_3^2v_{\Phi}^2&=\lambda_5+\lambda_6 \sin(2\theta_\Phi)+\frac{1}{4}\tan(\theta_\Delta)\left[\sin(2\theta_\Phi) \lambda_8+\lambda_9\right] +\frac{r^2_{\Delta}}{2}\left[\lambda_4\cos^2(\theta_\Delta)+\lambda_{13}\sin(\theta_\Delta)\right], \nonumber\\
    \mu_4^2v_{\Phi}^2&=\lambda_{11}+\lambda_{12}\sin(2\theta_\Phi)+\frac{\cot(\theta_\Delta)}{4}\left(\sin(2\theta_\Phi) \lambda_8+\lambda_9\right) +\frac{r^2_{\Delta}}{4}\left[(\lambda_{10}+\lambda_{13})+(\lambda_{13}-\lambda_{13})\cos(2\theta_\Delta)\right].\nonumber
\end{align}
The partition of the scalar sector into the four blocks $X^{B1}_{S},~X^{B2}_{S},~X^{B3}_{S},~X^{B4}_{S}$ is now justified, as each block has a well defined and unique charge under CP and $Z'_2$.  Thus, in this basis, the scalar mass-squared matrix $M^2_S$ is composed of four separate blocks: $\left(M^2_S\right)^{(a)}$, with $a=1,..,4$.
The block $a=4$ is already diagonal, as it is a null matrix. Before diagonalizing the remaining blocks ($a=1,2,3$), we will define some new variables for simplicity. First, we will exchange the inputs $(\lambda_8,\lambda_9)$ for $(M_{h_{\pm}},\phi_\lambda)$ via
\begin{equation}
    \lambda_8\pm \lambda_9=\mp 2\left(\frac{M_{h_{\pm}}}{v_\Phi} \right)^2 \frac{\sin 2(\theta_\Delta \pm \phi_\lambda)}{\cos 2\phi_\lambda (1 \pm \sin 2\theta_\Phi)}.
\end{equation}
The input variables $(\lambda_1,\lambda_3,\lambda_2,\lambda_7)$ can also be exchanged in favor of $(r_1,r_2,r_4,\theta_\lambda)$ via
\begin{align}\label{eq:theta-lambda}
\lambda_1&=\frac{1}{8\cos^2 2\theta_\phi}\left(\frac{M_{h^\pm}}{v_\Phi} \right)^2\left[\Delta r^2_{1+2}+\Delta r^2_{1-2}\sin 2(\theta_\lambda-\theta_\Phi) \right], \nonumber\\
\lambda_2+\lambda_7&=\frac{1}{8\cos^2 2\theta_\phi}\left(\frac{M_{h^\pm}}{v_\Phi} \right)^2\left[\Delta r^2_{1+2}-\Delta r^2_{1-2}\sin 2(\theta_\lambda+\theta_\Phi) \right], \\
\lambda_3&=\frac{1}{4\cos^2{2\theta_\phi}}\left(\frac{M_{h^\pm}}{v_\Phi} \right)^2\left[\Delta r^2_{1-2}\cos {2 \theta_\lambda}-\Delta r^2_{1+2} \sin {2\theta_\Phi} \right], \nonumber\\
\lambda_2-\lambda_7&=-\left(\frac{M_{h^\pm}}{v_\Phi}\right)^2\frac{r_4^2}{4},\nonumber
\end{align}
in which $\Delta r^2_{1\pm2}=r^2_1 \pm r^2_2$. Finally, we can exchange the inputs $(\lambda_5,\lambda_6)$ for $(\lambda_\alpha,\theta_\alpha)$, $(\lambda_{11},\lambda_{12})$ for $(\lambda_\beta,\theta_\beta)$, and $\lambda_4$ for $\tilde{M}^3$.  
We then have for the first block, $(M_S^2)^{(1)},$
\begin{equation} (M^{2}_{S})^{(1)}_{\text{Diag}}=M^2_{h^{\pm}}\mqty(\dmat[0]{r_1^2, r_2^2, r_{\Delta}^2 r^2_3,1}),
\end{equation}
corresponding to mass eigenstates $h_1$, $h_2$, $h_3$, and $h^1_{deg}$, with CP and $Z_2^\prime$ charges of $(+1,+1)$. The second block $(M_S^2)^{(2)},$  takes the form
\begin{equation}
(M^{2}_{S})^{(2)}_{\text{Diag}}=M^2_{h^{\pm}}\mqty(\dmat[0]{1,r_4^2,0,0}),
\end{equation}
consisting of the eigenstates $h^2_{deg}$ and $h_4$, with CP and $Z_2^\prime$ charges $(-1,+1)$. The third block $(M_S^2)^{(3)}$ is
\begin{equation}
(M^{2}_{S})^{(3)}_{\text{Diag}}=M^2_{h^{\pm}}\mqty(\dmat[0]{\cos^2 \theta_M,\sin^2 \theta_M ,0,0}),
\end{equation}
corresponding to the eigenstates $h_5$ and $h_6$, which have CP and $Z_2^\prime$ charges of $(+1,-1)$. In the above, the angle $\theta_M$ is given by  $c_{2\theta_M}=c_{2\theta_\Delta}/c_{2\phi_\lambda}$, and so we will exchange  $\phi_\lambda$ in favor of it. However, in order for the angle $\theta_M$ to be real, we must have $|c_{2 \theta_M}|>|c_{2 \theta_\Delta}|$ (so for example, if we require $\theta_M=\pi/4$, then the only allowable value is $\theta_\Delta = \pi/4$).

More explicitly, the scalar mass eigenstates consist of the six real scalar fields
\begin{align}\label{eq:scalar-eigenstates}
    h_1 &=\cos \theta_\lambda \left (\Re \Phi_0-\expval{\Re \Phi_0} \right)+\sin \theta_\lambda (\Re \Phi_z - \expval{\Re\Phi_z})+O(r_\Delta), 
    \nonumber
    \\
    h_2 &=-\sin \theta_\lambda \left (\Re \Phi_0-\expval{\Re \Phi_0} \right)+\cos \theta_\lambda (\Re \Phi_z - \expval{\Re\Phi_z})+O(r_\Delta),
\nonumber
    \\
    h_3&=\cos \theta_\Delta \left (\Delta^x_B -\expval{\Delta^x_B} \right)+\sin \theta_\Delta  \left (\Delta^x_A -\expval{\Delta^x_A} \right)+O(r_\Delta),    \\
    h_4&=-\sin \left( \theta_\Phi + \pi/4\right) \Delta^y_B+ \cos \left(\theta_\Phi + \pi/4 \right)\Delta^y_A  +O(r_\Delta),
    \nonumber
    \\
    h_5&=-\sin \phi_\lambda \Delta^z_B  +\cos \phi_\lambda \Delta^z_A +O(r_\Delta), 
    \nonumber
    \\
    h_6&=\cos \phi_\lambda \Delta^z_B + \sin \phi_\lambda \Delta^z_A +O(r_\Delta), \quad \quad \quad  \quad \quad \quad \quad\, \nonumber
\end{align}
with masses $M_{h_{1,2,4}}=r_{1,2,4} M_{h^\pm}$,  $m_{h_3}=r_\Delta r_3 M_{h^\pm}$, $M_{h_5}=\cos\theta_M M_{h^\pm}$, and $M_{h_6}=\sin\theta_M M_{h^\pm}$, 
and one complex scalar field
\begin{equation}
    h^\pm=-\frac{1}{\sqrt{2}}\cos \theta_\Delta \left(\Delta^x_A-\Delta^y_A - \expval{\Delta^x_A}  \right)+\frac{1}{\sqrt{2}}\sin \theta_\Delta\left(\Delta^x_B-\Delta^y_B - \expval{\Delta^x_B}  \right)+O(r_\Delta),
\end{equation}
with mass $M_{h^\pm}$. We see that, in this notation, the requirement for vacuum family (c) to be a relative minimum is equivalent to imposing positive scalar masses, as expected.

\section{Diboson Cross Sections in Scenario B}\label{Bappendix}
Here, we reproduce the full set of squared amplitudes for the diboson production processes considered in Section \ref{sec:diboson-B}. In keeping with the notation of Section \ref{sec:diboson-A}, we shall denote each cross section by as $|\mathcal{M}_{XY}|$, where $X, Y$ shall denote the two bosons being produced. We shall group these cross sections into three categories: Those which feature only the scalar bosons $h_1$, $h_2$, and $h_4$, those which feature the dark gauge boson $Z_D$, and finally those which feature the dark gauge bosons $W_{h,l}^\pm$. In the first category, we have
\begin{align}
    |\mathcal{M}_{h_1h_1}|^2&= \frac{e_D^4 (1-S_h)^2}{256 M_{Z_D}^4 s_{2D}^4 c_{2\phi}^4}\left(\frac{({M^-_L}^2- {M^+_L}^2 )^4 (t-u)^2 (tu-M_{h_1}^4)}{ {M^0_L}^4 ({M^0_L}^2 - t)^2 ({M^0_L}^2  -2u)^2 }+ (L \rightarrow E) \right),\nonumber
    \\
     |\mathcal{M}_{h_2h_2}|^2&= \frac{e_D^4 (1+ S_h)^2}{256 M_{Z_D}^4 s_{2D}^4 c_{2\phi}^4}\left(\frac{({M^-_L}^2- {M^+_L}^2 )^4 (t-u)^2 (tu-M_{h_2}^4)}{{M^0_L}^4 ({M^0_L}^2 - t)^2 ({M^0_L}^2  -2u)^2 }+ (L \rightarrow E) \right),
     \\ 
     |\mathcal{M}_{h_4h_4}|^2&= \frac{e_D^4 }{4 M_{Z_D}^4 s_{2D}^4 c_{2\phi}^4}\left(\frac{{M^0_L}^4 (t-u)^2 (tu-M_{h_4}^4)}{ ({M^0_L}^2 - t)^2 ({M^0_L}^2 - u)^2 }+ (L \rightarrow E) \right), \nonumber
     \\
    |\mathcal{M}_{h_1h_2}|^2&= \frac{e_D^4 (1- S_h^2)}{256M_{Z_D}^4 s_{2D}^4 c_{2\phi}^4}\left(\frac{({M^-_L}^2- {M^+_L}^2 )^4 (t-u)^2 (tu-M_{h_1}^2M_{h_2}^2)}{ {M^0_L}^4 ({M^0_L}^2 - t)^2 ({M^0_L}^2 - u)^2 }+ (L \rightarrow E) \right), \nonumber
\end{align}
where we follow the notation outlined in Eq.~(\ref{eq:B-g-2-3}).
Squared amplitudes for processes featuring final-state $Z_D$ are given by
\begin{align}
  |\mathcal{M}_{Z_DZ_D}|^2&=\frac{ e_D^4 {M^-_L}^4 {M^+_L}^4 }{4 M_{Z_D}^4 s_{2D}^4 {M^0_L}^8  }\Bigg(\frac{4 C_{Z_DZ_D,1}}{({M^0_L}^2 - t)^2} + \frac{4 C_{Z_DZ_D,2}}{({M^0_L}^2 - u)^2} + \frac{C_{Z_DZ_D,3}}{({M^0_L}^2 - u)^2({M^0_L}^2 - t)^2}  \Bigg)+ (L \rightarrow E),\nonumber\\
    |\mathcal{M}_{Z_Dh_1}|^2&=\frac{e_D^4 C_{\lambda\phi-}^2 {M^-_L}^2 {M^+_L}^2 ({M^-_L}^2- {M^+_L}^2 )^2  (2{M^0_L}^2-t-u )^2 C_{Z_Dh_1,1}}{32 M_{Z_D}^4 s_{2D}^4 c_{2\phi}^4 {M^0_L}^8 ({M^0_L}^2 - t)^2 ({M^0_L}^2 - u)^2}+ (L \rightarrow E),
    \\
   |\mathcal{M}_{Z_Dh_2}|^2&=\frac{e_D^4 C_{\lambda\phi+}^2 {M^-_L}^2 {M^+_L}^2 ({M^-_L}^2- {M^+_L}^2 )^2  (2{M^0_L}^2-t-u )^2 C_{Z_Dh_2,1}}{32 M_{Z_D}^4 s_{2D}^4 c_{2\phi}^4 {M^0_L}^8 ({M^0_L}^2 - t)^2 ({M^0_L}^2 - u)^2}+ (L \rightarrow E),\nonumber
\end{align}
where
\begin{align}
     C_{Z_DZ_D,1}&=-\frac{1}{4}\Bigl(4M_{Z_D}^8 -t^3u + 4 t^2 (t+u) M_{Z_D}^2 - t(7t+4u)M_{Z_D}^4 \Bigr), \nonumber\\ 
    C_{Z_DZ_D,2}&=-\frac{1}{4}\Bigl(4M_{Z_D}^8 -tu^3 + 4 u^2 (t+u) M_{Z_D}^2 - u(7u+4t)M_{Z_D}^4 \Bigr) \nonumber\\
    C_{Z_DZ_D,3}&=\frac{1}{2}\Bigl(16M_{Z_D}^8 -8(t+u) M_{Z_D}^6 - 7tuM_{Z_D}^4+ 4tu(t+u)M_{Z_D}^2 -t^2u^2 \Bigr)\\
    C_{Z_Dh_1,1}&=2M_{Z_D}^4+M_{h_1}^2M_{Z_D}^2 -2(t+u)M_{Z_D}^2 +tu, \nonumber\\
    C_{Z_Dh_2,1}&=2M_{Z_D}^4+M_{h_2}^2M_{Z_D}^2 -2(t+u)M_{Z_D}^2 +tu. \nonumber
\end{align}
Finally, the squared amplitudes for final states featuring $W_{h,l}^\pm$ are given by
\begin{align}
    |\mathcal{M}_{W_l^+ W_l^-}|^2&= -\frac{e_D^4}{128 M_{Z_D}^4 s_{lh}^4 {M_L^0}^4} \bigg( \frac{2 C_{W_l^+ W_l^-,1}}{({M_L^+}^2 - t)^2} + \frac{2 C_{W_l^+ W_l^-,2}}{({M_L^-}^2 - u)^2} + \frac{C_{W_l^+ W_l^-,3}}{({M_L^+}^2 - t)({M_L^+}^2 - u)}\bigg) + (L \rightarrow E)\nonumber\\
    |\mathcal{M}_{W_h^+ W_h^-}|^2&= -\frac{e_D^4}{128 M_{Z_D}^4 c_{lh}^4 {M_L^0}^4} \bigg( \frac{2 C_{W_h^+ W_h^-,1}}{({M_L^+}^2 - t)^2} + \frac{2 C_{W_h^+ W_h^-,2}}{({M_L^-}^2 - u)^2} + \frac{C_{W_h^+ W_h^-,3}}{({M_L^+}^2 - t)({M_L^+}^2 - u)}\bigg) + (L \rightarrow E)\\
    |\mathcal{M}_{W_h^+ W_l^-}|^2&= -\frac{e_D^4}{64 M_{Z_D}^4  c_{lh}^2 s_{lh}^2 {M_L^0}^4} \bigg( \frac{ C_{W_h^+ W_l^-,1}}{({M^+_L}^2-u)^2} + \frac{2C_{W_h^+ W_l^-,2}}{({M^-_L}^2-t)({M^+_L}^2-u)} + \frac{C_{W_h^+ W_l^-,3}}{({M^-_L}^2-t)^2} \bigg) + (L \rightarrow E)
  \nonumber\\
  |\mathcal{M}_{W_l^+ W_h^-}|^2&=|\mathcal{M}_{W_h^+ W_l^-}|^2 \Big|_{t\leftrightarrow u}, \nonumber
\end{align}
where
\begin{align}
    C_{W_l^+ W_l^-,1}&= \Bigl(4s_{lh}^8 M_{Z_D}^8-t(7t+4u)s_{lh}^4 M_{Z_D}^4 +4t^2(t+u)s_{lh}^2 M_{Z_D}^2 -t^3u\Bigr) \Big(H^-_{-1,-1}(M_{L}^+,M_{L}^-)\Big)^4,
\nonumber\\
C_{W_l^+ W_l^-,2}&= \Bigl(4s_{lh}^8 M_{Z_D}^8-u(7u+4t)s_{lh}^4 M_{Z_D}^4 +4u^2(t+u)s_{lh}^2 M_{Z_D}^2 -tu^3\Bigr) \Big(H^-_{-1,-1}(M_{L}^-,M_{L}^+)\Big)^4,
\nonumber\\
C_{W_l^+ W_l^-,3}&= \Bigl(16 s_{lh}^8 M_{Z_D}^8 -8(t+u) s_{lh}^6 M_{Z_D}^6-7tu s_{lh}^4 M_{Z_D}^4 +4tu(t+u) s_{lh}^2 M_{Z_D}^2 -t^2u^2  \Bigr)C_{W_l^+ W_l^-,3}^\prime ,
\nonumber\\
C_{W_l^+ W_l^-,3}^\prime &= \left(\frac{M_{L}^+M_{L}^- (c_{2\pm}+4)}{s_{2D}^2}  - \frac{1}{c_D }\left( \frac{2{M_L^0}^2 s_{2\pm} }{s_D}+M_{L}^+M_{L}^- \left(\frac{c_\pm^2}{c_D} - \frac{s_\pm^2}{c_D} \right)  \right) \right)^2,
\nonumber\\
C_{W_h^+ W_h^-,1}&= \Bigl(4c_{lh}^8 M_{Z_D}^8-t(7t+4u)c_{lh}^4 M_{Z_D}^4 +4t^2(t+u)c_{lh}^2 M_{Z_D}^2 -t^3u\Bigr)  \Big(J^+_{-1,-1}(M_{L}^-,M_{L}^+)\Big)^4 ,
\nonumber\\
C_{W_h^+ W_h^-,2}&= \Bigl(4c_{lh}^8 M_{Z_D}^8-u(7u+4t)c_{lh}^4 M_{Z_D}^4 +4u^2(t+u)c_{lh}^2 M_{Z_D}^2 -tu^3\Bigr)  \Big(J^+_{-1,-1}(M_{L}^+,M_{L}^-)\Big)^4 
\nonumber\\
C_{W_h^+ W_h^-,3}&= \Bigl(16 c_{lh}^8 M_{Z_D}^8 -8(t+u) c_{lh}^6 M_{Z_D}^6-7tu c_{lh}^4 M_{Z_D}^4 +4tu(t+u) c_{lh}^2 M_{Z_D}^2 -t^2u^2  \Bigr)C_{W_h^+ W_h^-,3}^\prime 
\nonumber\\
C_{W_h^+ W_h^-,3}^\prime &= \left(\frac{M_{L}^+M_{L}^- (-c_{2\pm}+4)}{s_{2D}^2} + \frac{1}{c_D }\left( \frac{2{M_L^0}^2 s_{2\pm} }{s_D}+M_{L}^+M_{L}^- \left(\frac{c_\pm^2}{c_D} - \frac{s_\pm^2}{c_D} \right)  \right) \right)^2,
 \\
C_{W_h^+ W_l^-,1} &=  \frac{(1-c_{8lh})M_{Z_D}^8+ 8u(c_{4lh}-1)M_{Z_D}^6 + \bigl(4t^2(c_{4lh}-1)+ 16tu(3-c_{4lh})\bigr)M_{Z_D}^4  -64t^2 u M_{Z_D}^2 +32t^3u}{32C_{W_h^+ W_l^-,1}^{\prime-1}}  ,
\nonumber\\
C_{W_h^+ W_l^-,1}^\prime &= \Big(J^+_{-1,-1}(M_{L}^-,M_{L}^+)\Big)^2 \Big(H^-_{-1,-1}(M_{L}^+,M_{L}^-)\Big)^2
\nonumber\\
C_{W_h^+ W_l^-,2} &= \frac{(1-c_{4lh}) M_{Z_D}^8+ (t+u)(c_{4lh}-1) M_{Z_D}^6+ tu (9-c_{4lh}) M_{Z_D}^4 -8tu(t+u)  M_{Z_D}^2 -8t^2u^2 }{8C_{W_h^+ W_l^-,2}^{\prime-1}},
\nonumber\\
C_{W_h^+ W_l^-,2}^{\prime}&=H^-_{-1,-1}(M_{L}^+,M_{L}^-)\,H^-_{-1,-1}(M_{L}^-,M_{L}^+)\,J^+_{-1,-1}(M_{L}^-,M_{L}^+)\,J^+_{-1,-1}(M_{L}^+,M_{L}^-)
\nonumber\\
C_{W_h^+ W_l^-,3} &=  \frac{(1-c_{8lh})M_{Z_D}^8+ 8t(c_{4lh}-1)M_{Z_D}^6 + \bigl(4u^2(c_{4lh}-1)+ 16tu(3-c_{4lh})\bigr)M_{Z_D}^4  -64u^2 t M_{Z_D}^2 +32u^3t}{32C_{W_h^+ W_l^-,3}^{\prime-1}}  ,
\nonumber\\
C_{W_h^+ W_l^-,3}^{\prime}&=\Big(J^+_{-1,-1}(M_{L}^+,M_{L}^-) \Big)^2\Big(H^-_{-1,-1}(M_{L}^-,M_{L}^+) \Big)^2 \nonumber
\end{align}
and 
\begin{align}
     H_{a,b}^\pm(A,B)\equiv Ac_\pm s_D^a \pm B s_\pm c_D^b, \\
      J_{a,b}^\pm(A,B)\equiv Ac_\pm c_D^a \pm B s_\pm s_D^b, \nonumber
\end{align}
\clearpage
\setlength{\bibsep}{3pt}
\bibliographystyle{JHEP}
\bibliography{main}

\end{document}